\documentstyle{mn}

   \title[IR mergers and IR QSOs with GWs. III. Mrk\,231]
   {IR mergers and IR QSOs with galactic winds. III. Mrk\,231:\\
   composite outflow with multiple expanding bubbles \\
   and BAL systems (an exploding young QSO?)  \\
   }

\author[L\'{\i}pari et al.]
   {S. L\'{\i}pari$^{1}$, R. Terlevich$^{2,3}$, W. Zheng$^{4}$,
   B. Garcia$^{5}$, S. Sanchez$^{6}$,  M. Bergmann$^{7}$\\  
$^{1}$ C\'ordoba Observatory and CONICET, Laprida 854, 5000 C\'ordoba, Argentina\\
$^{2}$ Institute of Astronomy, Madingley Road, Cambridge CB3 0HA\\
$^{3}$ Instituto Nacional de Astrofisica Optica y Electronica (INAOE), Puebla, Mexico\\
$^{4}$ Department of Physic and Astronomy, Univ. of John Hopkins, Baltimore MD 21218, USA\\
$^{5}$ Instituto de Astrofisica de Canarias, 38205 La Laguna, Tenerife, Canary Island, Spain.\\
$^{6}$ Calar Alto Observatory, C/Jesus Durban Remon 2-2, E-04004 Almeria, Spain\\
$^{7}$ NOAO Gemini Science Center, La Serena, Chile\\
}

\date{Received     ;
      in original form }

\pagerange{\pageref{firstpage}--\pageref{lastpage}}
\pubyear{2005}

\begin{document}

\maketitle

\label{firstpage}

\begin{abstract}
A study of outflow (OF) and BAL systems in Mrk\,231 (and in similar
IR QSOs) is presented.
This study is based mainly on 1D and 2D spectroscopy (obtained
at La Palma/WHT, HST, IUE, ESO/NTT, KPNO, APO and CASLEO observatories)
plus HST images.
For Mrk 231 we report  evidence that the extreme nuclear OF process has at
least 3 main components on different scales, which are probably associated
with: (i) the radio jet, at pc scale; and (ii) the extreme starburst at pc
and kpc scale. This OF  has generated 4 or more
concentric expanding bubbles and the BAL systems.

Specifically, inside and very close to the nucleus the 2D spectra show
the presence of an OF emission bump in the blend H$\alpha$+[N {\sc ii}],
with a peak at the same velocity of the main BAL--I system
($ V_{\rm Ejection BAL-I} \sim$ --4700\,km\,s$^{-1}$).
This bump was more clearly detected in the area located at 0\farcs6--1\farcs5
(490--1220 pc), to the south-west of the nucleus core, showing a strong
and broad peak.
In addition, in the same direction (at PA $\sim$60$^{\circ}$,
i.e. close to the PA of the small scale radio jet) at 1\farcs7--2\farcs5,
we also detected multiple narrow emission line components, with ``greatly"
enhanced [N\,{\sc ii}]/H$\alpha$ ratio (very similar to the spectra of
jets bow shocks). These results suggest that the BAL--I system
is generated in OF clouds associated with the parsec scale jet.

The HST images show 4 (or 5) nuclear bubbles or shells with radius
r $\sim$ 2.9, 1.5, 1.0, 0.6 and 0.2 kpc.
For these bubbles, the 2D H$\alpha$ velocity field (VF) map
and 2D spectra show:
(i) At the border of the more extended bubble (S1), a clear expansion of
the shell with blueshifted velocities (with circular shape
and at a radius r $\sim$ 5\farcs0).
This bubble shows a rupture arc --to the south--  
suggesting that the bubble is in the blowout phase. The axis of this rupture
or ejection (at PA $\sim$00$^{\circ}$) is coincident with the axis of the
intermediate and large scale structures detected at radio wavelengths.
(ii) In addition, in the 3 more external bubbles (S1, S2, S3), the
2D WHT spectra show multiple emission line components with OF velocities, of 
$\langle V_{\rm OF Bubble}\rangle$ S1, S2 and S3 $= [-(650-410) \pm 30],
[-500 \pm 30]$, and $[-230 \pm 30]$\,km\,s$^{-1}$.
(iii) In all the circumnuclear region (1\farcs8 $<$ r $<$ 5$''$), the
[N\,{\sc ii}]/H$\alpha$ and [S {\sc ii}]/H$\alpha$ narrow emission line
ratios show high values ($>$ 0.8), which are consistent with LINER/OF processes
associated with fast velocity shocks. Therefore, we suggest that these
giant bubbles are associated with the large scale nuclear OF
component, which is generated --at least in part-- by the extreme
nuclear starburst: i.e., type II SN explosions.

The variability of the short lived BAL--III Na ID system was studied,
covering almost all the period in which this system appeared (between
$\sim$1984--2004).
We found that the BAL-III light curve (LC) is clearly asymmetric with:
a steep increase, a clear maximum and an exponential fall
(similar to the shape of a SN LC).
The origin of this BAL-III system is discussed, mainly in
the frame work of an {\it extreme explosive event}, probably
associated with giant SNe/hypernova explosions.

Finally, the IR colours diagram and the UV-BAL systems of
IR+GW/OF+Fe {\sc ii} QSOs are analysed.
This study shows a new BAL IR QSO
and suggest/confirm that these objects could be {\it nearby young
BAL QSOs, similar to those detected recently at z $\sim$ 6.0}.
We propose that the {\it phase of young QSO} is associated with: accretion
of large amount of gas (by the SMBH) + extreme starbursts + extreme composite
OFs/BALs.
\end{abstract}


\section{Introduction}\label{introduction}

An important issue in astrophysics and cosmology  is the study of
{\it extreme star formation and galactic winds/out flows, in mergers
and QSOs} and their relation to the  early phases of the formation
of galaxies and AGNs (see for references Lipari et al. 2004a,b,c,d).

\subsection{Luminous IR galaxies, mergers and AGNs}\label{introirg}

Luminous  and ultraluminous IR galaxies (LIRGs:
L$_{IR} \geq 10^{11} L_{\odot}$ and ULIRGs:
L$_{IR} \geq 10^{12} L_{\odot}$, respectively) are
dusty, strong IR emitters where frequently a strong enhancement of star
formation is taking place (for references see L\'{\i}pari et al. 2004a,b,c,d).
Imaging surveys of LIRGs and ULIRGs show that a very
high proportion ( $\sim$70--95$\%$) are mergers or interacting
systems (Joseph \& Wright 1985; Rieke et al. 1985; Sanders et al. 1988a;
Melnick \& Mirabel 1990; Clements et al. 1996).
In luminous IR galaxies there is a clear increase of the nuclear activity
with the increase of the IR luminosity (Sanders et al. 1988a,b; Veilleux
et al. 1999, 2002).

Strong evidence indicates that virtually {\it all luminous IRAS galaxies}
are rich in interstellar gas, which is highly concentrated in their
nuclei (Sanders et al. 1987, 1988a; Sanders, Scoville \& Soifer 1991;
Scoville et al. 1991). Specifically,
there are observational evidence and theoretical works suggesting that in
IR mergers and IR QSOs tidal torque and loss of angular momentum drive large
amount of interstellar gas into the central regions, leading to {\it extreme
starburst processes} and probably fuelling {\it a supermassive black hole}
(see for references Lipari et al. 2004a,d)

An important, common phenomenon associated with the IR galaxies 
is the galactic wind: we found observational evidence for
galactic wind (GW) features from  starbursts and/or AGNs in luminous
IR systems (L\'{\i}pari et al.\ 2004a,b,c,d, 2003, 2000a,b, 1994, 1993).

\subsection{Galactic winds (associated with starbursts and AGNs)}\label{introgw}

Galactic winds and outflows have been observed mainly in starburst and Seyfert
galaxies (see Heckman et al. 2000, 1990; Cecil et al. 2002; Veilleux et al.
2002a, 1993; Lipari et al. 2004a,b,c,d, 2003, 2000b, 1994).
There is substantial theoretical literature about galactic winds associated
with both processes: starbursts (Larson 1974; Ostriker \& Cowie 1981;
Chevalier \& Clegg 1985; Ikeuchi \& Ostriker 1986; Tomisaka \& Ikeuchi 1988;
Norman \& Ikeuchi 1989; Suchkov et al. 1994; Strickland \& Stevens 2000;
and others) and with AGN (see for references Veilleux et al. 2002a).

There are clear evidence of galactic --and local-- winds, shells, arcs and
bubbles generated by multiple SN explosion and massive star winds
in starbursts.
Our understanding of the main phases of galactic winds associated with
starbursts was improved significantly by the use of theoretical and
numerical models (see Strickland \& Stevens 2000; Suchkov et
al. 1994; Mac Low, McCray \& Norman 1989; Tomisaka \& Ikeuchi 1988).
In general, good agreement has been found between these  models and the
observations.

On the other hand, for galactic winds associated with AGN the situation
is more complex, and very different models are proposed in order to explain
the observed data. In these models the OF could be generated by
jets driven thermal winds, accretion discs winds, X-ray heated torus
winds, etc. (see Veilleux et al. 2002a; Cecil et al. 2002; Capetti 2002;
Morganti et al. 2003).

However, there are increase evidence that nuclear galactic jets are one of
the main sources of out flows in QSOs/AGNs (Cecil et al. 2002; Veilleux et
al. 2002a; Capetti 2002).
In particular, there are clear detection and evidences of the interaction
of jets with various phases of the ISM, including the narrow
line region (NLR).

\subsection{IR Mergers with galactic winds}\label{introgwm}

IR mergers  often show strong starbursts with powerful out flow (OF) and
galactic winds (GW; L\'{\i}pari et al. 2003, 2004a,b,c,d).
Comparing our OF data base of IR mergers/QSOs (Lipari et al. 2004a, their
Table 1) with two samples of nearby IR mergers we found in both samples a
high proportion of  mergers with galactic winds: $\sim$75\%.
These results  suggest (or confirm) that: (i) GWs are  ``frequent
events'' in IR mergers, and (ii) extreme starbursts + GW and extreme IR
emission could be simultaneous processes, induced by merger events.
This last conclusion is also supported by a clear trend found in the plot
of OF velocity vs.\ $\log L_{\rm IR}$ (see their Fig. 13), in the sense
that extreme OF velocities are detected only in extreme IR emitters (ULIRGs).
We note that the reported trend is observed mainly
for IR mergers (Lipari et al. 2004a).
An explanation for this observed trend is that high values of OF velocity
and $L_{\rm IR}$  are both associated mainly with the same process:
``starburst + QSO'' events, probably induced by mergers (already
suggested by L\'{\i}pari et al.\ 2003; see also Genzel et al.\ 1998; Veilleux
et al.\ 1999; Sanders et al.\ 1988a; Rieke et al. 1985).

The low velocity OF process (LVOF, $V_{\rm LVOF} <$ 700\,km\,s$^{-1}$;
see L\'{\i}pari et al.\ 2004a,b, 2003, 2000a,b) observed in NGC 3256, 2623,
4039, 5514, etc --using 2D spectroscopy-- is consistent with those found
previously in Arp\,220, Mrk\,266, NGC\,1614,  NGC\,3690, and other IR mergers
with massive starbursts.
In addition, we found an interesting fact: starbursts and LINERs  are the
main sources of ionization in ``low velocity OF" IR mergers (L\'{\i}pari et
al.\ 2000b, 2003, 2004a).

\subsection{IR QSOs with galactic winds}\label{introgwq}

The discovery and study of IR QSOs (see for references L\'{\i}pari et al.
2004d, 2003; Zheng et al. 2002) raises several interesting
questions, in particular whether they are a special class of QSOs.
We found, or confirm, that a high percentage of IR QSOs show extreme
out flows with giant galactic arcs, merger features, BAL systems, and extreme
Fe {\sc ii} emission, and are radio quiet (L\'{\i}pari et al. 2004b, 2004a,
2003; L\'{\i}pari, Terlevich \& Macchetto 1993; L\'{\i}pari, Macchetto
\& Golombek 1991a).
We suggested that these objects could be {\it young IR$-$active
galaxies at the end phase of a strong starburst: i.e., composite and
transition QSOs} (L\'{\i}pari 1994).

Recently, for IR QSOs we found:
(i) extreme velocity OF (EVOF, $V_{\rm EVOF} >$ 700 km s$^{-1}$) objects
with a composite nuclear source: starbursts + AGNs/QSOs; for e.g.
in Mrk\,231, IRAS 19254-7245, 01003-2230, 13218+0552, 11119+3257, 14394+5332,
and others (L\'{\i}pari et al.\ 2003);
(ii) high resolution {\itshape HST\/} WFPC2 images of IR  + BAL +
Fe\ {\sc ii} QSOs show in practically all of these
objects arcs or shell features probably associated with OF 
or merger processes (L\'{\i}pari et al.\ 2003).

Recently, using our data base of GW/OF in IR mergers/QSOs we found
that in all the IR QSO candidates the H$\beta$ broad line
component is blushifted in relation to the narrow one, which
is clearly consistent
with the result obtained from the study of strong Fe {\sc ii} + BAL
emitters, by Boroson \& Meyer (1992). They proposed that blueshifted
offset/asymmetry detected in the H$\alpha$ broad  components--of IR QSOs
with strong Fe {\sc ii} + BAL systems--is probably due to
the emission of the out flowing material, associated with the BAL process.
Furthermore, our measurements are also consistent with the results
of the study of multiple emission line
components, in some starburst nucleus of galaxies (Taniguchi  1987).
Thus, a possible explanation for
``at least part" of these blushifted broad line systems in IR QSOs
is the high speed OF/GW generated in a burst of SN event near the
nuclear region (Heckman et al. 1990; Terlevich et al. 1992; Perry \&
Dyson 1992; Lipari et al. 2003, 1994, 2003, 2004a; Taniguchi et al. 1994;
Scoville \& Norman 1996; Lawrence et al. 1997; Collin \& Joly 2000; and
others).

\subsection{BALs in IR  mergers/QSOs}\label{introlink}

Some of the results obtained for \emph{nearby BAL QSOs}, such as
strong IR and Fe\,{\sc ii} emission,
strong blue asymmetry/OF in H$\alpha$, radio quietness, and
very weak [O\,{\sc iii}]$\lambda$5007 emission
(Low et al. 1989; Boroson \& Meyers 1992; L\'{\i}pari et al. 1993, 1994,
2003; Lipari 1994; Turnshek et al. 1997),
 can be explained in the framework of the starburst+AGN scenario.
In our study of Mrk\,231 and  IRAS\,0759+6559
(the nearest extreme IR + GW/OF + Fe\,{\sc ii} + BAL systems), we
detected typical characteristics of  young-starburst QSOs.
In our evolutive model for young and composite IR QSOs (see for
references L\'{\i}pari 1994) suggested that BAL systems could be linked
to violent supermassive starburst+AGN which can lead to a large-scale
expanding shells, often obscured by dust.
Several articles suggested that this evolutive model
shows a good agreement  with the observations (see Canalizo \&
Stockton 1997; Lawrence et al. 1997; Canalizo et al. 1998).

\subsection{Possible links between IR mergers and IR QSOs}\label{introlink}

The luminosities and space
densities of ULIRGs in the local Universe are similar to those of
quasistellar objects (QSOs; Soifer, Houck \& Neugebauer 1987).
In addition, at the highest IR luminosities, the presence of AGNs (and
mergers) in LIRGs becomes important.
Thus LIRGs probably represent an important stage
in the formation of QSOs and elliptical galaxies.
These results strongly suggest that it is important to
perform detailed studies of possible links among mergers, ULIRGs, QSOs
and elliptical galaxies (Toomre \& Toomre 1972; Larson 1974; Toomre
1977; Schweizer  1982; Joseph \& Wright 1985; Sanders et al. 1988a,b;
L\'{\i}pari et al. 1993, 1994, 2003; Sanders \& Mirabel 1996; Colina
et al. 2001).

Furthermore, the important detection of a  correlation  between the mass of
galactic bulges and the mass of supermasive black hole is a confirmation
that the formation and evolution of galaxies (bulges/ellipticals, mergers) and
supermassive black hole (AGNs and QSOs) are physically related to one another
(Magorrian et al. 1998; Ferrarese \& Merrit 2000; Gebhardt et al. 2000;
Kormendy 2000; Merrit \& Ferrarese 2001; Kormendy \& Richstone 1995).

In the last years, several possible \emph{links} between \emph{mergers,
starbursts, IR~QSOs and ellipticals} have been proposed. Specifically,
Joseph et al., Sanders et al. and L\'{\i}pari et al. suggested three
complementary sequences and evolutive--links:

(i) merger $\to$ giant shocks $\to$ super-starbursts + galactic
winds $\to$ elliptical galaxies;

(ii) merger $\to$ H$_2$-inflow (starbursts) $\to$ cold ULIRGs
$\to$  warm ULIRGs + QSOs;

(iii) merger{\bf /s} $\to$  extreme starburst + galactic-wind
(inflow + outflow) $\to$  IR + Fe\,{\sc ii} + BAL composite/transition QSOs
$\to$ standard QSOs and ellipticals.

Recently, several studies have confirmed the composite nuclear nature and
the presence of merger and galactic wind features in several IR QSOs
(Canalizo \& Stockton 2001).
In particular, Lipari et al. (2003, 2004a,d) found kinematical and
morphological evidence of galactic winds in nearby {\it ``IR + Fe\,{\sc ii}
+ BAL QSOs"}.

Thus, the study nearby {\it ``IR + Fe\,{\sc ii} + BAL
composite/transition QSOs"} is one observational
way to analyse the possible relation between IR mergers and IR QSOs. 
We have started detailed \emph{morphological, spectroscopic and
kinematics} studies of this kind of objects (since 1987:
see L\'{\i}pari et al. 1991a, 1993, 1994, 2003, 2004d; Lipari 1994).

\clearpage
\section{THE PROGRAMME AND OBSERVATIONS} \label{pobservations}

\subsection{The programme} \label{programme}

\subsubsection{IR mergers/QSOs with galactic winds (the main
programme)} \label{programme1}

A current key issue in astrophysics is to explore the evolution of the
star formation process, especially at high redshift, when the galaxies/QSOs
formed, and where it is expected that the star formation rate is very high.
With these aims in mind we began a project to study nearby
star forming + GW galaxies and distant Ly$\alpha$ emitters (see Lipari
et al. 2004a,b,c,d).
The first step in this project is to understand the star formation process
in nearby galaxies because we can obtain more detailed and unambiguous
information. Thus, our groups started a  study of nearby IR mergers/QSOs,
which are an excellent laboratory--at low redshift--for the analysis of
extreme star formation and GW processes (Lipari et al. 2004a, 2003, 2000a,b,
1994).

This study is based mainly on integral field spectroscopy, obtained at
the European Northern Observatory (ENO, La Palma--Spain), the European Southern
Observatory (ESO, Chile), the Complejo Astronomico El Leoncito (CASLEO,
Argentina), and Bosque Alegre (BALEGRE, Argentina) observatories, with the
4.2 m, 3.6 m, 2.15 m and 1.5 m telescopes, respectively.
The characteristics and goals of the programme have been
described in detail by L\'{\i}pari et al. (2004a).
In particular,
the main goals of this programme (at low redshift) is to analyse in
detail the properties of the different stages of extreme starbursts, galactic
winds, mergers, QSOs, and elliptical galaxies (and their interrelation).

In the present paper we present  new results from our programme of study of
 ``nearby" IR mergers/QSOs with galactic winds (EVOF): for Mrk\,231.

\subsubsection{BALs in IR + GW + Fe\,{\sc ii} mergers/QSOs} \label{programme2}

A study and search of UV (and optical) BAL systems in extreme
IR + GW/OF + Fe {\sc ii} QSOs have been started since 1993 (see Lipari 1994;
Zheng et al. in preparation), using mainly IUE and HST UV-spectra, plus
ESO NTT, KPNO, APO and CASLEO optical data. In general,
the role of BALs in IR+GW/OF+Fe{\sc ii} QSOs/mergers must be
carefully considered, since: (i) Low et al.
(1989) and Boroson \& Meyers (1992) found that IR selected QSOs
show a 27\% low-ionization BAL QSO fraction compared with 1.4\%
for the optically selected high-redshift QSOs sample (Weymann et
al. 1991); (ii) extreme IR galaxies (ULIRGs) are mainly mergers
(Section 1); (iii) very recently Maiolino et al. (2003) reported
also a high fraction of BAL QSOs at very high redshift (z $\sim$6).
The high percent of occurrence of broad absorption in extreme
IR + GW/OF + Fe{\sc ii} QSOs/mergers may be signals a fundamental relation
(rather than merely a coincidence), and deserves detailed studies.

L\'{\i}pari et al. (1993, 1994, 2003, 2004a,d); Scoville \& Norman (1996);
Egami et al. (1996); Lawrence et al. (1997) and others proposed that
the extreme IR + GW/OF + Fe\,{\sc ii} + BAL phenomena are related --at least
in part-- to the end phase of an ``extreme starburst+AGN" and the
associated ``powerful bubble/galactic--wind". At the final stage of a
strong starburst, i.e., type II SN phase
([8-60]\,$\times10^{6}$\,yr from the initial burst; Terlevich et
al. 1992; Norman \& Ikeuchi 1989; Suchkov et al. 1994) giant
galactic arcs and extreme Fe\,{\sc ii}+BAL systems can appear.

Specifically, the search of UV broad absorption line was focused mainly
in the IR objects of our data base of IR + GW/OF + Fe{\sc ii} QSOs/mergers
(see Table 4), and studying the redshifted lines: Ly$\alpha$, Si~IV, C~IV,
and Mg II. Lipari (1994) reported a first study of the UV BAL systems in
IRAS 0759+651, with projected ejection velocity of about 10000~${\rm km
s^{-1}}$. Previously, it was also reported UV BAL systems in
Mrk 231 (IRAS 12540+5708), IRAS 17002+5153, and IRAS 14026+4341.

It is important to note that in Table 4 several examples clearly show
the consistency of the OF data presented.  In particular,
for IRAS F05024-1941, our study of EVOF in the ULIRG 1 Jy Sample
(using ``emission" lines), we found that this object is a candidate
to EVOF, with a velocity close to 1500 km s$^{-1}$ (Lipari et al.
2000b). From a study of narrow ``absorption" lines Rupke et al. (2002)
obtained for this object an OF value of 1600 km s$^{-1}$.
We have verified that for Mrk 231 the results using the offset method
(by Zheng et al. 2002) gave the same value of OF: --1000 km s$^{-1}$,
to that already obtained from the detection of two emission line
systems in [O {\sc ii}]$\lambda$3727 (by Lipari et al. 1994).
A similar agreement between the values of the OF obtained from multiple
emission lines (Lipari et al. 2000b) and the offset method (Zheng et
al. 2002) was found, for the cases of IRAS 11119+3257 and
IRAS 15462-0450 (see Table 4).

In this paper we report new results from this programme. In particular,
the detection of a new IR + GW/OF + Fe {\sc ii} QSO with BAL system.

\subsection{Mrk 231: the nearest extreme  IR + GW + BAL + Fe\,{\sc ii}
merger/QSO (of the programme)} \label{m231}

Since Markarian 231 was discovered (by its strong ultraviolet continuum;
Markarian 1979) it has been extensively  studied and has been recognised
as a remarkable galaxy: it is the most luminous galaxy in the local
universe (z $<$ 0.1), the nuclear spectrum shows extreme broad emission
lines and strong BAL systems, the morphology is associated with a merger, etc
(Arakelian et al. 1991; Adams 1972; Adams \& Weedman 1972;
Rieke \& Low 1972, 1975; Boksemberg et al. 1977; and others).
In particular, it is important to remark that Sanders et al. (1987);
Bryant \& Scoville (1996) and Downes \& Solomon (1998) found
{\it a very high concentration of CO molecular gas in the nuclear region}
(3 $\times$ 10$^9$ M$_{\odot}$, within a diameter of 1\farcs0, 814 pc).

Before the IRAS was lunched, Mrk 231 was known to have
an extreme IR luminosity  with M$_K$ = -24.7 
(Rieke \& Low 1972, 1975; Cutri, Rieke \& Lebofsky 1984).
Even after IRAS expand the known population of ULIRG, Mrk 231
remain one of the most luminous object in the local
universe, with L$_{IR[8-1000 \mu m]} = 3.56 \times 10^{12} L_{\odot}$,
and L$_{IR}$/L$_{B}$ = 32.
Consequently, the bolometric luminosity (dominated by the continuum
IR emission) places the nucleus of this IR merger among the QSOs.
Weedman (1973) already proposed that this
extreme IR luminosity is associated with ultraviolet (UV) emission
re--radiate thermally --in the IR-- by dust.

Mrk 231 also shows very interesting spectral characteristics, dominated
in the optical by extremely strong Fe {\sc ii} and broad Balmer emission
lines at Z$_{em}$ = 0.042 (Arakelian et al. 1971; Adam \& Weedman 1972;
Boksemberg et al. 1977).
In addition, it shows remarkable absorption line systems:
a clear stellar absorption at Z$_{abs}$ = 0.042 plus at least three
strong broad absorption line (BAL) systems.
These strong BAL systems show the following velocity of ejection:
V$_{eject}$ of BAL I, II and III $\sim$4700, $\sim$6000,
and $\sim$8000 km s$^{-1}$.
The last BAL-III system sudden appear sometime between 1984 December and
1988 May (Boroson et al. 1991)

The extreme IR luminosity of Mrk 231  is associated  with two main sources
of energy:
(i) an AGN detected clearly at radio and X-ray wavelengths (see for
references Condon et al. 1991; Ulvestad et al. 1999; Turner 1999; Gallagher et
al. 2002), and (ii) an extreme nuclear + circumnuclear starburst observed
almost at all wavelengths (see Downes \& Solomon 1998; Taylor et al. 1999;
Gallagher et al. 2002). In addition, Lipari et al. (1994) reported
kinematics, morphological  and  physical evidence of a powerful nuclear
outflow process: {\it i.e. a ``galactic superwind with a broken shell, in
the blowout phase"}.
Even Lipari et al. (1994, 1993) suggested that Mrk 231 is an evolving
{\it ``young and composite IR QSO"}.

Thus, Mrk\,231 is the nearest extreme  IR + GW/OF + BAL + Fe\,{\sc ii} QSOs.
In this paper, new results of the OF process in Mrk 231 are reported.
Throughout the paper, a Hubble constant of H$_{0}$ = 75 km~s$^{-1}$
Mpc$^{-1}$ will be assumed. 
For Mrk\,231 we adopted the distance of $\sim$168 Mpc (cz = 12465 
$\pm$10 km~s$^{-1}$), and thus, the angular scale is 1$'' \approx$814 pc.

\subsection{La Palma/WHT + INTEGRAL 2D spectroscopy} \label{whtobservations}

Two-Dimensional (2D) optical spectroscopy of  Mrk\,231 was obtained during a
 photometric night in  April 2001 at the 4.2 m telescope at Observatorio
Roque de los Muchachos on the Canary island of La Palma. 
The 4.2 m William Herschel Telescope (WHT) was used with the INTEGRAL fibre
system (Arribas et al. 1998) and the WYFFOS spectrograph (Bingham et al.
1994). The seeing was approximately 1\farcs0.

INTEGRAL links the f/11 Naysmith focus of WHT with the slit of WYFFOS
via three optical fibre bundles. A detailed technical description of INTEGRAL
is provide by Arribas et al. (1998); here we only recall its main
characteristics. The three bundles have different spatial
configuration on the focal plane and can be interchanged depending the
scientific programme or the seeing conditions. At the focal plane the fibres
of each bundles are arranged in two groups, one forming a rectangle and the
other an outer ring (for collecting background light, in the case of
small-sized objects).
The data analysed in this paper were obtained with the standard bundle 2:
which consistent of 219 fibres, each of 0\farcs9 in diameter in the sky.
The central rectangle is formed by 189 fibres, covering a field of view of
16\farcs4 $\times$ 12\farcs3; and the other 30 fibres form a ring 90$''$
in diameter.

The WYFOS spectrograph was equipped with  a 1200 lines mm$^{-1}$ grating,
covering the $\lambda\lambda$6200--7600 \AA range. A Tex CCD array of
1124 $\times$ 1124 pixels of 24 $\mu$m size was used, giving a linear
dispersion of about $\sim$1.45 \AA\ pixel$^{-1}$ ($\sim$2.5\,\AA\, effective
 resolution, $\sim$ 70 km s$^{-1}$). With this configuration and pointing 
to the central region of  Mrk\,231, we took three exposure of 1800 s.

\subsection{HST ACS, WFPC2 and NICMOS broad band images} \label{hstobservations}

Near--UV  {\itshape HST\/} Advanced Camera for Surveys (ACS) archival images
of Mrk 231 were analysed, obtained with the High Resolution Channel (HRC).
They include images with the filter F330W (3354 \AA, $\Delta\lambda$ 588 \AA).
The scale is 0\farcs027\,pixel$^{-1}$.

Optical {\itshape HST\/} Wide Field Planetary Camera 2 (WFPC2) observations
were analysed, which include broad-band images positioned on the
Planetary Camera (PC) chip with scale of 0\farcs046\,pixel$^{-1}$,
using the filters
F439W (4283 \AA, $\Delta\lambda$ 464 \AA, $\sim$B Cousin filter), and
F814W (8203 \AA, $\Delta\lambda$ 1758 \AA, $\sim$I).

{\itshape HST\/} Near IR Camera and Multi Object
Spectrometer (NICMOS) archival data
were studied, which include mainly images with the filter 
F160W (1.60 $\mu$m, $\Delta\lambda$ 0.4 $\mu$m), 
using the camera 2 with a scale of 0\farcs076\,pixel$^{-1}$.

For details of the {\itshape HST\/}  observations see also Table 1.

\subsection{HST UV--FOS and UV-STIS 1D Spectroscopy} \label{hst2observations}

{\itshape HST\/} FOS aperture spectroscopy of the main optical nucleus 
of Mrk 231 was obtained, from the HST archive (at ESO Garching).
The spectra were taken with the
G190L ($\lambda\lambda$1150--2300), G190H ($\lambda\lambda$1575--2320) and
G270H ($\lambda\lambda$2225--3290)  gratings and the blue detector.
The G190H and G270H observations were made with the effective aperture of
4\farcs3 $\times$ 1\farcs4; and these spectra have a resolution of $\sim$3
and 4 \AA, FWHM respectively.
The G160L observations were made with the aperture of 1\farcs0,
and the spectra have a resolution of $\sim$8, FWHM.

In addition, {\itshape HST\/} FOS spectra of extreme IR + Fe {\sc ii}
QSOs with GW were analysed.
These data were used in our study/search of UV--BAL systems (for details
see Table 1).

{\itshape HST\/} STIS long slit spectra of IR + GW/OF + Fe {\sc ii}
QSOs were obtained (mainly from the HST archive).
The spectra were taken with the
G750M grating, with the slit  0\farcs1 $\times$ 52\farcs0,
giving a resolution of 50 km s$^{-1}$.
The STIS CCD  has a scale of 0\farcs05 pixel$^{-1}$.
%The slit was aligned at PA 91$^{\circ}$.

\subsection{La Palma/NOT optical broad band images} \label{whtobservations}

Optical V broad band optical image of  Mrk\,231 was obtained during a
photometric night in  1991 May at the 2.5 m Nordic Optical
Telescope (NOT) at Observatorio
Roque de los Muchachos on the Canary island of La Palma. 
A TEX CCD chip with scale  0\farcs197 per pixel was used.
The observations were carried out in excellent seeing conditions,
with a values of 0\farcs7 (FWHM).

\subsection{KPNO long-slit 1D spectroscopy} \label{kplsobs}

Optical long slit spectra of Mrk 231 and IR + GW/OF + Fe {\sc ii}  QSOs
were obtained on the KPNO 2.15\,m telescope
with the Gold Cam Spectrograph, by using two grating of
300\,grooves\,mm$^{-1}$ (7\,\AA\, $\sim$300 km s$^{-1}$ resolution)
covering the wavelength ranges 3350--5200 and 5100--7100\,\AA.
The observations were taken during one photometric night
in February 1991 (Table 1). The spectra were obtained with
a slit width of 1.5 arcsec. The seeing was in the range 1.0--1.5
arcsec (FWHM).

The data of the {\it IRAS} 1 Jy ULIRGs sample were obtained with the
Gold Cam Spectrograph on the KPNO 2.15\,m telescope by using a
grating of 300\,grooves\,mm$^{-1}$ (8\,\AA\,resolution); and
covering the wavelength range 4500--9000\,\AA.

\subsection{IUE UV 1D spectroscopy} \label{kpls}

UV International Ultraviolet Explorer (IUE) spectra 
%(of IRAS 07598+6508 and PG 1700+518)
were obtained for our programme of search of UV--BAL systems
in extreme IR + GW/OF + Fe {\sc ii} QSOs (Table 1). In addition, data from
the IUE archival --of this type of objects-- were analysed.
The IUE LWP and SWP  spectra  cover the wavelength ranges 1150--1980 and
1900--3290\,\AA,; with  a resolution  of $\sim$8\,\AA\,
($\sim$300 km s$^{-1}$).

\subsection{ESO NTT and 3.6 m telescope spectra} \label{nttobs}

The {\itshape ESO\/} Faint Object Spectrograph and Camera II (EFOSC II)
on the 3.5 m new technology telescope (NTT) at La Silla was used, to
obtain long slit spectra  of IR + GW/OF + Fe {\sc ii}  QSOs.
These long-slit observations were taken during 1 photometric night
in July 1993  (see Table 1).
The spectra cover the wavelength ranges $\sim$3560--5200\,\AA,
with a resolution of $\sim$7\,\AA\, ($\sim$300 km s$^{-1}$)

The ESO Faint Object Spectrograph and Camera (EFOSC) on the 3.6\,m
telescope at La Silla was used to obtain long-slit spectra and
high resolution images. Medium-resolution spectra were obtained
with the B150, O150, and R150 grisms, which provide a resolution
of $\sim7$\,\AA\, in the wavelength range 3600-9800\,\AA\,\,
(during 3 photometric nights in 1995 January and 1991 July; see Table 1).

\subsection{Apache Point Observatory (APO) 1D spectroscopy} \label{APOobs}

The {\itshape APO\/} spectra  of IR GW/OF + Fe {\sc ii}  QSOs
were taken with the 3.5 m telescope.
Optical long-slit observations were obtained during 1 photometric night
in September 2004  (see Table 1).
The medium resolution spectra give an effective resolution of
$\sim$6 \AA\ ($\sim$290 km s$^{-1}$), covering the wavelength ranges
$\lambda\lambda$3500--5600 and 5300--10000 \AA.
The seeing was  $\sim$1\farcs5 (FWHM).

\subsection{CASLEO 1D spectroscopy} \label{CASLEOobs}

Spectrophotometric observations of IR + GW/OF + Fe {\sc ii} QSOs
were taken at CASLEO (San Juan, Argentina) with the 2.15 m
Ritchey-Chr\'etien telescope.
Optical long-slit and aperture spectroscopy 
observations were obtained during 3 photometric nights in June 1989,
July 1993, March 1997  (see Table 1).
Long-slit  spectroscopic observations with medium 
resolution were obtained with the University of Columbia spectrograph (UCS;
e.g. L\'{\i}pari et al. 1997).
The medium resolution spectra were made using a 600 line mm$^{-1}$ grating,
a slit width of 2\farcs5,
which gives an effective resolution of $\sim$6 \AA\ ($\sim$290 km s$^{-1}$)
and a dispersion of  120 \AA\ mm$^{-1}$, covering the wavelength range
$\lambda\lambda$4000--7500 \AA.
Aperture spectroscopic data were obtained with the `Z-machine'
(e.g. L\'{\i}pari et al. 1991a,b).
These aperture spectra were made using a 600 line mm$^{-1}$ grating,
giving a dispersion of 130 \AA\ mm$^{-1}$ and an effective resolution
of $\sim$8 \AA\ ($\sim$300 km s$^{-1}$)
covering the wavelength range $\lambda\lambda$4700--7200.
The seeing was in the range 1\farcs5--2\farcs5 (FWHM).

\subsection{Reduction} \label{reductions}

The {\sc IRAF}\footnote{{\sc IRAF} is the imaging analysis software facility
developed by NOAO} software package was used to reduce and analyse the data.
The reduction of the 2D spectroscopic observations consist of two main
steps: (1) reduction of the spectra (for each 219 fibres), and (2)
generation of 2D maps by interpolated the selected parameter (e.g.,
emission-line intensity, continuum intensity, radial velocity, etc) from the
grid values defined by the fibre bundle.
The step (1) was basically done in the same way as for long-slit spectroscopy,
including bias subtraction, aperture definition and trace, stray-light
subtraction, the extraction of the spectra, wavelength calibration, throughput
correction and cosmic-ray rejection.
We obtained typical wavelength calibration errors of 0.1 \AA, which give
velocity uncertainties of 5 km s$_{-1}$.
For step (2) we used the software package {\sc Integral}\footnote{{\sc
Integral} is the imaging and spectroscopic analysis software facility
developed by the Instituto de Astrofisica de Canarias (IAC)},
with 2D interpolation routines.
In particular we transformed ASCII files with the positions of
the fibres and the corresponding spectral features, into regularly spaced
standard FITS files. Maps generated in this way are presented in the next
sections.

The IRAF and STSDAS\footnote{STSDAS is the reduction
and analysis software facility developed by STScI}
software packages were used to analyse the {\itshape HST\/},
{\itshape ESO\/}, IUE, ESO, KPNO, NOT, APO, CASLEO imaging and
spectrophotometric data.

The {\itshape HST\/} data were reduced at the Space Telescope Science
Institute (STScI), using the standard process.
All the bands were calibrated according to the procedures detailed in
Lipari et al. (2004a).
Specifically,  photometric calibration of
the {\itshape HST\/} data was performed using
published photometric solution (Lipari et al. 2004a).

For the study of the kinematics data we used the software ADHOC\footnote{ADHOC
is the analysis 2D/3D kinematics software facility developed by Marseille
Observatory}

\clearpage
% ******************************************************************
% ******************************************************************
\section{RESULTS} \label{results}

This section focuses on presenting (for Mrk 231):
(i) high resolution HST near-UV, optical and near--IR images
of the main body;
(ii) 2D spectroscopy mainly for the central regions;
(iii) 1D spectroscopy for the nuclear region;  
(iv) a study of variability of the Na ID BAL--III system (based
in 1D spectroscopy).
In particular, we study the extreme nuclear OF with
multiple expanding concentric shells, giant shocks and BAL systems.
In addition, new observations of UV-BAL systems in IR+GW/OF+Fe {\sc ii}
QSOs are analysed.

% 3.1 ******************************************************************
\subsection{Near-UV, optical and near--IR  HST and NOT images:
multiple concentric nuclear bubbles (and knots)}
\label{results-mermor}

Here,  a multiwavelength morphological study (using mainly high resolution
HST near-UV, optical and near-IR  broad band images) of the nuclear and
circumnuclear structures of Mrk 231 is presented.

The NOT V image (Fig. 1) shows almost the whole merger.
This galaxy consist of a nearly elliptical main body (R $\sim$ 10 kpc)
with a compact nucleus, plus two faint tidal tails (see also
Lipari et al. 1994; Hamilton \& Keel 1987; Neff \& Ulvestad 1988).
These are typical features observed in advanced mergers.

It is important to note, that previously, in the circumnuclear region of Mrk
231 L\'{\i}pari et al. (1994) found clear evidence of a powerful nuclear
galactic--wind and they proposed that this OF process (in part associated
with a starburst) has generated the circumnuclear blue arc/shell (at
$r\sim3$\,kpc, and to the south of the nucleus).
However, Armus et al. (1994) suggested that this arc is associated
with an obscured second nucleus (they also suggested that
in this blue region or ``shell" there is no evidence of star-formation
process). HST--WFPC2 observations of Mrk\,231 confirmed
that this blue arc is a \emph{``dense shell of star-forming
knots"} (see Surace et al. 1998; Lipari et al. 2003).
In the next sections we will expand these previous studies.

\subsubsection{The multiple concentric nuclear bubbles/shells}
\label{results-mermor1}

Figs. 2, 3 and 4, show  four (or five) concentric supergiant galactic shells
or bubbles with centre in the nucleus and showing several bright knots.
These figures present high resolution HST broad-band images
obtained in the near--UV, optical and near--IR
wavelengths through the filters F330W ($\sim U$), F439W ($\sim B$),
F814W ($\sim I$), and F160W ($\sim H$).
We note that Figs. 3a,b,c,d were obtained from the subtraction of
a smooth image (for each filter) of the main body of Mrk 231, in order to
study the morphological residuals.

These figures show intereting structures, in particular:

\begin{enumerate}

\item
{\it Bubble or Shell--S1:}
This is the more external and well defined shell detected in Mrk 231,
with a radius $R_{S1}\,=3\farcs5=2.9$\,kpc.
Lipari et al. (1994) already noted that this shell shows a clear elongation
to the south (of the nucleus). At the border of this elongation there
is a blue arc with several knots. They associated this arc with typical
structures generated in the blowout phase of a nuclear
galactic wind (with giant shocks). These  shocks
probably generated a new extranuclear star formation process, with
knotty structure.

Fig. 2 shows interesting details of this bubble + arc in the
{\itshape HST\/} ACS, WFPC2 and NICMOS images. In particular,
the  {\it HST\/} WFPC2/B image depicts clearly the
complete shell S1. However the {\it HST\/} ACS/U image
shows remarkable details of the blue compact knots located
at the border of the southern arc.

\item
{\it Bubble--S2:}
The next internal bubble, at $R_{S2}\,= 1\farcs8 = 1.5$\,kpc,
shows also a well defined circular shape.
However, this more internal bubble is clearly obscured
in the south-east region by reddening, associated with
the presence of dust.
Again, the  {\it HST\/} WFPC2/B image depicts more clearly the
partial shell S2. It is important to remark that
this shell shows a very interesting knot--2 in the
south-west direction (see Table 2 and the next sections for
details about the properties of this knot).

\item
{\it Shell--S3:}
The next bubble, with a radius $R_{S3}\,= 1\farcs2 = 1.0$\,kpc,
depicts also a circular shape, and their emission
is strong mainly in the south area. 
This structure is well defined in the {\itshape HST\/} ACS/U image,
which shows also a knotty structure.
This shell depicts a bright knot--3, to the south--east. 

\item
{\it Shell--S4:}
This shell is clearly observed mainly  in the {\itshape HST\/}
ACS/U image (Fig. 2a; we note that this ACS image has the best
spatial resolution) and also in the ACS/U, WFPC2/B residual images (Figs.
3a,b). The radius  of this shell S4 is $R_{S4}\,= 0\farcs7 = 0.6$\,kpc.

\item
{\it Possible shell--S5:}
This possible shell is only observed in the  HST optical (I-B) colour image
(Fig. 4) and very close to the nucleus, with a radius
$R_{S5}\,= 0\farcs3 = 0.2$\,kpc.

\item
Close to the nucleus,  the HST images show radial
filaments. These structures are more clear in
the near-IR image since at these wavelengths the dust absorption is
significantly smaller. In particular, it was found that these 
filaments are more larger and clear in the direction north--south (PA
$\sim$00$^{\circ}$) and at the position angle of the radio jet
(PA $\sim$65$^{\circ}$; see sections 4.2 and 4.3 for details about this jet).
We note that these PAs are coincident with
the directions of the two main OF processes in Mrk 231 (see the next
sections).
In our programme of study of galactic winds in IR mergers/QSOs, similar
filaments were detected associated with the OF processes of NGC 3256,
NGC 2623 and NGC 5514.

\end{enumerate}

This type of {\it circular and concentric ``bubbles or shells"} could
be associated mainly with symmetric explosive events.
In sections 3.2 and 4.3 the kinematics, age, energy and possible origin
of these structures will be discussed.

\subsubsection{The circumnuclear knots}
\label{results-mermor2}

Table 2 presents the location and properties  of the main
knots, in the nuclear and circumnuclear areas of Mrk 231.
In addition, the positions of these knots are labeled in Fig. 3b.
Surace et al. (1998) already studied in detail these knots using
 HST photometry, in the B and I bands. In this paper,  we will study mainly their
properties in relation to the process of OFs and their location
in the shells.

From the data of Table 2, it is clear that in the southern part of the
shell S1, the knots 5, 6, 11 and 12 show
blue colours, probably associated with new star formation
events. In particular, these knots emit a strong UV continuum,
due to the presence of a large number of massive O and B stars.
These massive stars are usually responsible for the rupture phase of the
galactic bubble and also are the progenitors of core--collapse SNe.
In the supergiant bubble of NGC 5514 we also detected the presence of several
knots with massive WR stars (inside the areas of rupture of the external
shell).
The radius of this bubble in NGC 5514 --just detected in the rupture phase--
is also of $\sim$3 kpc, i.e. the same value of the shell S1 in Mrk 231.
We already noted that the results of hydrodynamical models for galactic
winds (associated mainly with extreme starbursts)
suggest that 3 kpc is the typical value of radius for the beginning
of the blowout phase (after $\sim$8 Myr, from the initial starburst;
Suchkov et al. 1994).

The remaining knots show different colours (see Table 2). Some of the
knots depict red colours, which could be explained --in part--
by the presence of dust in star formation areas.
In particular, the HST NICMOS image shows that the knot 14 is bright at
the H-band, suggesting that the red colour in this knot (located in
the southern arc of the shell S1) is due mainly to reddening by dust.
However, for another two knots with very red colours, the 2D
spectroscopy suggest another origin for the red colours. For the knot 2
(located in the shell S2) the spectrum shows properties typical of bow shocks,
similar to those found in radio jets
(see for details the next section).
In addition, the knot 29 (positioned at the north-west area of the
shell S1) also shows very red colour, and it is located in another OF region.
Where the [S {\sc ii}]/H$\alpha$ and [N {\sc ii}]/H$\alpha$ maps depict
extended radial filament, probably
emerging from the bubble (see section 3.4).  For this area also Krabbe et
al. (1997) reported OF ejection (of $\sim$1400 km s$^{-1}$ in Pa$\alpha$;
 from a study of 2D near--IR spectroscopy).

In the next section, the kinematics and physical properties of the
multiple concentric shells and the circumnuclear knots will be analysed 
(using mainly WHT 2D spectroscopy).

%\clearpage
% 3.2 ******************************************************************
\subsection{Mapping the outflows in the nucleus and in the multiple
supergiant bubbles (2D Spectra)}
\label{results-gw}

The H$\alpha$+[N {\sc ii}] emission lines profiles obtained
{\it simultaneously} with INTEGRAL for the central region of Mrk\,231
($16''\times$12$''$; 13.0 kpc$\times$10.0 kpc) are shown in Fig. 5.
These are reduced data, but represent only a small part ($\sim$100 \AA)
of the full spectral coverage (6200--7600 \AA) at each of the 189 fibre
location. The nucleus was posioned close to the centre of the field
(at fibre 107), where it is evident the presence of broad emission lines.

First, the OF at small scale will be analysed, in the
region close to the nucleus (R $<$ 1--2 kpc). In particular,
it is important to remark the following main results:

\begin{enumerate}

\item
In the fibre 101 (adjacent to the fibre of the nucleus) the 2D spectrum shows
the presence of an strong OF emission bump in the blend
 H$\alpha$+[N {\sc ii}], with the peak corresponding to the same velocity
 of the main BAL--I system ($ V_{\rm Ejection BAL-I} \sim$
 --4700\,km\,s$^{-1}$).

This bump was detected in the area covered by this fibre,  at
0\farcs6--1\farcs5 (490--1220 pc), to the south-west of the
nucleus core at PA $\sim$60$^{\circ}$, showing a strong and
broad peak (see Fig. 6b).

It is important to note that the direction/PA of fibre 101 (from
the main nucleus) is the same that the PA of the small scale radio jet
(PA$_{jet}$ $\sim$60--70$^{\circ}$).

\item
Following the same direction --at PA $\sim$60$^{\circ}$-- 
in the next fibre (91; Fig. 6c) we found multiple narrow emission line
components, with ``greatly" enhanced [N\,{\sc ii}]/H$\alpha$ ratio,
very similar to the spectra of jets bow shocks found in the 
Circinus galaxy (by Veilleux  \& Bland--Hawthorn 1997).
This feature was detected at 1\farcs7--2\farcs5 (1340--2030 pc), inside of
the shell S2 and coincident with the knot 2

\item
For the nucleus (i.e., fibre 107, r $<$0.45''$ \sim$ 366 pc) the blue OF bump was
observed, but with very weak emission (Fig. 6a).

\end{enumerate}

Therefore, these results are consistent with the interesting fact that
the blue OF H$\alpha$ bump plus the BAL--I system could be
related with OF clouds associated with the small scale jet.
A discussion of this point will be presented in section 4.3.

Now,  the OF kinematics of the multiple shells will be analysed, for the
central region (R $<$ 6 kpc).
In  the 3  external bubbles S1, S2 and S3 the 2D WHT spectra  show multiple
emission line components (see Fig. 6), specially in the bright knots.
Where the following OF velocities values were measured:

\begin{enumerate}

\item
For the shell S1,

\begin{itemize}
\item
to the south, in the knot 14 (fibre 065),\\
$\langle V_{\rm OF Shell-S1/South-F065}\rangle = [-420 \pm 30]$\,km\,s$^{-1}$\\

\item
to the east, in the knot 4 (fibre 084),\\
$\langle V_{\rm OF Shell-S1/East-F084}\rangle = [-485 \pm 30]$\,km\,s$^{-1}$\\

\item
to the north--west, in the knot 29 (fibre 131),\\
$\langle V_{\rm OF Shell-S1/NorthWest-F131}\rangle = [-500 \pm 30]$\,km\,s$^{-1}$\\

\item
to the west, in the fibre 077,\\
$\langle V_{\rm OF Shell-S1/West-F077}\rangle = [-650 \pm 30]$\,km\,s$^{-1}$\\
\end{itemize}

\item
For the shell S2, in the knot 2 (fibre 091),\\

$\langle V_{\rm OF Shell-S2/F091}\rangle = [-500 \pm 30]$\,km\,s$^{-1}$, and\\

\item
For the shell S3, in the knot 3 (fibres 103, 102)\\

$\langle V_{\rm OF Shell-S3/F103}\rangle = [-230 \pm 40]$\,km\,s$^{-1}$.

\end{enumerate}

These values are consistent with expanding bubbles, and their properties
and origin will be discussed in section 4.3

%\clearpage
% 3.3 ******************************************************************
\subsection{Velocity field of the ionized gas, in the nuclear
and central regions (2D Spectra)}
\label{results-k}

In order to study the kinematics of the ionized gas, in the central region
of Mrk 231, we measured the velocities from the centroids of the stronger
emission lines H$\alpha$, [N {\sc ii}]$\lambda$6584 and [S {\sc ii}]$\lambda
\lambda$6717--6731, fitting gaussians (with the software SPLOT and
SPECFIT; for details see section 2).
In Mrk\,231, first we study  the main emission line component; 
which is strong in all the central region. 
In the shells the presence of multiple OF components required a
detailed study.

Fig. 7 shows  -for the ionized gas- the H$\alpha$ velocity field
for the central region $\sim$16$''\times$12$''$ $\sim$13 kpc$\times 10\,kpc$,
 with a spatial resolution/sampling of 0\farcs9.
This map was constructed using the techniques described in section 2,
and for the main component of the H$\alpha$ emission lines.
The errors vary from approximately
$\pm10\,$km~s$^{-1}$ in the nuclear and central regions (where the emission
lines are strong), to $\sim$20 km~s$^{-1}$ for the weakest lines away
from the nuclear areas.
Using the 2D H$\alpha$ velocity field (VF) we have obtained a mean value
of the systemic velocity V$_{Syst}$ = 12645$\pm10\,$ km s$^{-1}$,
this value was defined as the zero of the VF.

The H$\alpha$ isovelocity colour map (Fig. 7)  shows 
interesting structures:

\begin{enumerate}

\item
Coincident with almost the border of the more extended bubble or shell (S1),
parts of a ring with blueshifted values of velocities and circular shape
was detected. We note, that this structure reach the south border of the
field, and probably part of this ring (to the south) is located outside of our
INTEGRAL field. The radius of this structure/ring is
r $\sim$ 5\farcs0 $\sim$ 4 kpc.

\item
Close to the centre of this ring (which it is located near to the nucleus) 
there is a small and circular lobe with blueshifted velocities.

In addition, inside of the ring 4 symmetric redshifted lobes were
also detected.

\item
To the north-west, part of an elongated blueshifted lobe
was found. In this structure we have measured high values of velocities of
$\Delta$ V $\sim$--200 km s$^{-1}$.

\end{enumerate}

These results are clearly consistent with those found in section 3.2.
In particular, the detection of a blueshifted ring coincident with
the position of the shell S1 confirm the OF/expansion of this shell
(even in the main component of the emission lines).
The presence of an elongate blueshifted structure to the north--east
(of the INTEGRAL VF) is also consistent with the detection --in this area--
of filaments in the [S {\sc ii}]/H$\alpha$ and [N {\sc ii}]/H$\alpha$
maps (see section 3.4) and
the previous detection of  OF ejections (Krabbe et al. 1997).

%\clearpage
% 3.4 ******************************************************************
\subsection{Mapping the emission line ratios and the ionization structure:
 large scale galactic shocks}
\label{results-r}

The set of 2D spectroscopic data, which cover the main structures of the
central region of Mrk 231, allows the investigation of the ionization
structure and the physical conditions in the gaseous phases.
Figs. 8 (a) and (b)
show 2D maps (of 16\farcs4 $\times$ 12\farcs3, with 0\farcs9
spatial sampling) of the emission line ratios
[N {\sc ii}]\ $\lambda$6583/H$\alpha$, and
[S {\sc ii}]\ ${\lambda 6717 + 31}$/H$\alpha$.
These maps were constructed using the techniques described in section 2
and are based on the main component of the emission lines.
Figs. 8(c) and (d) show the values of these ratios for each fibre
(which were used in the generation of the maps). 

Figs. 8(a) and (b)  show interesting features. We note the following:

\begin{enumerate}

\item
For the area where we measured the narrow emission line components, i.e. 
in almost all the circumnuclear region (1\farcs8 $<$ r $<$ 5$''$; i.e.
including the 2 more external shells/bubbles S1 and S2), the
[N\,{\sc ii}]/H$\alpha$ emission line ratios
show high values ($>$ 0.8).
This result, obtained from the 2D map (Fig. 8a) was verified
using the individual spectrum of each fibre (Fig. 8c).

\item
The [S {\sc ii}]\,${\lambda 6717 + 31}$/H$\alpha$  map (Fig. 8b)
also shows high values in the circumnuclear areas.
In particular, we found a range of values for this ratio:
3 $>$ [S {\sc ii}]\,${\lambda 6717 + 31}$/H$\alpha$ $>$ 0.8 (see Figs. 8d,b).

\item
Specially, to the north--west the [S {\sc ii}]\,${\lambda 6717 + 31}$/H$\alpha$ and
[N {\sc ii}]/H$\alpha$ maps
show filaments, which are probably associated with ejection
from the bubble. Lipari et al. (2004d) already discussed in detail the
fact that the [S {\sc ii}]/H$\alpha$ map is one of the best tracer
of filaments associated with OF and shock processes.

\end{enumerate}

Thus, in almost all the circumnuclear and central regions of
Mrk\,231 (including the shells) these
INTEGRAL  emission line ratio maps show
high values ($>$ 0.8), which are consistent with
an ionization process produced mainly by  shock-heating in a
in the outflowing gas of the expanding shells
(Lipari et al. 2004a,d; Dopita \& Sutherland 1995;
Heckman 1980, 1996; Heckman et al.\ 1987, 1990; Dopita 1994).
Furthermore, these ratios (for the main emission line components) are
located in the area of fast shock velocities of
$\sim$400--500\,km\,s$^{-1}$ in the upper part of the
[N {\sc ii}]\,$\lambda$6583/H$\alpha$
vs.\ [S {\sc ii}]\,${\lambda 6717 + 31}$/H$\alpha$ diagram (published by
Dopita \& Sutherland 1995, their Fig. 3a).

Similar results were obtained in the 2D studies of the bubble NGC 3079
(Veilleux et al. 1994) and the OF nebula and bubble  of NGC 2623 and NGC 5514
(L\'{\i}pari et al. 2004a,d). They found that
the ratios [N {\sc ii}]$\lambda$6583/H$\alpha$ and
[S {\sc ii}]$\lambda$ 6717 + 31/H$\alpha$ are $>$ 1, in  almost all the
bubbles and the OF regions.
They associated these results to the presence of large scale OF + shocks.

%\clearpage
% 3.5 ******************************************************************

\subsection{The optical BAL systems and the nuclear spectrum
of Mrk 231 (1D--spectra)}
\label{results-1d}

\subsubsection{Optical BAL systems: the variability of the Na {\sc I}D BAL
III systems}
\label{results-opbal1}

It is  important to study in detail the optical absorption and emission  
features of this IR QSOs, in order to detect similarities and differences 
between this object and similar extreme IR + GW/OF + Fe {\sc ii} emitters.
The optical and UV spectrum of Mrk 231 (see Fig. 9a)
presents the features that are typical of extremely strong optical Fe {\sc ii} 
emitters with BALs: i.e., strong Na I D$\lambda\lambda$5889-5895 absorption,
strong unresolved blends of optical Fe {\sc ii} lines, and very weak
high--excitation forbidden emission lines. It is important to
remark the strong fall in the continuum flux at the UV wavelengths
(this is also a typical feature of dusty luminous infrared galaxies;
see section 3.6).

Boksemberg et al. (1977) detected two nonstellar absorption lines systems
(I and II) with velocities of 6250 and 8000 km s$^{-1}$, respectively.
This correspond to ejection velocities of -6650 and -4950  km s$^{-1}$,
with respect to the systemic velocity of 12900 km s$^{-1}$. Boroson et al.
(1991) detected a new absorption system III in the Na ID Ca H,K and He I,
in their optical spectra of 1988. This system was not present in the spectra
of Boksemberg et al. (1977), taken in 1975. With an observed radial
velocity of 4660 km s$^{-1}$, this new system has the highest ejection
velocity -8240 km s$^{-1}$.
The origin of these absorption--line systems have been
discussed by Boroson et al. (1991); Kollatschny, Dietrich \&
Hagen (1992); Boroson \& Meyers (1992); Lipari et al. (1994);
Foster, Rich \& McCarty  (1995); Smith et al. (1995); Rupke et al. (2002),
and others.

It is interesting to recall about the detection of another absorption system
(IV), which correspond to the systemic velocity and has the narrower
absorption width.
This system was already associated by Boksemberg et al. (1977) with
hot stars in the nucleus of Mrk 231.

Fig. 9b shows the  BAL systems at the line Na ID.
The corresponding width of these absorptions systems are:
FWHM of BAL system I, II and III =  700, 130, 80 km s$^{-1}$, respectively.
These  are in general low values for BAL systems. Thus,
these values are more consistent with those of mini-BALs or
associated absorption lines (AAL; de Kool et al. 2001, 2002).

Now, we will expand the previous study of variability
of the Na ID BAL III system  (covering almost all the period
in which this system appeared).
In order to study the behaviour of the variability of this  absorption
system, the measured values of the equivalent width ratio  of
BAL III/(BAL I + II) were included in Table 3. 
These data were obtained from our observation and also derived from published
spectra (together with data previously published by Kollatschny et al. 1992;
and Forster et al. 1995). We note that the previous study of variability
cover the period 1980--1992 (with only few point after the maximum).
The observation of Table 3 cover the period between $\sim$1980--2000.
Figs. 10a,b  show
the shape of the BAL-III light curve (LC), which is clearly asymmetric with:
a steep increase, a clear maximum and an exponential fall.
In particular Fig. 10b shows a very good fit of the LC fall, using
an exponential function. In general, the shape of the BAL III system LC
is similar to those LC of  SNe flux, emission line, etc.

The probable origin of this BAL III system will be discussed in section 4.3.
Specifically, we will discuss 4 possible scenarios. One of this scenario
is an explosive event associated with explosions of giant SNe/hypernova
(with very massive progenitors and located close to the AGN and/or in 
 accretion disks of AGNs).

It is important to note that previously it was assumed that this light
curve (of BAL III) is symmetric. This type of variability was associated
mainly with absorbing clouds crossing the central continuum light source
(Kollatschny et al. 1992).

\subsubsection{Out flow from the nuclear optical emission line}
\label{results-opof}

For the nuclear region of Mrk 231, a new detailed study of multiple
emission line components was performed.
In particular, for the line [O {\sc ii}]$\lambda$3727
the presence of at least 3 components were detected (see Fig. 9c):
with OF1 and OF2 velocities of --800 and --1300 km s$^{-1}$.
The value of the OF previously reported by Lipari et al. (1994)
was obtained from the blend of the OF1+OF2.
In addition,  in the line Ni {\sc ii} an OF component was
measured, and it was measured a value similar to the OF1,
of the line [O {\sc ii}].

It is interesting to note that Schmidt \& Miller (1985) from a
spectropolarimetry study detected an OF velocity (for the dust) of
700 km s$^{-1}$. Which is a value very close --within the errors--
to our OF1 velocity.

A discussion of the extreme optical Fe {\sc ii} emission of this IR QSO,
was reported by Boksemberg et al. (1978) and Lipari et al. (1994). 
A ratio of Fe {\sc II}$_{OPT}$/H$\beta \sim$8, [O {\sc III}$5007$/H$\beta
\sim$0.01, Na ID/H$\alpha \sim$ 0.03 and H$\alpha$/H$\beta \sim$5
were measured. These results are
in agreement with the typical values obtained for extreme Fe {\sc II}$_{OPT}$
emitters, like IRAS 07598+6508, PG 1700+518, IRAS 18508-7815 (Lipari 1994).

% 3.6 ******************************************************************

\subsection{The UV BAL systems and spectrum of Mrk 231 and
IR + GW/OF + Fe {\sc ii} QSOs}
\label{results-uv1d}

\subsubsection{UV spectrum and BALs in Mrk 231}
\label{results-uv1dm231}

Significant reddening in luminous infrared sources hampers the study of their
UV properties.
For Mrk 231, Figs. 11a,b  and 9a show --at UV wavelengths-- a strong
reddening, with weak Mg {\sc ii}, C {\sc iv} and Ly$\alpha$
emission lines. UV absorption associated with the BAL system I are
present in the Mg {\sc ii} and C {\sc iv}  lines (Figs. 11a,b). In particular,
The low--ionization emission line Mg II is observed superposed to the BAL 
system I (Fig. 11a). We measured for this absorption line a width (FWHM)
of $\sim$ 700 km s$^{-1}$. This value is equal to the that measured
--in this system-- in the Na ID optical line.
In addition, in the region between $\lambda\lambda$2000--2800
there are weak absorption due to UV Fe II.

It is important to study in detail the UV (plus optical) continuum
of extreme IR--Fe {\sc ii} emitters and specifically  BAL--QSOs because
they show a clear deficit in the  UV flux as compared to the
non-BAL QSOs (see Weymann et al. 1991; Sprayberry
\& Foltz 1992). 
Fig. 9a  shows a clear drop in the level of the UV  (and even in
the B) continuum of Mrk 231, for $\lambda$ shorter than 4500 \AA.
We found that an SMC--like reddening of E(B--V) $\sim$ 0.13 accounts
for the differences between the  UV spectrum of Mrk 231 and the
non--BAL QSO composite/mean spectrum (Reichard et al. 2003).

\subsubsection{UV-BALs in IR + GW/OF + Fe {\sc ii} QSOs}
\label{results-uv1dqso}

From our programme of study and search of UV and optical BAL systems in
extreme IR + GW/OF + Fe{\sc ii} QSOs  we already found interesting results in
relation with the UV low ionization BAL system detected in 
IRAS 07598+6508 (Lipari 1994).

We recall that
the standard definition of BAL QSO (Weymann et al. 1991) is based in
the measurement of  the equivalent width of the C {\sc iv} resonance
absorption line system (called balcity index: BI).
The QSOs with BI $>$ 0 km s$^{-1}$ are considered BAL QSO.
In part by this reason we observed the BAL candidates at
UV (for C {\sc iv} and Mg {\sc ii} lines) and
optical (for Na ID absorption) wavelengths.

From this programme, the following main results
were found:

\begin{enumerate}

\item
The UV--{\it IUE} spectrum of IRAS 21219-1757  (Figs. 12a,b) clearly
shows a blend of emission and absorption features --at low S/N--
in the C {\sc iv} line.
In these two figures the UV spectrum of IRAS 21219-1757 is superposed
with the spectra of standard low ionization BAL  and non-BALs QSO,
respectively (obtained from Reichard et al. 2003).
From these figures it is clear that the profile of the line C {\sc iv}
is very similar to the spectrum of low-ionization BAL QSO
(and different to the spectra of non-BAL QSOs).

In general, the UV and optical spectral features of IRAS 21219-1757 are
similar to those observed in nearby low--ionization BAL
QSOs: like IRAS 07598+6508 and PG 1700 +518. These objects  are also
extreme IR + Fe {\sc ii} emitters
(Lipari 1994; Wampler 1985;  Lanzeta et al. 1993; Hines \& Wills 1995).

In the IUE UV spectrum of IRAS 21219-1757 the following features were also
observed:

\begin{itemize}

\item
The absorption feature in the C {\sc iv}$\lambda$1549 line
 extends from $\sim$1719 to $\sim$1665 \AA, implying 
outflow/ejection velocities of $\sim$1500  to $\sim$10400 km s$^{-1}$. 

The emission line of C {\sc iv}$\lambda$1549 is also weak, as has been observed
in BAL--QSO (Weymann et al. 1991).

\item
Figs. 12a,b also show the presence of weak low-- and high--ionization 
``emission" lines.
In particular, Ly$\alpha$ and N V$\lambda$1240 are weak.

\item
There are also weak emissions from O I$\lambda$1303 and  Si IV+O IV
$\lambda$1400. 

\end{itemize}

We note that the archive HST STIS spectra of IRAS 21219-1757 
does not cover  the region where we detected the BAL
in this IR QSO (the line C {\sc iv}$\lambda$1549).

\item
In order to compare this  BAL IR QSO, with the previous observed
BAL IR QSOs, we have also included in Fig. 13 the UV spectra of
IRAS 07598+6508, PG 1700+518, IRAS 14026+4341.

Furthermore, it is important to note
that the UV--BAL features observed in these  IR + GW/OF + Fe {\sc ii}
QSOs are very similar to
those observed at high redshift in low--ionization BAL--QSOs 
(at z $>$ 6.0; see Maiolino et al. 2003).

\item

In Fig. 13d the IUE spectrum of IRAS 18508-7815 is presented. In this extreme IR
+ GW/OF + Fe {\sc ii} QSO there is only evidence of a possible narrow
absorption system in the line C {\sc iv} (and close to the systemic velocity).

\item
For PHL 1092 (another extreme IR + GW/OF + Fe {\sc ii} QSO) we have
studied mainly the region of the Mg {\sc ii}+UV--Fe {\sc ii}.
Fig. 13e does not show evidence of Mg II BAL systems, in PHL 1092.
However, the fit of the emission line in this area required to include
an UV Fe {\sc ii} template (we used the UV Fe II emission of I Zw 1,
from the HST spectra) and also a blue OF component. For
this blue component a OF value of --8565 km s$^{-1}$ was measured.

At the optical/red wavelengths, this OF (with very high velocity) was
not detected at the line H$\alpha$.
However, we found  3 blue OF componets with relativelly
low velocities. For these blue components the
following OF values were measured, OF1 = -1500, OF2 = -2500 and OF3 =
-3100 km s$^{-1}$ (see Fig. 13f). These velocities are almost the
same to those obtained for the OFs detected in H$\beta$ (using the
offset method; Lipari et al. 2004d: their Table 8).

It is  important to note that Dietrich et al. (2002) and Barth et al.
(2003) already discussed that in order to obtain a good fit of the UV
lines Mg {\sc ii} + Fe {\sc ii} in very high redshift QSOs,
they need to include a strong blushifted component (they explain that
this component was used without a physical explanation).
In particular, Barth et al. (2003) in their Fig. 2 show
this strong blue OF component in the line Mg {\sc ii}, for
the Fe {\sc ii}-QSO SDSS J114816.64+525150.3. This object is
one of the younger QSO known, with a redshift z =  6.4.
They measured for the Mg {\sc ii} OF component --of this distant/young
QSO-- a value of $\sim$ --5700 km s$^{-1}$ (Barth et al. 2003: their
Fig. 2), which is almost 1/3 smaller than the value obtained for PHL 1092.

Therefore,
we strongly suggest that this type of blue component observed
in the Mg {\sc ii} emission line --in very high redshift QSOs-- is associated
with extreme OF processes.
Again, the results obtained from QSOs  at very high redshift
(by Dietrich et al. 2002 and Barth et al. 2003) are very similar to our
results for  IR + GW + Fe {\sc ii} QSOs at low
redshift (where we detected blue OF components in H$\alpha$,
H$\beta$, [O {\sc iii}], [O {\sc ii}] and Mg {\sc ii} emission lines).
However,
these OF emissions are probably associated with very
different OF processes, similar to those found in this paper for Mrk 231.

\end{enumerate}

% ******************************************************************
% ******************************************************************
\clearpage
\section{DISCUSSION} \label{discussion}

In this section we discuss the properties of Mrk 231 as:
(i) an evolving elliptical galaxy;
(ii) a merger with composite nuclear energy; 
(iii) an EVOF QSO, with also a composite nature;
(iv) as an exploding QSO.
In addition, we analyse the IR colours diagram for a large sample of
 IR mergers/QSOs with galactic winds (including BAL QSOs and Mrk 231).

% 4.1 ******************************************************************
\subsection{The merger Mrk\,231 as an evolving elliptical galaxy
with galactic wind}
\label{discussion-elipt}

Hamilton \& Kell (1987) already found that the optical $R$ broad band surface
brightness profiles of Mrk 231 follow the $r^{1/4}$ law (in the
range 1.0 kpc $\leq r \leq$ 10 kpc).
Lipari et al.\ (1994) from high resolution {\itshape NOT\/} 
V broad band images found that the nucleus is compact at scales of
0\farcs7 ($\sim$500 pc).
Soifer et al.\ (2000) using Keck mid-IR 7.9 to 19.7 $\mu$m broad band images
found that the nucleus is compact and unresolved at 0\farcs13 (100 pc).
At radio wavelengths, Condon et al.\ (1991) also detected an unresolved
radio core, smaller than 0\farcs25 (at 8.44 GHz), in Mrk 231.
{\itshape HST\/} ACS/U, WFPC2/I and NICMOS/H broad band images (see section
3 and Quillen et al. 2001) confirmed that at scales better than 0\farcs1
($\sim$80 pc, FWHM), the bright compact nuclear peak is unresolved.
These results suggest that the nuclei of the
original colliding galaxies have coalesced into a common nucleus with an AGN,
and that the merger is in a very advanced phase (a relaxed
system), probably evolving to an elliptical galaxy.

Therefore Mrk 231 is another candidates for a
proto-elliptical galaxy. It is important to note
that from the theoretical point of view there is a wealth of literature
proposing a relevant role for galactic wind and OF in the formation
and evolution of elliptical galaxies. We can highlight that: (i)
Mathews \& Baker (1971) suggested that a galactic wind can explain
the deficiency of observed gas as compared to that returned by
stars in ellipticals; (ii) Larson (1974), Vader (1986) and Kauffmann
\& Charlot (1998) proposed for ellipticals a galactic wind
associated with the early star formation episode to explain their
colour--magnitude and mass--metallicity relation;
(iii) Larson (1974), Vader (1987) and
Dekel \& Silk (1986) proposed that dwarf ellipticals evolved from
initially more massive elliptical galaxies that suffered
substantial gas loss through supernova-driven galactic winds.

The results obtained for Mrk 231 (and previous studies of
mergers with OF) suggest that extreme starbursts plus AGN with galactic
winds play an important role in galaxy evolution, e.g.\ giant expanding
shells in composite galactic winds could produce BAL systems, massive
starbursts could generate AGNs/QSOs, successive extreme starburst processes
could transform the observed high gas-density (in the ISM of mergers) to
high stellar-density, etc. Furthermore, these extreme
starburst+AGN with galactic winds --induced by mergers-- could be a
nearby analogy for processes that occurred at high redshift, when
the galaxies and QSOs formed, i.e.\ similar to the
first massive star formation episodes (population III stars) in hierarchical
mergers and/or primordial collapses  (Larson 2003, 1999, 1998; Bromm \&
Loeb 2003; Bromm, Coppi, \& Larson 1999; Scannapieco \& Broadhurst 2001).

% 4.2 ******************************************************************
\subsection{The composite nucleus of Mrk 231: an AGN/QSO embedded in an
extreme starburst}
\label{discussion-nuc}

In order to study the nature and origin of the complex nuclear outflow
process detected in Mrk\,231, it is important to
analyse --previously-- the  composite nature of the nuclear energy.

\subsubsection{The  AGN/QSO }
\label{discussion-nuc1}

There is clear and interesting evidence of the presence of an AGN/QSO
in the nucleus of Mrk 231.
Specifically, at radio wavelength Preuss \& Fosbury (1983) already detected
a very compact nuclear radio source, at scale $<$1 pc (using VLBI
interferometric observations).
Recently, using the Very Long Baseline Array (VLBA) Ulvestad et al.
(1999a,b) found a parsec scale radio jet, with an extension of $\sim$2.5 pc.

At X-ray wavelength, using ASCSA data Turner (1999)  found that the
optical-to-X ray spectrum of Mrk 231 is typical of a QSO. In addition,
using Chandra observations Gallagher et al. (2002)  detected that the
bulk of the X-ray luminosity is emitted from an unresolved
nuclear point source and the spectrum is remarkably hard.

In section 3, we found clear evidence that --at least--  the Na ID BAL
system I is associated with the presence of the AGN+jet.

\subsubsection{The  extreme starburst}
\label{discussion-nuc2}

Recently, the star formation in the nucleus of Mrk 231 was
studied at very high resolution by Bryant \& Scoville (1996) and
Downes \& Solomon (1998), observing the CO gas. In particular,
Bryant \& Scoville (1996) found that the CO distribution is elongated with
kinematics properties consistent with a disk oriented in the east-west
direction, and containing a mass of 3 $\times$ 10$^9$ M$_{\odot}$ of
molecular gas (within a diameter of 1\farcs0, 814 pc).
Downes \& Solomon (1998) using CO interferometer data also detected a
1\farcs2 diameter inner disk plus a 3\farcs0 diameter outer disk. They found
{\it extreme star formation} processes in the nuclear CO disks or rings of
Mrk 231. Confirming previous results that
suggested the presence of a strong nuclear starburst in this IR merger
(Hamilton \& Keel 1987; Hutching \& Neff 1987, and others). In particular
Lipari et al. (1994) also suggested the presence of an extreme
nuclear starburst, as the main source of the strong galactic wind,
detected in Mrk 231.
These works suggested that at least 1/3 of the total source of the
nuclear energy in Mrk 231 is generated by the extreme starburst.

On the other hand,
continuum observations at 21 cm with VLBA (Carrilli, Wrobel \& Ulvestad
1998) show emission from an extended disk of 370 pc, that exhibits
absorption against the radio emission from the inner disk (r $<$ 160 pc).
Further radio continuum observations reported by Taylor et al. (1999)
trace this emission out to 1\farcs0. They interpret this disk radio emission
 also as due to {\it strong star formation} activity with a SFR of 60--200
M$_{\odot}$ yr$^{-1}$.
At near-IR wavelengths, the nuclear spectrum of this galaxy shows an
anomalously large Pa$\alpha$/Br$\gamma$ ratio (Cutri et al.\ 1984),
which imply that the electron density in the broad line clouds of Mrk
231 is higher than the standard values of the BLR in AGNs/QSOs.

In section 3.4, it was found that extended shocks (associated with multiple
expanding bubbles) are also an important source of
energy and ionization, in the nuclear and circumnuclear regions.
In addition, we found evidence of a multiple explosive event in Mrk 231,
which is in part associated with the extreme starburst, probably generated
during the merger process.

At least two strong starbursts were generated in the merger process of
Mrk 231: one associated with the origin of an extended post-starburst
population (associated with the underlain hot stellar population in
the optical spectrum), plus a young and dusty nuclear extreme starburst.
%(of $\sim$10--xx? Myr).

% 4.3 ******************************************************************
\subsection{The composite nuclear outflow + BAL systems in Mrk 231:
an exploding QSO?}
\label{discussion-ofbal}

In section 3  very interesting results, about the OF and BAL systems in
Mrk 231 were found. Specifically, it was detected: (i) multiple concentric
expanding bubbles, (ii) a blue H$\alpha$-emission bump at the ejection
velocity  of the BAL I, and (iii) a light curve for the
BAL-III system showing a steep increase and exponential
fall (very similar to the shape of a SN light curve).

In order to analyse these new results, it is important to remark
previous studies at radio wavelengths, which are relevant for the
discussion of the OF in Mrk 231. In particular:

\begin{enumerate}

\item
VLBI images at 1.7 Ghz (Neff \& Ulvestad 1988) and  VLBA images at 2.3 Ghz
(Ulvestad et al. 1999a) show a north-south triple structure with a central
unresolved core and two symmetric resolved lobes; with a total extension
of $\sim$40 pc.
This radio structure is elongated (at PA $\sim$00$^{\circ}$) in the direction 
perpendicular to a 350 pc HI + CO starburst disk/ring.

\item
VLA images at 4.9 Ghz (Baum et al. 1993) and at 1.5 Ghz (Ulvestad et al.
1999a) show a very large structure of $\sim$35--50 kpc, also elongated
in the direction north-south.

\item
Using the VLBA at 15.3 GHz, Ulvestad et al. (1999a,b) found a parsec
scale radio jet, with an extension of $\sim$2.5 pc, low apparent speed,
and at the position angle PA $\sim$65$^{\circ}$.

\end{enumerate}

\subsubsection{Multiple expanding shells}
\label{discussion-ofbal1}

The presence of multiple concentric expanding bubbles/shells, with centre in
the nucleus and with highly symmetric circular shape could be associated
mainly with giant {\it symmetric} explosive events.
These giant explosive events could be explained in a composite
scenario:
where mainly the interaction between the starburst and the AGN could
generate giant explosive events.
In particular, Artymowicz, Lin, \& Wampler (1993) and Collin \& Zahan (1999)
already analysed the evolution of the star formation (SF) close to  super
massive black hole (SMBH) and inside of accretion disks. They suggested that
the condition of the SF close to the AGNs could be similar to those of the
early/first SF events, where it is expected giant explosive process generated
by hyper--novae (with very massive progenitors: M $\sim$100--200 M$_{\odot}$).

For the shell S1, Lipari et al. (1994) already proposed that the morphology
and kinematics of the blue arc (located to the south of the nucleus) are
in agreement with the results of glactic winds hydrodynamic models.
Specifically, the starburst GW models of Suchkov et al. (1994) predict
after 8 Myr in the blowout phase: blue arcs with radius r $\sim$
3 kpc. We noted that after 8 Myr the starburst is in the type II SN phase.

It is important to remark that exactly in the direction of this arc
(north-south from the nucleus, PA $\sim$00$^{\circ}$)
Neff \& Ulvestad (1988) and Ulvestad et al. (1999a) found a  elongate
structure of $\sim$40 pc, which is perpendicular to the CO starburst
disk/ring.
Furthermore, Baum et al. (1993) and Ulvestad et al. (1999a) detected a 
large structure (of $\sim$35--50 kpc), also elongated
in the direction north-south. These results are all clearly consistent with
the blowout or rupture phase of the bubble S1.

Following our study of the superbubble in NGC 5514 (Lipari et al. 2004d)
the dynamical timescale of  the more extended shells S1 and S2 in Mrk 231
were derived, using the relation:\\

t$_{dyn} =$ 1.0 $\times 10^6$ R$_{bubble, kpc}$ V$^{-1}_{bubble, 1000}$ yr \\

where  R$_{bubble, kpc}$ and V$^{-1}_{bubble, 1000}$ are the linear dimension
of the bubble and the velocity of the entrained material (in units of kpc and
1000 km s$^{-1}$, respectively). The following values were obtained for the
shells S1 and S2:

\begin{itemize}
\item
t$_{dyn-S1} \sim$ 4.8 $\times$ 10$^6$ yr;
\item
t$_{dyn-S2} \sim$ 3.0 $\times$ 10$^6$ yr.
\end{itemize}

Then a period of $\sim$1.8 $\times$ 10$^6$ yr is the separation between  the
two explosive events that probably generate the shells S1 and S2.

From the 2D spectra a mean value of  H$\alpha$ flux
of  0.5 $\times$ 10$^{-13}$ erg cm$^{-2}$ s$^{-1}$ was obtained,
for the OF associated with the shell S1.
Using the relations given by Mendes de Olivera et al.\ (1998) and
Colina et al.\ (1991)  a value for the ionized-gas mass, in the
OF associated with the shell S1, of 2.0 $\times$ 10$^{6}$ $M_{\odot}$ was found.
Using this value of ionized-gas mass, we derived for the kinetic energy of
the OF--S1,
E$_{\rm KIN OF-S1} \sim$  0.5 $\times$ M$_{\rm OF-IG}$ $\times$
$\langle V_{\rm OF-S1}\rangle^{2}$ = 2.0 $\times$ 10$^{54}$ erg. This
high level of energy, obviously required the presence of multiple SNe events,
or an unusual type of ``giant SN" or hypernovae.
Heiles (1979) already suggested that this
last alternative need to be considered seriously, in order to explain
supershells with enegy $>$ 3.0 $\times$10$^{52}$ erg (even if these energies
of supershells are hundred of times larger than that available from a
single and ``standard" SNe).

In section 4.3.4 the origin of this multiple shells will be discussed in
detail.

\subsubsection{The blue H$\alpha$-emission bump at velocity of BAL I system}
\label{discussion-ofbal2}

In section 3 another interesting results was found:
a blue H$\alpha$-emission bump at the ejection velocity of the BAL I
(and in a area located very close to the nucleus and at the same PA of
the radio jet). Furthermore, in the same direction where we detected
this bump (at PA $\sim$60--70$^{\circ}$) at 1\farcs3--2\farcs2, we also
 found multiple narrow emission line components, with ``greatly" enhanced
[N\,{\sc ii}]/H$\alpha$ ratio (very similar to the spectra of jets bow
shocks). These results are consistent with a scenario where the BAL--I
system is generated in OF clouds associated with the parsec scale jet.

It is important to note, that
de Kool et al. (2001, 2002) found a very interesting result
--and physically similar to those found for Mrk 231--
in their Keck high resolution spectroscopic study of AAL and BAL, in
the QSOs FIRST J104459.6+365605 and FBQS 0840+3633.
The absorption lines of these QSOs cover a ranges of velocities of --200
to --1200,  --3400 to --5200 km s$^{-1}$; and the width of the
individual absorption lines ranges from 50 to 1000 km s$^{-1}$
(these are values similar to those measured in Mrk 231).
They found that the distances between the AGN and the region where the
OF gas generate the AAL and BAL line are $\sim$700 and $\sim$230 pc,
respectively.
Therefore, for Mrk 231 (BAL system--I), J104459.6+365605 and FBQS 0840+3633
the distances found between BAL or AAL forming regions and the
continuum source (AGN) are large, of: $\sim$ 200--1000 pc (BALs and AALs
are generally thought to be formed in OF at a much smaller distance
from the nucleus; de Kool et al. 2001).

\subsubsection{The origin of the variability of the BAL III system}
\label{discussion-ofbal2}

Boroson et al. (1991) and Kollatschny et al. (1992) already explained
some physical mechanism that could explain variations in the absorption line
strengths. Here we re-analyse these and new mechanisms, according to the
results obtained in section 3.

The principal mechanisms that can explain the variability of the NA ID
BAL III system, can be enumerate as follow:

\begin{enumerate}

\item
Changes in the intensity of the continuum source (and the ionizing parameter
in the absorbing region). Since the continuum flux remained constant in the
period in which the system III varied, the first possible explanation
can be excluded.

\item
Changes in the ionizing parameters and density in the absorbing clouds when
they move away from the centre. This second possibility have been excluded by
Boroson et al. (1991), since the relative change in distance can not explain
the strong change and the systematic increasing and decreasing epochs.

\item
Motion of absorption clouds with traverse velocity occulting the continuum
source. This was in the past the best explanation for part of the
light curve of BAL III, observed between 1980 and 1992. However,
the new observation --presented in section 3.5-- exclude this explanation,
sincethe LC fall is clearly asymmetric/exponential.

\item
An explosive scenario for the origin of the BAL III system could explain:
the shape of the light curve variability, and also the presence of
multiple concentric expanding bubbles/shells (with circular shape).
In the next section we will discuss in detail this explosive scenario
and the probable origin.

\end{enumerate}

\subsubsection{The exploding scenario for Mrk 231 and  BAL systems}
\label{discussion-ofbal1}

In section 3  strong evidence for a composite origin of
the OF in the nuclear region of Mrk 231 were detected.
Furthermore, for IR mergers/QSOs with EVOF a similar result was found 
(see for details section 1 and Lipari et al. 2004a,d).
Thus it is probably that also very energetic explosive events are
originated in the interaction between  extreme starbursts and  AGNs.
We already explained that Artymowicz et al. (1993) and Collin \& Zahan (1999)
suggested that the conditions of the SF close to the AGNs+accretion disks
could be similar to those of the early/first SF events, where it is expected
giant explosive processes generated by giant SNe.

On the other hand, different works also proposed that the BLR could be
associated with OF processes in accretion disks, extended stellar envelopes,
etc (see for references/review Sulentic, Marziani, \& Dultzin--Hacyan 2000;
Scoville \& Norman 1996; Dyson \& Perry 1992; Terlevich et al. 1992).
Thus an interesting questions is: could an explosive event
generate a very unusual spectrum similar to that detected in
the nuclear region of Mrk 231?.
Figure 14 shows the superposition of the spectrum of the unusual radio
SN of type II 1979c (observed in 1979 June 26.18; Branch et al. 1981).
Only using colours we can distinguish each spectrum, since they are
almost identical.
Thus, an more constant OF (than  a single/standard SN) could
explain even the spectrum of the BLR in Mrk 231.

In the composite (starburst+AGN) scenarios, two main
theoretical models for the origin of BAL systems were proposed:
(i) for IR dusty QSOs/galaxies, in the outflowing gas+dust
material the presence of discrete trails of debris (shed by
individual mass-loss stars) produce the BAL features (Scoville \&
Norman 1996); and (ii) in SN ejecta, which are shock
heated when a fast forward shock moves out into the ISM (with a
velocity roughly equal to the ejecta) and a reverse shock accelerates
back and moves towards the explosion centre; the suppression
of red-shifted absorption lines arise since SN debris moving
toward the central source are slowed down much more rapidly -by
the wind- than is material moving away (Perry \& Dyson
1992; Perry 1992).
The presence of large galactic-scale OF,  shells or
rings in IR QSOs could be a third explanation -in a 
composite scenario- for the origin of BAL systems, in these
objects.

It is interesting to recall that in their study  of PG 1700+518,
Hazard et al. (1984)  already suggested that: high redshift
low--ionization BAL--QSOs could be explained by a violent ejection during
the first onset of the QSO activity, similar to a {\it ``giant SN explosion"}.
This approach is very similar to that proposed in this paper for Mrk 231
(where {\it ``giant expandig superbubbles"} were foundd).
Therefore, it is important to study in detail the contribution of
extended outflow associated with superwind/superbubble regions in order
to explain the origin of some BAL systems and even the BLR of Mrk 231  
(see Lipari 1994; Lipari et al. 1994; Terlevich et al. 1992; Dyson,
Perry \& Williams 1992;  Voit, Weymann \& Korista 1993).
In general,  the results and conclusions presented in
Section 3 appear to be in good agreement with this unusual explosive scenario 
proposed in this paper for Mrk 231:
where the strong IR+Fe II emission and BAL system could be linked --in
part-- to a violent extreme--starburst+AGN.

Finally, the serendipitous finding that a high number of QSOs show
BAL systems at low redshift (in IR + Fe {\sc ii} emitters)
and at very high redshift (in Fe {\sc ii} emitters; Maiolino et al. 2003,
2004) is very interesting. On the other hand,
the relation between BAL, IR, GW and Fe II emission is
not well understood. Fe II emission (with low ionization level) could be 
originate in warm regions, and thus obscured from the
direct ionizing UV photons. In the case of strong Fe {\sc ii} emitters,
the obscuring material may be in the form of an expanding shell. 
If so, BAL+Fe {\sc ii}+IR objects are not ordinary AGN, rather, they may be 
associated with early starburst activities.
If some absorptions are the result of  outbursts, the mass
involved should be much larger than a typical supernova, since the Fe II 
emission and``some" BAL systems do not fall in years.
However, in order to understand this type of giant outbursts (i.e., from
hypernova, SN explosion from population III stars, SN in accretion discks
of AGNs, etc) it is required more detailed theoretical studies.
The results found in this paper for Mrk 231 (and similar IR QSOs)
support the reality of these giant explosive processes.

% 4.4 ******************************************************************
\subsection{Mrk 231 and extreme IR + GW/OF + Fe {\sc ii} + BAL QSOs as
transition/young QSOs}
\label{discussion-transi}

In this section, we will analyse the properties of Mrk 231
together with the other members of the interesting  group of
{\it IR + GW/OF + Fe {\sc ii} + BAL transition/young QSOs}.

\subsubsection{The IR colours diagram for a small sample of
IR + GW/OF + Fe {\sc ii} + BAL  QSOs}
\label{discussion-transi1}

In general, the IRAS colour-colour diagrams have been used as an important 
tool to detect and discriminate different types of activity in the 
nuclear/circumnuclear regions of galaxies (e.g., Seyfert and starburst
activity; see de Gripj et al. 1985, 1987; Sekiguchi 1987; Rowan-Robinson
\& Crawford 1989).
Lipari (1994) already found that the IR colours (i.e., IR energy
distribution) of  $\sim$10 {\it extreme}  IR +
Fe {\sc ii} QSOs (e.g., Mrk 231, IRAS~07598+6508, IRAS 17002+5153,
IRAS 14026+4341, IRAS~00275-2859, I Zw 1, IRAS 18508-7815, IRAS 21219-1757,
Mrk 507, Mrk 957, etc) are distributed between the power law (PL) and the
black-body (BB) regions: i.e., the {\it transition area}.
On the other hand, the {\it low and strong} IR--Fe II emitters
are  located mainly in the PL region. 

In particular, we detected that  Mrk 231 and IRAS~07598+6508 
(the nearest IR + GW/OF + Fe {\sc ii} + BAL QSOs) have a
close position in this diagram: near
to the BB area; thus showing both systems strong starburst components.
Recently, Canalizo \& Stocton (2001) confirmed that the host galaxies of
both QSOs have strong starburst populations (using Keck spectroscopy).
In general, the extreme Fe II + IR  emitters show weaker ``Seyfert/AGN"
components than the low Fe II QSOs.

Furthermore, we found that the IR properties/colours of narrow--line
Seyfert 1 IR AGN with extreme Fe II emission (like Mrk 507 and Mrk 957),
lie also in the {\it transition} area between the BB and PL regions,
and close to standard disk galaxies (Rowan-Robinson \& Crawford 1989)
and starburst objects (Sekiguchi 1987).

It is important to remark
that of a total of $\sim$10 IR transition objects of this original
sample, the first 4 systems are BAL IR QSOs. Therefore, we
already suggested that BALs IR QSOs (like Mrk 231, IRAS~07598+6508,
IRAS 17002+5153 and IRAS 14026+4341) could be associated with the {\it
young phase of the QSO activity}.

\subsubsection{The IR colours diagram for a large sample of
IR mergers and IR QSOs with galactic winds}
\label{discussion-trannsi2}

In this paper, using our data base of more than 50 IR Mergers and QSOs
with galactic winds and using for comparison the large sample of standard
PG QSO (from Boroson \& Green 1992) we will expand our previous study.
Figure 15  and Table 4 show the IR energy distribution [spectral indexes 
$\alpha$${(60,25)}$ vs. $\alpha$${(100,60)}$; where $\alpha$${(\lambda 2,
\lambda 1)}$ = --log[F($\lambda 2$)/F($\lambda 1$)]/log[$\lambda 2$/$\lambda 
1$]] for:
(i)  IR mergers and IR QSO with GW (originally 51 IR systems, from Table 4);
(ii) standard QSOs from the PG QSOs sample of Boroson \& Green (1992;
originally 87 PG QSOs that have $z\leq$0.5).

IR fluxes--densities, in the bands of 12, 25, 60 and 100 $\mu$m, were obtained
from the IRAS and ISO Archival Catalogue (using NED).
Only objects with a good detection in the three required bands have been
included. Also, the localisation of the three main regions in 
this colour-colour diagram (i.e., the QSOs/Seyferts, Starbursts, and
powerful IR galaxies areas) have been plotted.
An inspection of this diagram clearly shows the following:

\begin{enumerate}

\item
All the IR mergers with LVOF are located very close to the BB and
starburst area.

\item
Almost all the IR QSOs with EVOF are located in the transition region.

The only object, NGC 3079, not following this trend is the
only spiral galaxy in this sample.

\item
The standard QSOs and radio QSOs are located around the PL region.

\item
All the BAL IR QSOs are located in the transition region, in almost a clear
sequence: from
Mrk 231 (close to the BB area) $\to$  IRAS 07598+6508 $\to$ IRAS 21219-1757
$\to$ IRAS/PG 17072+5153 and IRAS 14026+4341 (close to the PL area) $\to$
standard QSOs.

\end{enumerate}

Therefore, these results confirm our previous finding (obtained from
a small sample of IR galaxies): in the sense that
{\it IR QSOs are probably ``transition" objects},
between IR mergers and standard QSOs.
However, it is important to remark that it is required more
detailed studies of individual IR systems (like Mrk 231) in order to determine
the different paths of possible evolution (see Farrah et al. 2001;
Rigopoulow et al. 1999).

\subsubsection{The possible relation between  IR + GW/OF + Fe {\sc ii} + BAL
transition/young QSOs and very high redshift BAL QSOs}
\label{discussion-transi3}

It has been proposed that  extreme starburst + galactic wind processes
associated with IR mergers could play a relevant role in the formation and
evolution of galaxies and QSOs/AGNs, i.e. in their structure, kinematics,
metallicity, etc.\ (see sections 1 for references). Furthermore,
recent detailed observations and theoretical studies have confirmed that 
OF, galactic winds, BAL, large amount of gas+dust and strong Fe {\sc ii}
emission are important components and processes at high redshift
($z$ $\sim$ 4--6), when the galaxies and QSOs formed
(Frye, Broadhurst, \& Benitez 2002; Ajiki et al. 2002;
Pettini et al. 2001; Dawson et al. 2002; Carilli et al. 2004a,b;
Solomon et al. 2003; Taniguchi \& Sioya 2000; Barth et al. 2003;
Iwamuro et al. 2002; Freudling, Corbin, \& Korista 2003; Bertoldi et
al. 2003a,b; Omont et al. 2001; Maiolino et al. 2004a,b).

Specifically, cm and mm observations of thermal molecular line emission
from high redshift QSOs reveal massive gas reservoirs (10$^{10}$ to
10$^{11}$ M$_{\odot}$) required to fuel extreme star formation
(Carilli et al. 2004a,b; Solomon et al. 2003; Carilli \& Blain 2002;
Carilli, Menten \& Yung 1999).
In addition, cm and mm observations also reveal in high redshift QSOs
very large amount of dust of  10$^{8}$ M$_{\odot}$. The presence
of this large mass of dust is associated mainly with massive
starbursts concurrent with AGNs, which implied star formation rates
greater than 10$^3$ M$_{\odot}$ yr$^{-1}$ (Carilli et al. 2000;
Bertoldi et al. 2003a,b; Omont et al. 2001; Cox et al. 2002).

On the other hand,
Maiolino et al. (2004a,b, 2003)  presented near-IR spectra
of eight of the more distant QSO (at 4.9 $<$ z $<$ 6.4). Half of these
QSOs are characterised by strong UV BAL systems (at C IV, Mg II, Si IV, Al
III lines).
Although the sample is small, the large fraction of BAL QSOs suggest that
the accretion of gas, the amount of dust and the presence of OF process
are more larger (in these objects) than in standard QSO at z $<$ 4.0.
They also suggested that the very high amount of dust was generated by
early explosions of SNe (Maiolino et al. 2004b).

Finally, it is important to remark the similar properties  found in
IR + GW/OF + Fe {\sc ii} + BAL QSOs at low redshift (like Mrk 231)
and  very high redshift BAL QSOs (at z $\sim$ 6.0; Maiolino et al. 2003,
2004a,b; Carilli et al. 2004a,b). According to these similarities,
 we propose that the {\it phase of young QSO} could be associated
 with the following main processes:

\begin{enumerate}

\item
In young QSOs with extremely large amount of gas (concentrate in their
nuclear region), the accretion rate of gas --by the SMBH-- could be
extremely high (see Maiolino et al. 2004a). 

\item
In addition this extremely large amount of molecular gas
 could generate extreme starbursts; and the presence of AGNs
 could increase the SF close to the nucleus. Specially in the
 accretion disks, with properties --of the SF-- similar to the
 population III of stars.
In these extreme starbursts --close associated with QSOs/AGNs-- it
is expected  giant SN explosions.

\item
In young and distant QSOs the very high number of BAL detections
suggest that composite OFs (or EVOFs) play a main role in
their evolution.

\end{enumerate}

Furthermore, this scenario is in close agreement with the results of
the study of proto-QSO evolution and black hole growth (using
hydrodynamic models). In particular,
Kawakatu et al. (2003) found that a ULIRG phase (where the host
galaxy is the dominant source of luminosity: i.e. IR galaxies/mergers with
starbursts) precedes a galactic wind epoch: i.e., {\it composite young
IR+OF/GW+BAL mergers/QSOs}. Which would be a transition state to the
AGN--dominated phase: i.e., the standard QSO phase.
This evolutive path is almost identical to the sequence found for
BAL IR+Fe {\sc ii} QSOs, in Figure 15.

\clearpage
% ******************************************************************
% ******************************************************************
\section{Summary and Conclusions}

We have presented in this paper optical 2D and 1D spectroscopy
(obtained at La Palma/WHT, HST, IUE, ESO/NTT, KPNO, APO and CASLEO
observatories) and deep HST broad
band images of Mrk 231 and similar IR + GW/OF + Fe {\sc ii} QSOs.
The main results and conclusions can be summarised as follows:

\begin{enumerate}

\item
A detailed study of the kinematics, morphology, physical conditions and
ionization structure of the  extreme galactic superwind + bubble
(detected previously  in the nuclear and central regions
of the IR merger/QSO Mrk\,231) is presented.
This study is based mainly on INTEGRAL two-dimensional INTEGRAL
spectroscopy (obtained at La Palma 4.2 m William Herschel Telescope) plus HST
near--UV/ACS, optical/WFPC2 and near--IR/NICMOS broad band images.

\item
Inside and very close to the nucleus the 2D spectra show
the presence of an OF emission bump in the blend H$\alpha$+[N {\sc ii}],
with a peak at the same velocity of the main BAL--I system
($ V_{\rm Ejection BAL-I} \sim$ --4700\,km\,s$^{-1}$).
This bump was more clearly detected in the area located at 0\farcs6--1\farcs5
(490--1220 pc), to the south-west of the nucleus core, showing a strong
and broad peak.
In addition, in the same direction (at PA $\sim$60, i.e. close to
the PA of the small scale radio jet) at 1\farcs7--2\farcs5 (1340--2030 pc),
we also detected
multiple narrow emission line components, with ``greatly" enhanced
[N\,{\sc ii}]/H$\alpha$ ratio (very similar to the spectra of bow shocks in
jets, already found in the Circinus galaxy).
Thus, these results suggest that probably the BAL--I system
is generated in OF clouds associated with the parsec scale jet.

\item
The HST images show 4 (or 5) nuclear bubbles or shells with radius
r $\sim$ 2.9, 1.5, 1.0, 0.6 and 0.2 kpc
(3\farcs5, 1\farcs8, 1\farcs2, 0\farcs7 and 0\farcs3).
For these bubbles, the 2D H$\alpha$ velocity field (VF) map
and 2D spectra show:
(i) At the border of the more extended bubble or shell (S1), a clear expansion of
the shell with blueshifted velocities (with circular shape
and at a radius r $\sim$ 3\farcs0).
This bubble shows a rupture arc --to the south--  
suggesting that the bubble is in the blowout phase. The axis of this rupture
or ejection (at PA $\sim$00$^{\circ}$) is coincident with the axis of the
intermediate and large scale OF detected at radio wavelengths
(r $\sim$ 0\farcs6 $\sim$ 50.0 pc at 2.3 GHz, and r $\sim$ 30\farcs0
$\sim$ 25.0 kpc at 1.5 GHz).
(ii) In addition in the 3 more external bubbles (S1, S2, S3), the
2D WHT spectra show multiple emission line components with OF velocities, of 
$\langle V_{\rm OF Bubble-S1}\rangle = [-(410-650) \pm 30]$\,km\,s$^{-1}$,
$\langle V_{\rm OF Bubble-S2}\rangle = [-500 \pm 30]$\,km\,s$^{-1}$, and
$\langle V_{\rm OF Bubble-S3}\rangle = [-230 \pm 30]$\,km\,s$^{-1}$.
(iii) In all the circumnuclear region (1\farcs8 $<$ r $<$ 5$''$), the
[N\,{\sc ii}]/H$\alpha$ and [S {\sc ii}]/H$\alpha$ narrow emission line
ratios show high values ($>$ 0.8), which are consistent with LINER/OF processes
associated with fast velocity shocks. Therefore, we suggest that these
giant bubbles are associated with the large scale nuclear OF
component, which is generated --at least in part-- by the extreme
nuclear starburst: i.e., type II SN explosions.

\item
High resolution {\itshape HST\/} ACS, WFPC2 and NICMOS  broad band images
(using the filters F330W, F439W $\sim$$B$, F814W $\sim$$I$, and F160W
$\sim$$H$)  were combined with the 2D WHT spectra to study the physical
properties of the multiple bubbles/shells. Specially, we study the
properties of several blue/young star forming regions, located mainly
at the border of these shells.

\item
Finally, the variability of the short lived BAL--III Na ID system was studied,
covering almost all the period in which this system appeared (between
$\sim$1984--2004).
We found that the BAL-III light curve (LC) is clearly asymmetric with:
a steep increase, a clear maximum and an exponential fall
(similar to the shape of a SN LC).
Previously the nature of this LC was discussed assuming a symmetric or Gaussian
form.
In this paper we discuss the origin of this BAL-III system, mainly in
the frame work of an {\it extreme explosive event} (at parsec scale), probably
associated with giant SNe or hyper-nova, with very massive star progenitors
and located close to the AGN and/or in an accretion disk.

\end{enumerate}

Using our data base for more than 50 IR Mergers and QSOs
with galactic winds and using for comparison the large sample of standard
PG QSO (from Boroson \& Green 1992), we confirm our previous finding
(obtained from a small sample of IR galaxies): in the sense that
{\it IR QSOs are probably ``transition" objects},
between IR mergers and standard QSOs.

Finally, new observations of UV-BAL systems of IR + GW/OF + Fe {\sc ii} QSOs
were analysed. This study shows a new BAL IR QSO
and suggest/confirm that these objects could be {\it nearby young
BAL QSOs, similar to those detected recently at z $\sim$ 6.0}.
We propose that the {\it phase of young QSO} is associated with: accretion
of large amount of gas (by the SMBH) + extreme starbursts + extreme composite
OFs/BALs.

\section*{Acknowledgments}

The authors thank J. Acosta, J. Ahumada, H. Dottori, R. Diaz, J. C. Forte
 E. Mediavilla and Z. Tsvetanov for  assistance and constructive discussions.
We wish to thank T. Boroson, R. Green, D. Kim, D. Macchetto, D. Sanders,
Z. Tsvetanov and S. Veilleux for their 1D spectra, kindly made available to us. 
The 4.2 m William Herschel Telescope is operated by the Issac Newton Group
at the Observatorio de Roque de los Muchachos of the Instituto de Astrofisica
de Canarias. The authors thanks all the staff at La Palma, ESO KPNO, APO
and CASLEO observatories, for their kind support.
This work was based on observations made using the
NASA and ESA HST satellite, obtained from archive data at
ESO-Garching and STScI-Baltimore. This work was made using the
NASA Extragalactic Database NED; which  is operated by the Jet
Propulsion Laboratory, California Inst. of Technology, under
contract with NASA. This work was supported in part by Grants from
Conicet, SeCyT-UNC, and Fundaci\'on Antorchas (Argentina).
%Finally, we wish to thank the referee for constructive and
%valuable comments, which helped to improve the content
%and presentation of the paper.

\clearpage
\begin{table}
\footnotesize \caption{Journal of  observations of Mrk\,231 and
similar extreme IR + GW/OF + Fe {\sc ii} QSOs}
\label{tobser}

\begin{tabular}{llllcl}
\hline
\hline
Object & Date & Telescope/ & Spectral region &Expos.\ time   & Comments \\
       &      & instrument &                 & [s]           & \\
\hline

Mrk 213& 1991 Feb 15& 2.15\,m KPNO/GoldCam&$\lambda\lambda$3350--5200 \AA&900$\times$2&
PA = 90, slit width = 1\farcs5 \\
       &            & 2.15\,m KPNO/GoldCam&$\lambda\lambda$5100--7100 \AA&900$\times$2&
" \\
       &      &              &                           &         & \\
Mrk 231& 1991 May 11& 2.5\,m NOT   &V                          &1200$\times$3&
$\langle$FWHM$\rangle$ = 0\farcs7 \\
       &       &              &                           &         & \\

Mrk 231& 1992 Nov 27&{\itshape HST\/}/FOS& G190H, $\lambda\lambda$1275--2320 \AA & 5760&
 archival\\
Mrk 231& 1992 Nov 27 &{\itshape HST\/}/FOS& G270H, $\lambda \lambda$2225--3295 \AA & 2880&    "\\
Mrk 231& 1996 Nov 21 &{\itshape HST\/}/FOS& G160L, $\lambda \lambda$1150--2300 \AA &  770&    "\\
       &    &            &                           &         & \\

Mrk 231& 1995 Oct 23&{\itshape HST\/}/WFPC2& F439W, $\lambda\lambda$4283/464 \AA & 2226&
$\langle$FWHM$\rangle$ = 0\farcs1, archive\\
Mrk 231& 1995 Oct 23 &{\itshape HST\/}/WFPC2& F814W, $\lambda \lambda$8203/1758 \AA & 712&    "\\
       &    &            &                           &         & \\

Mrk 231& 2003 Mar 17&{\itshape HST\/}/ACS  & F330W, $\lambda \lambda$3354/588 \AA & 1140&
$\langle$FWHM$\rangle$ = 0\farcs1, archival, NIC2\\
       &    &            &                           &         & \\

Mrk 231& 2003 Sep 09 &{\itshape HST\/}/NICMOS  & F160W, $\lambda \lambda$1.60/0.40 $\mu$m& 640&
$\langle$FWHM$\rangle$ = 0\farcs22, archival, NIC2\\

       &    &            &                           &         & \\
I21219-1757&1989 Jun 29   &2.15m CASLEO/Z-machine&$\lambda \lambda$4700-7200 \AA&1500$\times$3&
aperture: 3$''\times$6$''$ \\
I21219-1757&1993 Jul 13   &2.15m CASLEO/UCS &$\lambda \lambda$6500-7000 \AA&1800$\times$2&
PA 90$^{\circ}$ slit width= 2\farcs0 \\
I21219-1757&1997 Mar 13    &2.15m CASLEO/UCS&$\lambda \lambda$4000-7500 \AA&1800$\times$2&
PA 90$^{\circ}$ slit width= 1\farcs5 \\

I21219-1757&1995 Oct 24&2.15\,m KPNO/GoldCam&$\lambda\lambda$4500--9000 \AA&1800$\times$2&
PA = 90, slit width = 1\farcs5 \\

I21219-1757&1995 May 26&IUE/LWP30773&$\lambda\lambda$1150--1980 \AA&9000 &\\
I21219-1757&1995 May 26&IUE/SWP54757&$\lambda\lambda$1900--3290 \AA&112000&\\

I21219-1757&1999 Jun 03&{\itshape HST\/}/STIS& G160L, $\lambda \lambda$1150--1550 \AA &  2420&archive\\

       &    &            &                           &         & \\
I07598+6508&1991 Feb 15&2.15\,m KPNO/GoldCam&$\lambda\lambda$3350--5200 \AA&1800$\times$2&
PA = 90, slit width = 1\farcs5 \\
I07598+6508&1991 Feb 15& 2.15\,m KPNO/GoldCam&$\lambda\lambda$5100--7100 \AA&1800$\times$2&
" \\

I07598+6508&1998 Dec 10&{\itshape HST\/}/STIS& G140L, $\lambda \lambda$1000--1400 \AA &  2570&archive\\

I07598+6508&          &IUE/SWP37199&$\lambda\lambda$1150--1980 \AA&   &archive\\
I07598+6508&          &IUE/SWP40376&$\lambda\lambda$1150--1980 \AA&   &archive\\
I07598+6508&          &IUE/LWP16445&$\lambda\lambda$1900--3290 \AA&   &archive\\
I07598+6508&          &IUE/LWP19437&$\lambda\lambda$1900--3290 \AA&   &archive\\
       &    &            &                           &         & \\
I14026+4341&1990 Sep 19&2.15\,m KPNO/GoldCam&$\lambda\lambda$4350--5700 \AA&1800$\times$2&
PA = 90, slit width = 1\farcs5 \\

I14026+4341&1994 Jul 17&{\itshape HST\/}/FOS& G190H, $\lambda \lambda$1570--2300 \AA &  978&archive\\

I14026+4341&           &IUE/LWP20600&$\lambda\lambda$1150--1980 \AA&   &archive \\
I14026+4341&           &IUE/SWP36278&$\lambda\lambda$1900--3290 \AA&   &archive \\
       &    &            &                           &         & \\
I17002+5153&1990 Sep 19&2.15\,m KPNO/GoldCam&$\lambda\lambda$4350--5700 \AA&1800$\times$2&
PA = 90, slit width = 1\farcs5 \\

I17002+5153&1993 Mar 10&{\itshape HST\/}/FOS& G190H, $\lambda \lambda$1600--2300 \AA & 2026&archive\\

I17002+5153&           &IUE/LWP4596&$\lambda\lambda$1150--1980 \AA&   & archive\\
I17002+5153&           &IUE/LWP6655&$\lambda\lambda$1150--1980 \AA&   & archive\\
       &    &            &                           &         & \\

PHL1092& 1993 Jul 15 & 3.5\,m ESO NTT&$\lambda\lambda$3200--5000 \AA&1500$\times$3&
$\langle$FWHM$\rangle$ = 1\farcs0 \\
PHL1092& 2004 Sep 08 & 3.5\,m APO&$\lambda\lambda$3500--5600 \AA&1800$\times$2&
$\langle$FWHM$\rangle$ = 1\farcs0 \\
PHL1092& 2004 Sep 08 & 3.5\,m APO&$\lambda\lambda$5300--10000 \AA&1800$\times$2&\\
       &    &            &                           &         & \\
I18508-7815&1991 Jul 15&3.6 m ESO/EFOSC&$\lambda\lambda$3600--5590 \AA&1800$\times$2&
PA = 90, slit width = 1\farcs5 \\
I18508-7815&1991 Jul 15&3.6 m ESO/EFOSC&$\lambda\lambda$5050--7000 \AA&1800$\times$2&\\
I18508-7815&1991 Jul 15&3.6 m ESO/EFOSC&$\lambda\lambda$6600--9800 \AA&1800$\times$2&\\
I18508-7815&1991 Jul 18&IUE/LWP20841&$\lambda\lambda$1150--1980 \AA&25200&\\
I18508-7815&1995 May 27&IUE/SWP54772&$\lambda\lambda$1900--3290 \AA&24300&\\

       &    &            &                           &         & \\

\hline

\end{tabular}

\end{table}

\clearpage

\begin{table}
\footnotesize \caption{Main properties of the circumnuclear knots
(associated with the shells)}
\label{tknots}

\begin{tabular}{lrrcclrl}
\hline
\hline
Knots &$\alpha$&$\delta$&B$_{F439W}^a$&I$_{F814W}^a$&B-I & R$_{eff}^a$ & Comments \\
      & [$''$] & [$''$] & [mag]       & [mag]       &    & [pc]        &    \\
\hline

  &     &     &       &       &      &      &     \\
1 & 1.10& 1.68& 22.23 & 21.12 & 1.11 &    66& Shell S1\\
2 &-1.52&-0.90& 23.27 & 20.86 & 2.41 & $<$20& Shell S2\\
3 & 1.10&-0.16& 22.38 & 20.86 & 1.52 &    62& Shell S3\\
4 & 2.12&-1.45& 23.98 & 22.60 & 1.38 &    42& Shell S1\\
5 & 1.70&-2.97& 21.31 & 20.50 & 0.81 & $<$20& Shell S1\\
6 & 1.56&-2.74& 22.97 & 22.28 & 0.69 &    71& Shell S1\\
7 & 1.01&-3.70& 22.69 & 21.47 & 1.22 &    38& Shell S1\\
11&-0.46&-3.66& 21.17 & 20.66 & 0.51 &   136& Shell S1\\
12&-0.69&-3.61& 21.46 & 20.70 & 0.76 &    31& Shell S1\\
14&-1.20&-3.15& 21.79 & 20.33 & 1.46 &    40& Shell S1\\
29&-2.30& 1.82& 23.26 & 21.06 & 2.20 &    46& Shell S1\\
  &     &     &       &       &      &      &     \\

\hline

\end{tabular}

\end{table}

Notes of Table 2:\\
$^a$: From Surace et al. (1998).\\

\begin{table}
\footnotesize \caption{Relative variability of the Na ID BAL--III  system}
\label{tbal4}

\begin{tabular}{lcrrl}
\hline
\hline
 Date      & Julian Date & Intens. Ratio & $\sigma$ & References \\
           & 2442507+    &               &          &           \\
\hline
           &             &               &          &           \\
Previous Data&           &               &          &           \\
04/05/1975 & 0000        &    0.000      & 0.000    & (1) and (2) \\
05/01/1981 & 2218        & $<$0.005      & 0.000    & (1) and (3) \\
12/21/1984 & 3548        &    0.040      & 0.010    & (1) and (4) \\
02/02/1987 & 4321        &    0.079      & 0.006    & (1) \\
07/06/1988 & 4841        &    0.090      & 0.004    & (5) and (1) \\
01/10/1992 & 6124        &    0.052      & 0.007    & (1) \\
05/21/1992 & 6257        &    0.050      & 0.003    & (1)  \\

           &             &               &          &           \\
New Data   &             &               &          &           \\

02/01/1991 & 5781        &    0.062      & 0.008    & (6) and (7) \\
02/15/1991 & 5795        &    0.060      & 0.007    & (6) and (8) \\
04/22/1994 & 6558        &    0.035      & 0.003    & (6) and (9) \\
02/28/2001 & 9461        &    0.011      & 0.004    & (6) and (10) \\

           &             &               &          &           \\

\hline

\end{tabular}

\end{table}

Notes of Table 3:\\
References: (1) Kollatschny et al. (1992),
            (2) Bocksemberg et al. (1975),
            (3) Hamilton \& Keel (1987),
            (4) Schmidt \& Miller (1985),
            (5) Boroson et al. (1991),
            (6) This paper,
            (7) Boroson \& Mayers (1992),
            (8) Lipari et al. (1994),
            (9) Forster et al. (1995),
            (10) Rupke et al. (2002).\\

\clearpage

\begin{table}
\footnotesize \caption{IR Mergers and IR QSOs/AGN with low and
high/extreme velocity outflow} \label{gwmergersqso}
\begin{tabular}{lcccccccccccl}
\hline \hline

Object       & $V_{\rm OF\,1}$ &  $z$&$L_{\rm FIR}$&$L_{\rm IR}$/$L_{B}$&
Nuclear &Morph&RF&BAL&$\alpha_{60,25}$ & $\alpha_{100,60}$ & OF &OF  \\
&km/s  & &$\log$L/L$_{\odot}$ & &
activity&Type &  &   &                 &                   &Type &Reference\\
\hline

{\it Low velocity OF}& &      &     &   &     &    &   &   &     &     &  & \\

Arp 220       & $-$450 &0.0183&12.18& 87&L+SB & PM &---&---&-2.94&-0.16&EL&Heckman et al.1987\\
IRAS00182-7112& $-$450 &0.3270&12.90&200&L+SB &(M) &---&---&-2.51&+0.05&EL&Heckman et al.1990\\
IRAS03250+1606& $-$431 &0.1290&12.06&---&L    & M  &---&---&---  &-0.49&AL&Rupke et al.2002\\
IRAS03514+1546& $-$200 &0.0210&11.10&---&  SB & M  &---&---&-2.32&-0.79&AL&Heckman et al.2000\\
IRAS09039+0503& $-$656 &0.1250&12.07&---&L    & M  &---&---&---  &-0.65&AL&Rupke et al.2002\\
IRAS11387+4116& $-$511 &0.1490&12.18&---&  SB & OM &---&---&---  &-0.76&AL&"\\
IRAS23128-5919& $-$300 &0.0449&12.60&  8&L+SB & M  &---&---&-2.18&-0.03&EL&L\'{\i}pari et al.2000b\\
Mrk 266       & $-$300 &0.0290&11.37&  8&  SB & PM &---&---&-2.14&-0.80&EL&Wang et al. 1997\\
Mrk 273       & $-$600 &0.0385&12.14& 36&L+SB & PM &---&---&-2.58&+0.03&EL&Colina et al.1999\\
NGC 1614      & $-$400 &0.0155&11.61& 18&L+SB & M  &---&---&-1.70&-0.02&EL&Ulrich 1972\\
NGC 2623      & $-$405 &0.0185&11.55& 17&L+SB & M  &---&---&-2.96&-0.37&EL&L\'{\i}pari et al.2004d\\
NGC 3256      & $-$370 &0.0094&11.57&  9&  SB & MM &---&---&-1.95&-0.52&EL&L\'{\i}pari et al.2000a\\
NGC 3690      & $-$300 &0.0103&11.91& 23&  SB & PM &---&---&-1.71&-0.07&EL&Heckman et al.1990\\
NGC\,4039     & $-$365 &0.0056&10.99&  6&L+SB & PM &---&---&---  &---  &EL&L\'{\i}pari et al.2000b\\
NGC\,5514     & $-$320 &0.0245&10.70&  1&L+SB & PM &---&---&-1.96&-1.14&EL&L\'{\i}pari et al.2004d\\
{\it Extreme OF}&      &      &     &   &     &    &   &   &     &     &  &\\

IRAS01003-2238& $-$770 &0.1180&12.24& 67&QSO+SB&OM &---&---&-1.42&+0.48&EL&L\'{\i}pari et al.2000b\\
IRAS05024-1941& $-$1150&0.1920&12.43&---&S2    & M &---&---&-2.30&-0.47&EL&Lipari et al.2000b\\
IRAS05024-1941& $-$1676&0.1920&12.43&---&S2    & M &---&---&-2.30&-0.47&AL&Rupke et al.2002\\
IRAS05189-2524& $-$849 &0.0420&12.07&---&S2    & M &---&---&-1.56&+0.36&AL&Rupke et al.2002\\
IRAS10378+1108& $-$1517&0.1360&12.26&---&L     & M &---&---&---  &---  &AL&"\\
IRAS11119+3257& $-$1300&0.1885&12.64&---&S1+SB & M&1.12&---&-1.73&+0.80&EL&L\'{\i}pari et al.2000b\\
IRAS11119+3257& $-$950 &0.1885&12.64&---&S1+SB & M&1.12&---&-1.73&+0.08&OSM&Zheng et al.2002\\
IRAS13218+0552& $-$1800&0.2048&12.63& 96&QSO+SB&OM &---&---&-1.22&+0.98&EL&L\'{\i}pari et al.2000b\\
IRAS14394+5332& $-$880 &0.1050&12.04&---&S2    &MM &---&---&-1.98&-0.40&EL&" \\
IRAS15130-1958& $-$780 &0.1093&12.09&---&S2    & M &---&---&-1.82&-0.36&EL&"\\
IRAS15462-0450& $-$1000&0.1001&12.16&---&S1    & M &---&---&-2.13&-0.05&EL&"\\
IRAS15462-0450& $-$1110&0.1010&12.35&---&S1    & M&1.32&---&-2.13&-0.05&OSM&Zheng et al.2002\\
IRAS19254-7245& $-$800 &0.0597&12.04& 30&QSO+SB& PM&---&---&-1.70&-0.11&EL&Colina et al.1991\\
Mrk\,231      & $-$1000&0.0422&12.53& 32&QSO+SB& M&1.83&Yes&-1.49&+0.11&EL&L\'{\i}pari et al.1994\\
Mrk\,231      & $-$1100&0.0422&12.53& 32&QSO+SB& M&1.83&Yes&-1.49&+0.11&OSM&Zheng et al.2002\\
NGC\,3079     & $-$1600&0.0040&10.49&  2& L+SB &Sp &---&---&-3.01&-1.41&EL&Heckman et al.1990\\
NGC\,6240     & $-$930 &0.0243&11.83& 15& L+SB &PM &---&---&-2.16&-0.40&EL&"\\
{\it IR QSOoffset-OF}& &      &     &   &      &   &   &   &     &     &  &\\
IRAS00275-2859&  $-$730&0.2792&12.64&  9&QSO+SB&PM&1.47&---&-1.58&-0.12&OSM&Zheng et al.2002\\
IRAS00509+1225& $-$1110&0.0600&11.87& 20&QSO+SB&PM&1.40&---&-0.70&-0.32&OSM&L\'{\i}pari et al.2004d\\
IRASZ01373+0604&$-$1660&0.3964&12.30&  3&QSO+SB&--&2.33&---&-0.50&-1.59&OSM&"\\
IRAS02054+0835& $-$1355&0.3450&12.97&---&QSO   &--&2.42&---&---  &---  &OSM&Zheng et al.2002\\
IRAS02065+4705&  $-$500&0.1320&12.27&---&QSO   &--&1.65&---&-1.96&---  &OSM&"\\
IRAS04415+1215&  $-$875&0.0890&12.41&---&QSO   &--&1.74&---&---  &---  &OSM&"\\
IRAS04505-2958& $-$1700&0.2863&12.55& 20&QSO+SB&PM&1.33&---&-1.41&-0.32&OSM&L\'{\i}pari et al.2004d\\
IRAS06269-0543&  $-$550&0.1170&12.49&---&QSO   &--&0.57&---&-1.36&+0.19&OSM&Zheng et al.2002\\
IRAS07598+6508& $-$2030&0.1483&12.45&  5&QSO+SB& M&2.75&Yes&-1.32&-0.04&OSM&"\\
IRAS07598+6508& $-$1920&0.1483&12.45&  5&QSO+SB& M&2.75&Yes&-1.32&-0.04&OSM&L\'{\i}pari et al.2004d\\
IRAS09427+1929& $-$1640&0.2840&12.61&---&QSO   &--&2.98&---&---  &---  &OSM&Zheng et al.2002\\
IRAS10026+4347&  $-$650&0.1780&12.20&---&QSO   &--&2.07&---&-1.26&-0.88&OSM&"\\
IRASZ11598-0112& $-$970&0.1510&11.91&---&QSO   &--&1.71&---&---  &---  &OSM&"\\
IRAS13305-1739&  $-$730&0.1480&12.21&---&S2    &OM&--- &---&-1.24&+0.21&OSM&L\'{\i}pari et al.2004d\\
IRAS13342+3932&  $-$920&0.1790&12.49&---&QSO   &--&0.73&---&---  &---  &OSM&Zheng et al.2002\\
IRAS13451+1232& $-$1210&0.1220&12.28&---&S2    &PM&--- &---&-1.20&-0.14&OSM&L\'{\i}pari et al.2004d\\
IRAS14026+4341& $-$1500&0.3200&12.55& 40&QSO   & M&1.12&Yes&-0.89&-0.92&OSM&"\\
IRAS17002+5153& $-$1050&0.2923&12.58&  5&QSO+SB&PM&1.56&Yes&-0.65&-0.06&OSM&"\\
IRAS18508-7815& $-$1250&0.1620&12.00&  8&QSO   & M&2.40&---&-1.63&-0.84&OSM&"\\
IRAS20036-1547&  $-$400&0.1930&12.70&---&QSO   &--&2.74&---&---  &---  &OSM&Zheng et al.2002\\
IRAS20520-2329&  $-$660&0.2060&12.52&---&QSO   &--&2.02&---&---  &-1.27&OSM&"\\
IRAS21219-1757&  $-$460&0.1100&11.98& 89&QSO   &OM&1.82&Yes&-1.00&-0.20&OSM&L\'{\i}pari et al.2004d\\
IRAS22419-6049&  $-$390&0.1133&11.30&---&QSO+SB& M&--- &---&-0.33&---  &OSM&"\\
IRAS23389+0300& $-$1640&0.1450&12.09&---&S2    &PM&--- &---&---  &+0.10&OSM&"\\
%IR merger-OFcand&     &      &     &   &      &  &    &   &     &     &   &\\
%Arp 193       &   --- &0.022 &11.60& 17&   SB &PM&--- &   &     &     &EL &L\'{\i}pari et al.2004d\\
%NGC 520       &   --- &0.00x &10.xx&---&   SB & M&--- &   &     &     &EL &"\\
%NGC 1222      &   --- &0.009 &10.46&  6&   SB &(M)&---&   &     &     &EL &Heckman et al.1990\\
%NGC\,4194     &   --- &0.009 &10.87&  6&   SB & M &---&   &     &     &EL &"\\

\hline

\end{tabular}

\end{table}

\clearpage

Notes.
Col.\ 2: OF values obtained from the references included in Col.\ 9.
Values between parentheses are possible detections.
[O {\sc iii}] MC indicates emission line                                          
[O {\sc iii}]\,$\lambda$5007 with multiple components.

Col.\ 4: The L$_{\rm IR}$ were obtained for [8--1000 $\mu$m], using the
relation given by Sanders \& Mirabel (1996).

Col.\ 6: The properties of the nuclear activity were obtained from Veilleux
et al. (1999), Canalizo \& Stockton (2001) and L\'{\i}pari et al.\ (2003). S1:
Seyfert 1, S2: Seyfert 2, L: Liners and SB: starburst.

Col.\ 7: For the morphological or interaction type we used the classification
criteria of Veilleux, Kim \& Sanders (2002b) and Sourace (1998); PM:
pre--merger, M: merger, OM: old merger, MM: multiple merger, Sp: spirals.

Col.\ 8: RF is the ratio of the emission line
Fe {\sc ii}$\lambda$4570/H$\beta$

Cols.\ 9 and 10: $\alpha$ are the IR colour indexes (for 3 IR bands).

Col.\ 11: EL, AL and OSM indicate OF derived from emission, absorption lines,
and offset emission line method (see the text), respectively.

\begin{figure*}
\vspace{12.0 cm}
\begin{tabular}{c}
\includegraphics{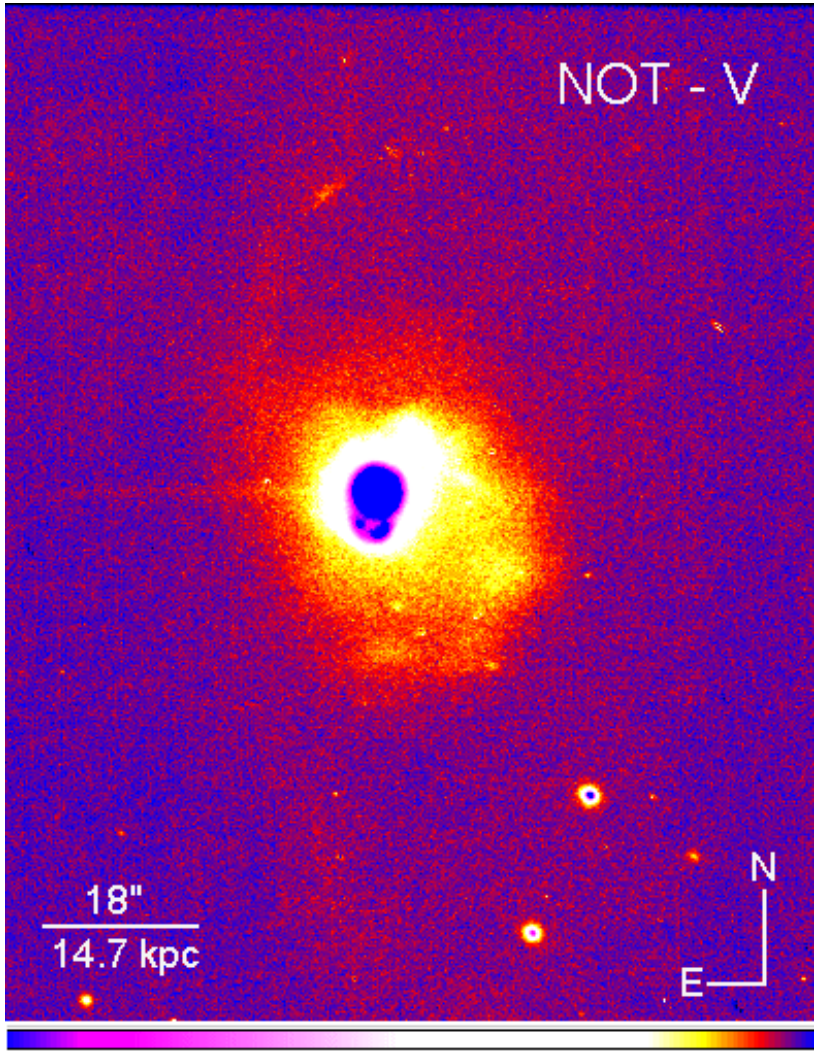}\cr
\end{tabular}
\vspace{8.0 cm}
\caption {V broad band image of the main body and the faint tails of Mrk 231
(obtained at the 2.5 NOT telescope, La Palma Spain).
}
\label{f1not}
\end{figure*}

\clearpage

\begin{figure*}
\vspace{12.0 cm}
\begin{tabular}{cc}
\includegraphics{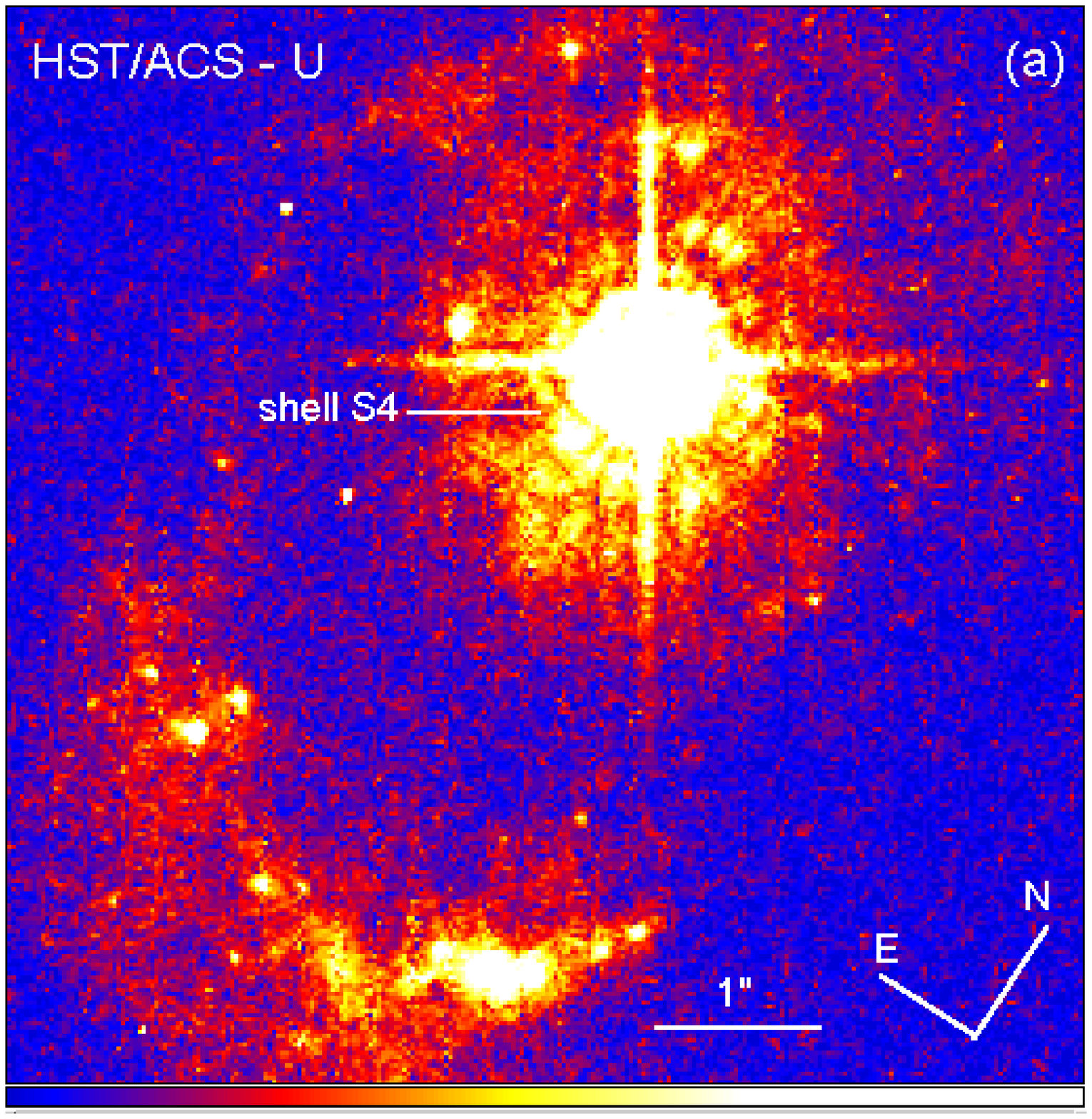}&
\includegraphics{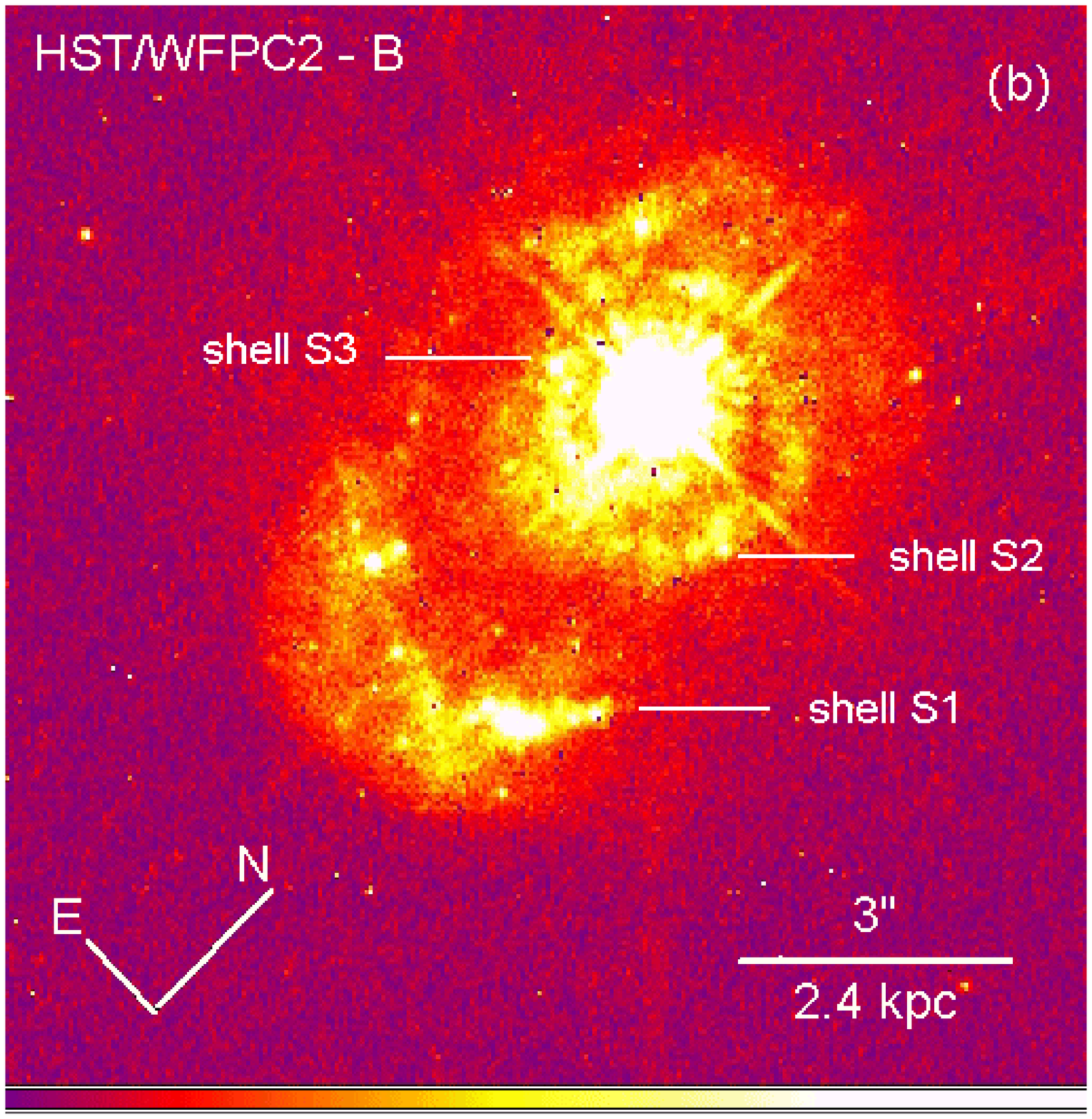} \cr
\includegraphics{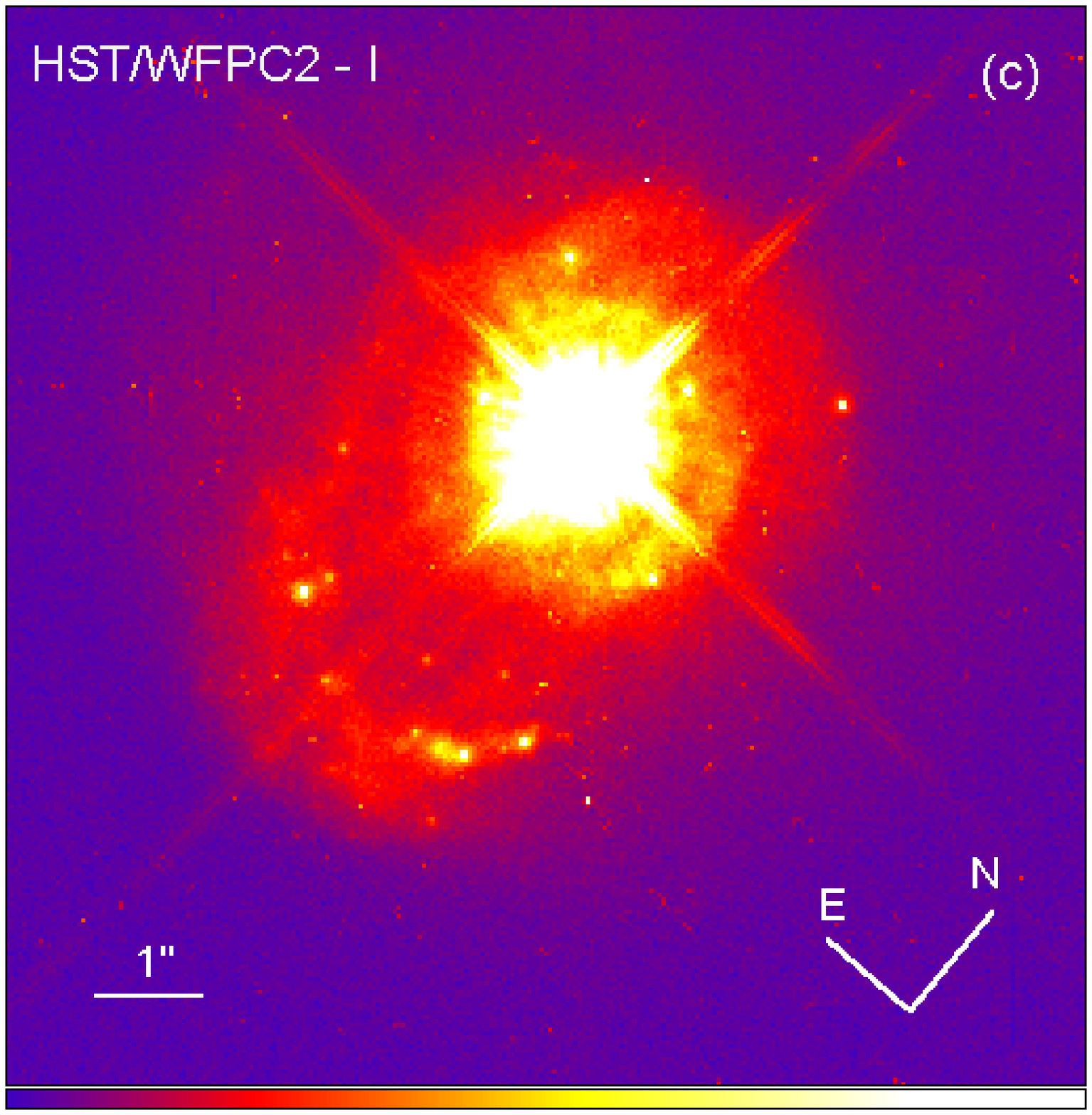}&
\includegraphics{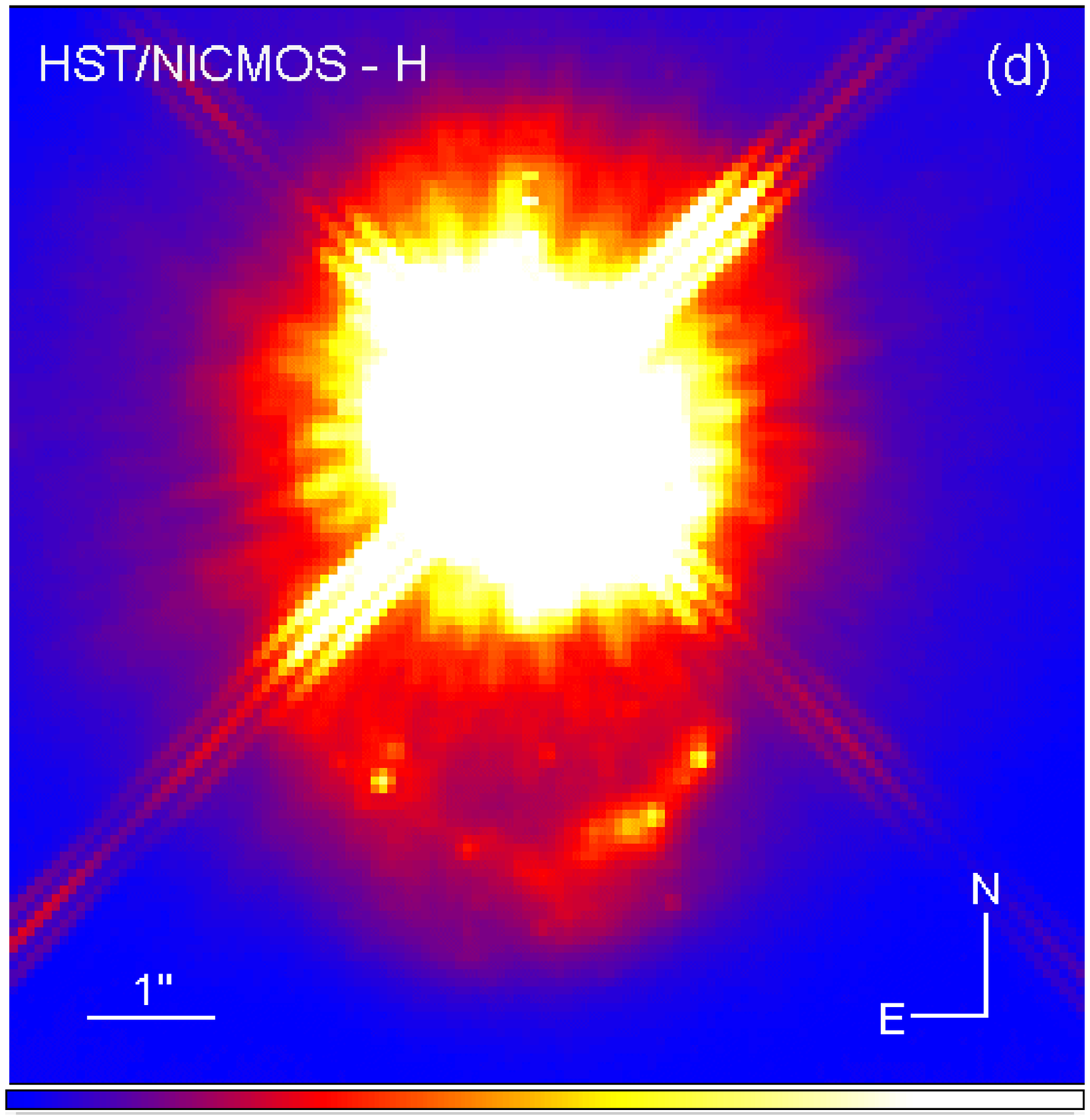} \cr
\end{tabular}
\vspace{8.0 cm}
\caption {Near-UV, optical and near-IR HST broad band images of the
circumnuclear regions and the multiple shells, in Mrk 231}
\label{f2shells}
\end{figure*}

\clearpage

\begin{figure*}
\vspace{12.0 cm}
\begin{tabular}{cc}
\includegraphics{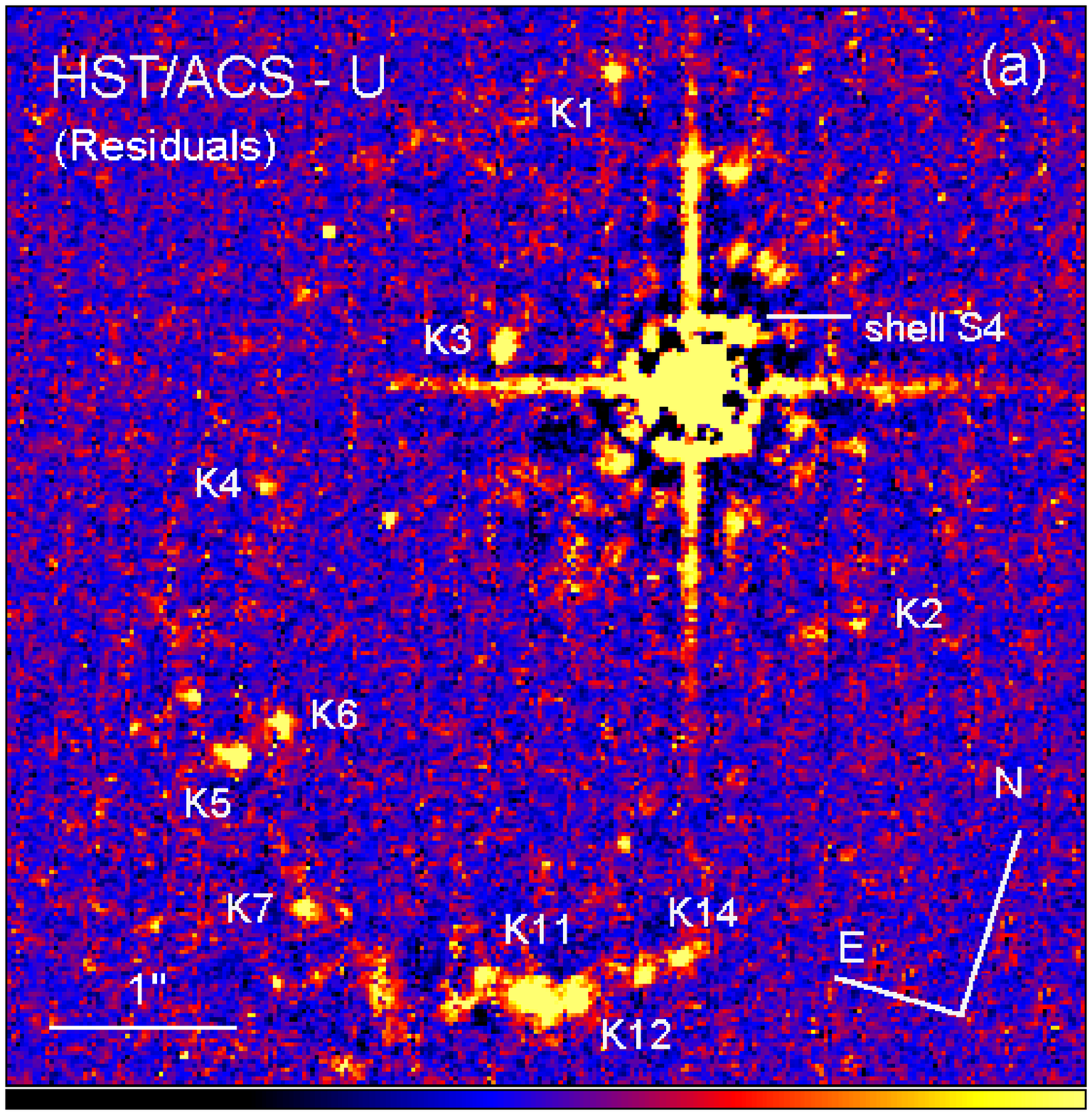}&
\includegraphics{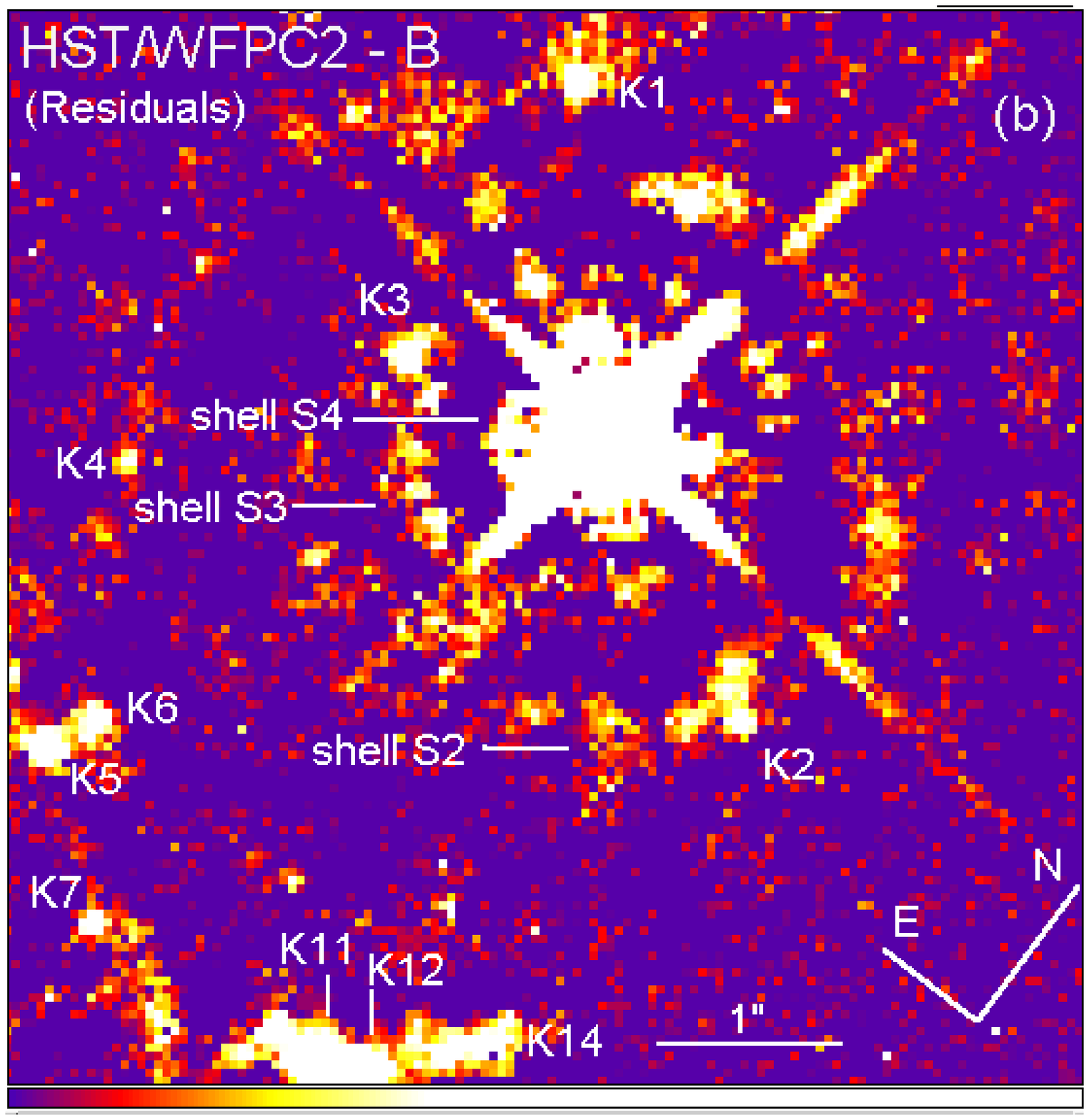} \cr
\includegraphics{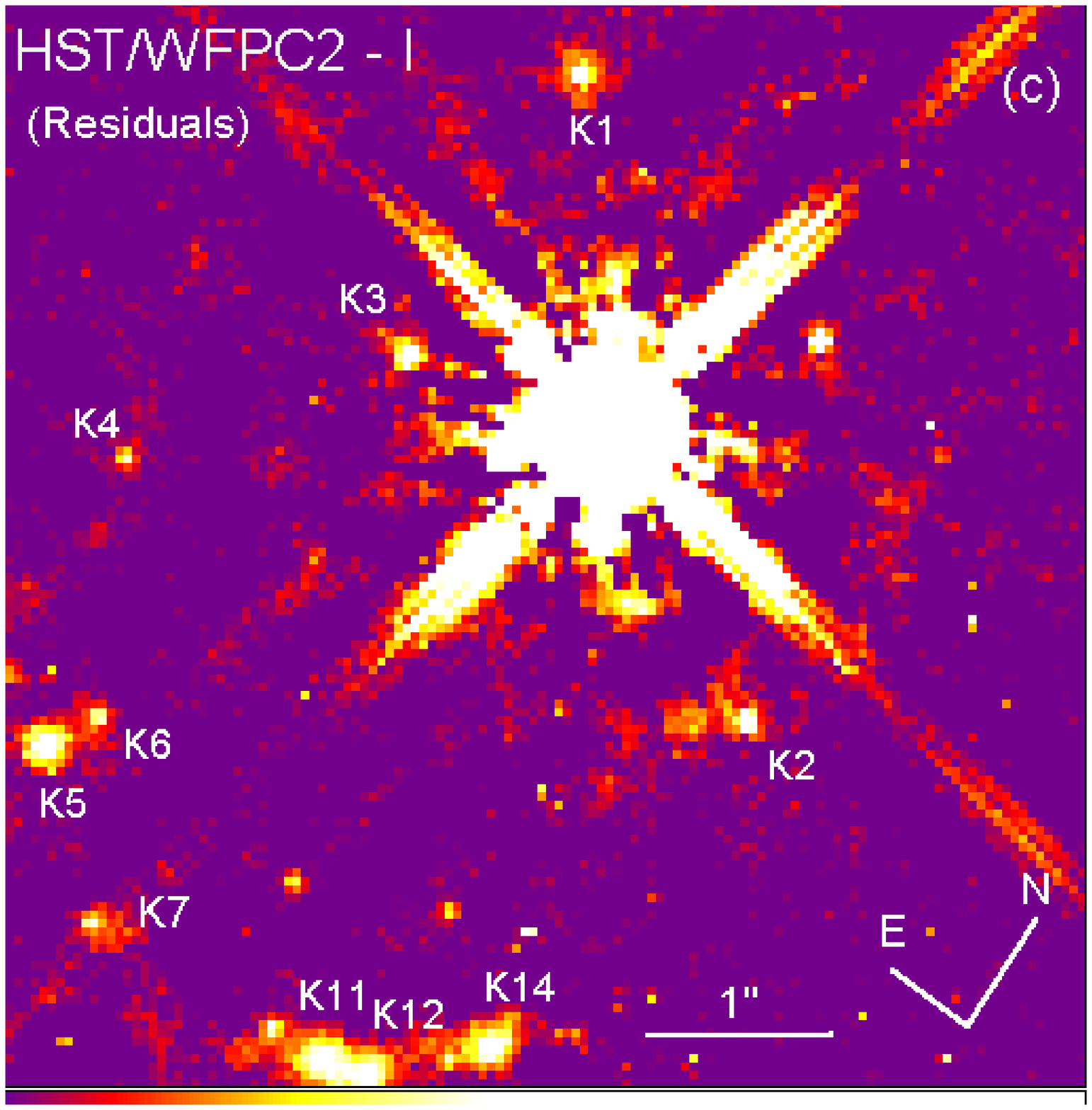}&
\includegraphics{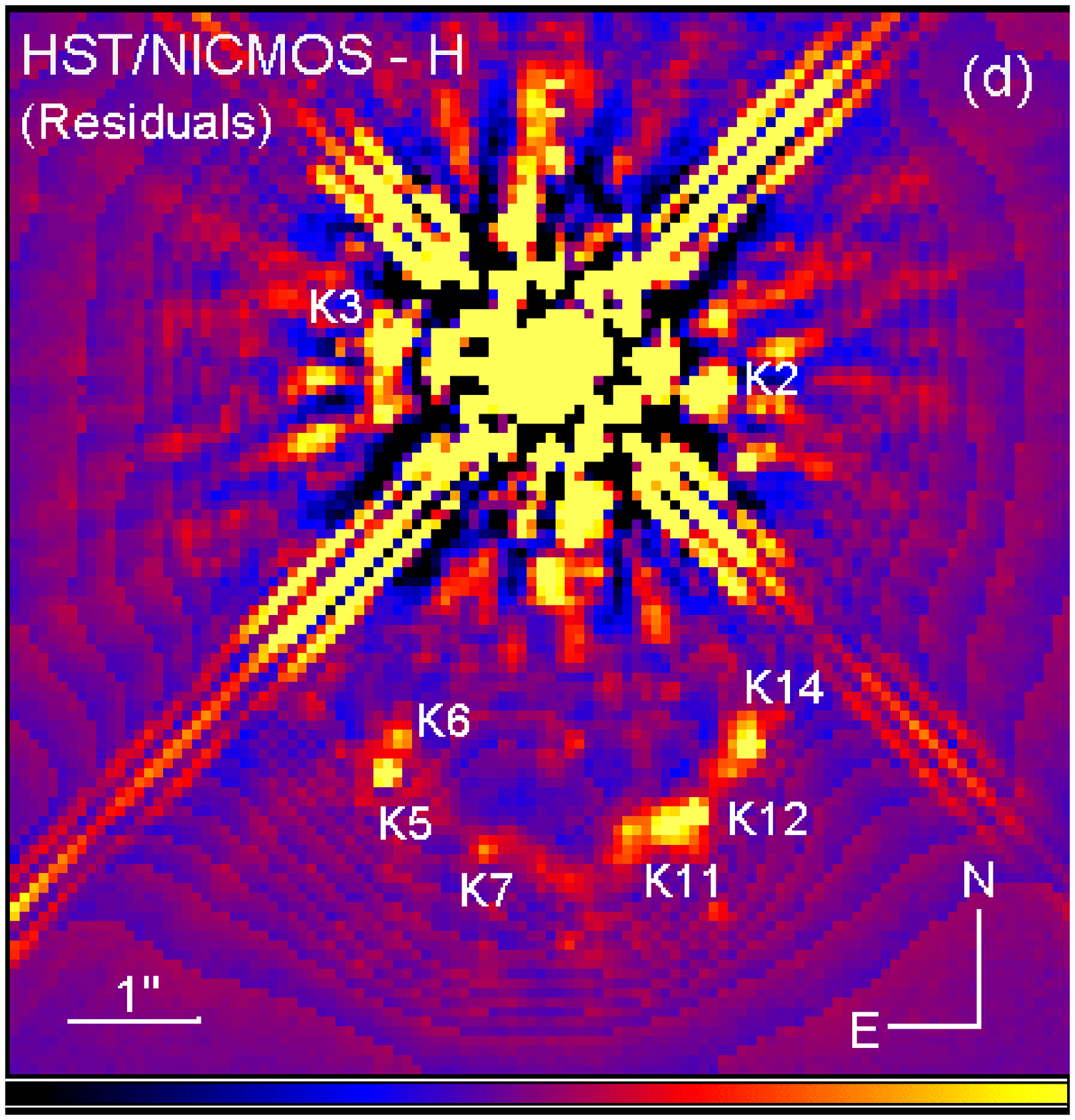} \cr
\end{tabular}
\vspace{8.0 cm}
\caption {Near-UV, optical and near-IR HST broad band residuals images of the
circumnuclear region and multiple shells, in Mrk 231}
\label{f3knots}
\end{figure*}

\clearpage

\begin{figure*}
\vspace{12.0 cm}
\begin{tabular}{c}
\includegraphics{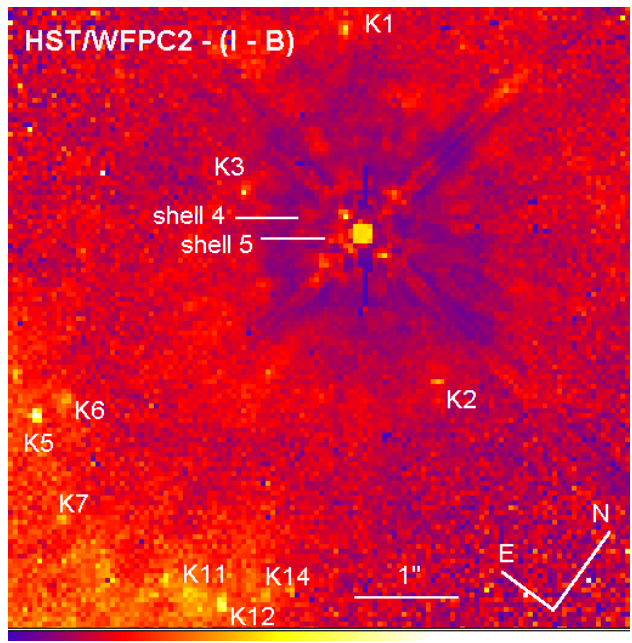}\cr
\end{tabular}
\vspace{6.0 cm}
\caption {HST WFPC2 (I - B) colour images from the optical F814W and F439W
filters.
}
\label{f4colour}
\end{figure*}

\begin{figure*}
\vspace{12.0 cm}
\begin{tabular}{c}
\includegraphics{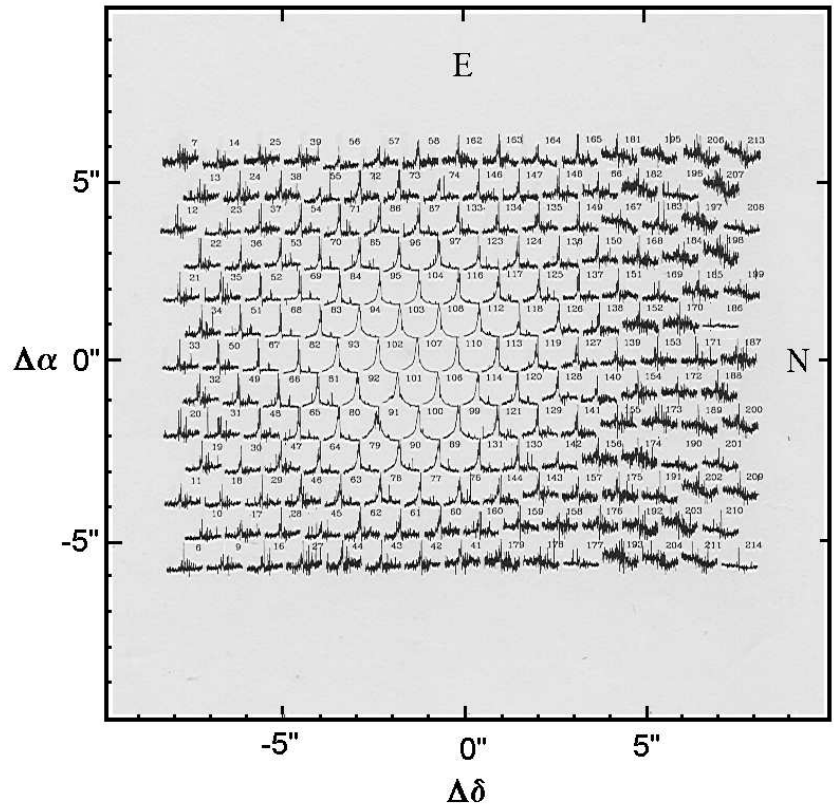}\cr
\end{tabular}
\vspace{6.0 cm}
\caption {WHT+INTEGRAL 2D optical spectrum of Mrk 231,
for the H$\alpha$+[N {\sc ii}]$\lambda$6848 blend.
}
\label{f5spec2d}
\end{figure*}

\clearpage

\begin{figure*}
\vspace{12.0 cm}
\begin{tabular}{cc}
\includegraphics{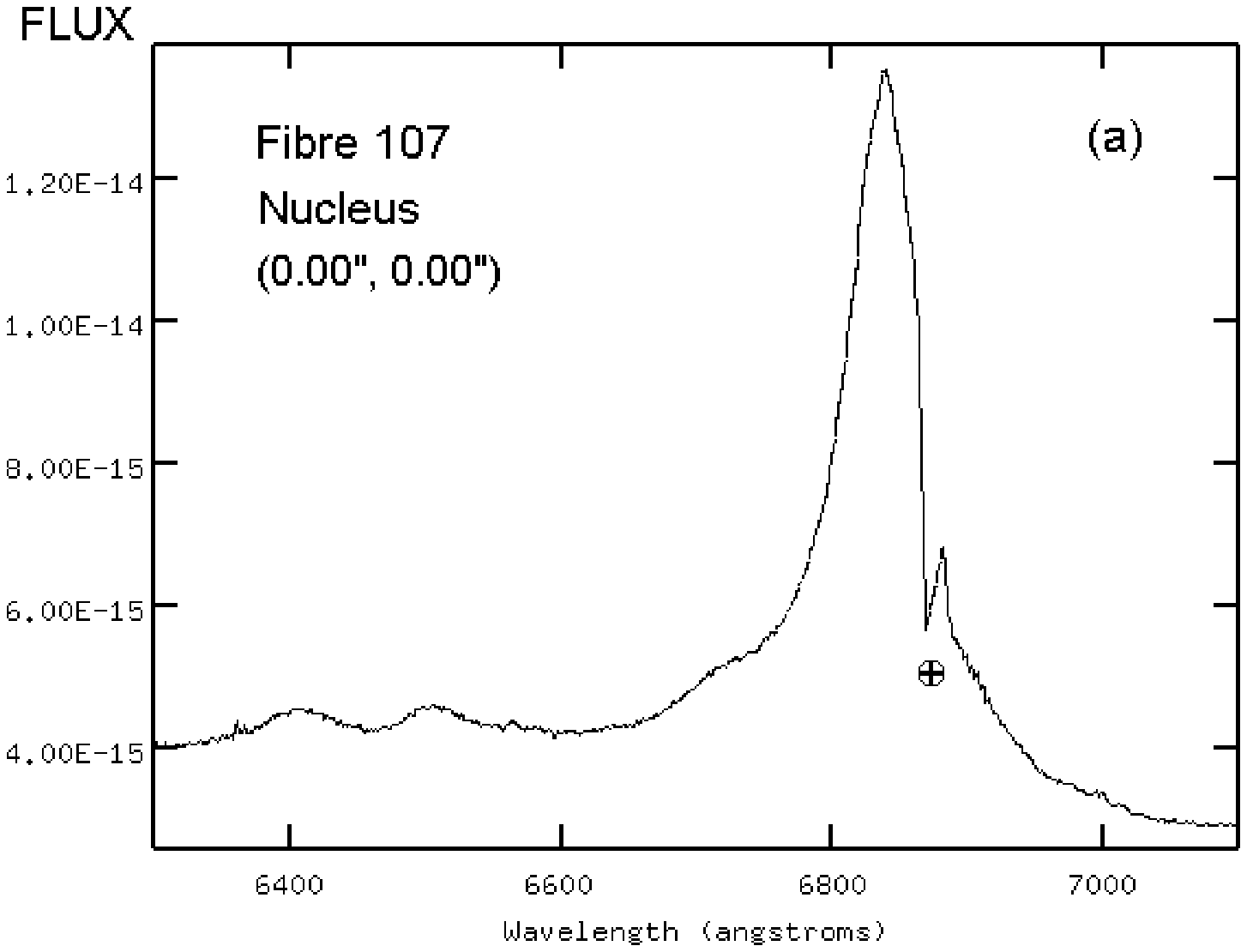}&
\includegraphics{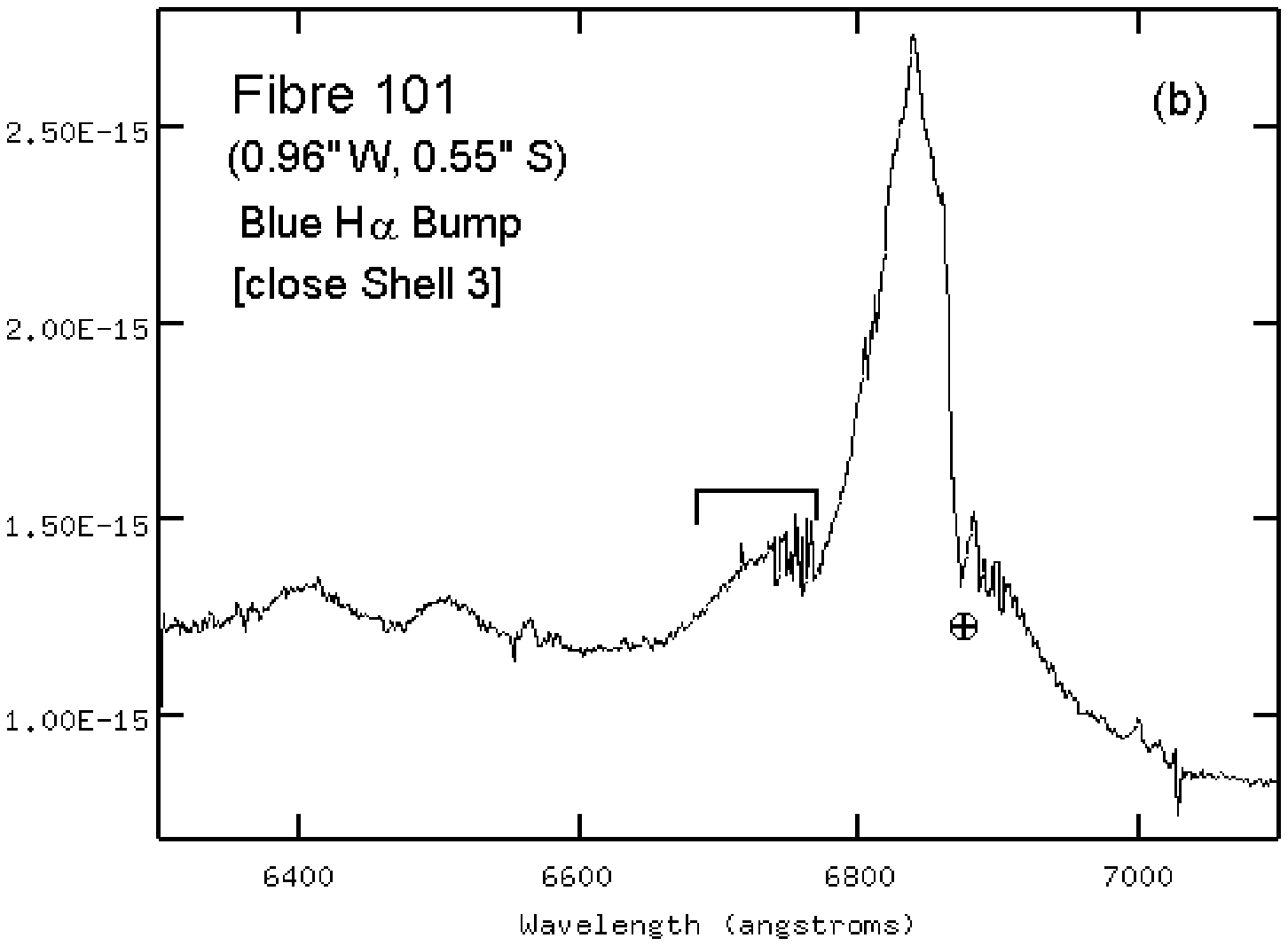} \cr
\includegraphics{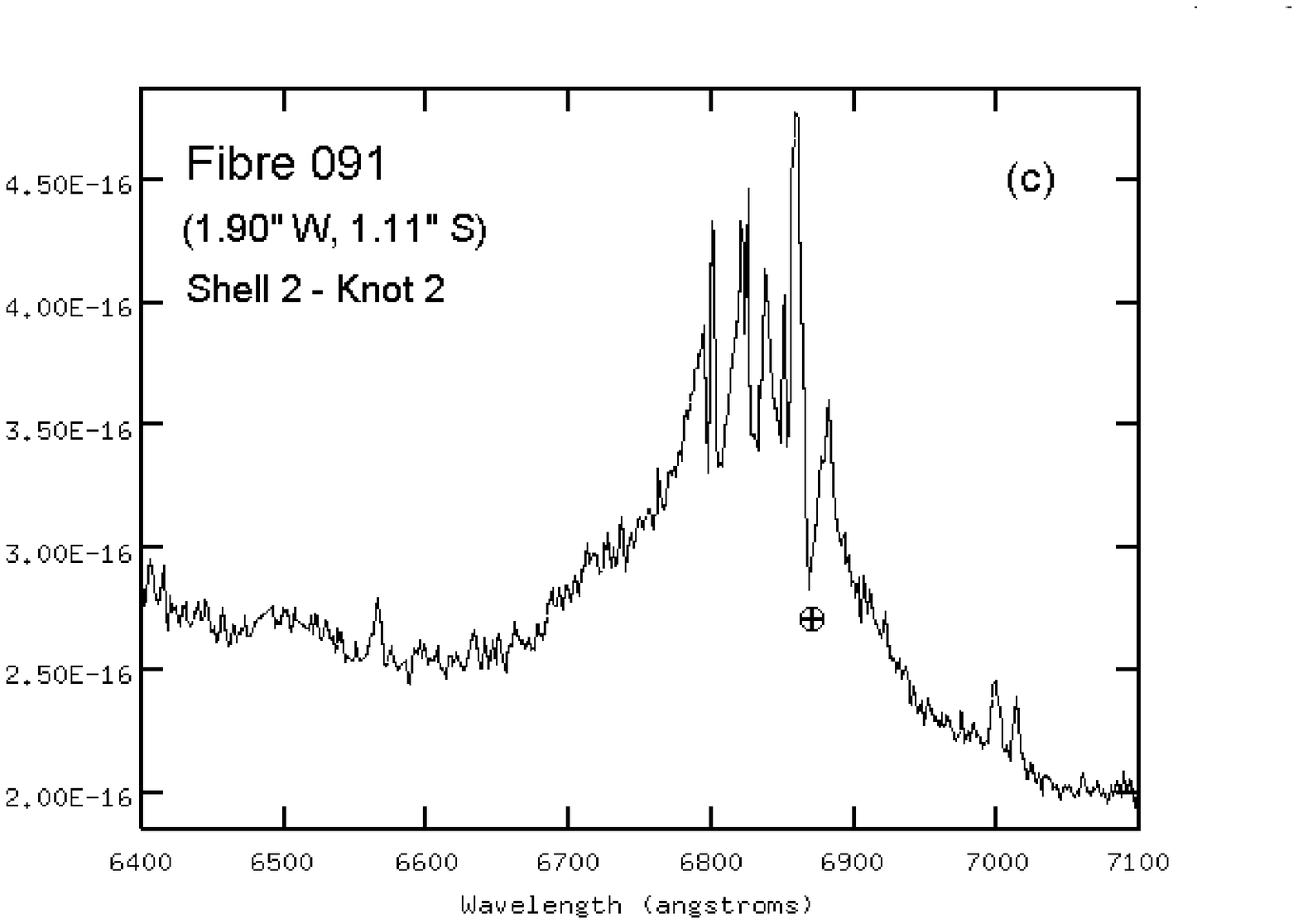}&
\includegraphics{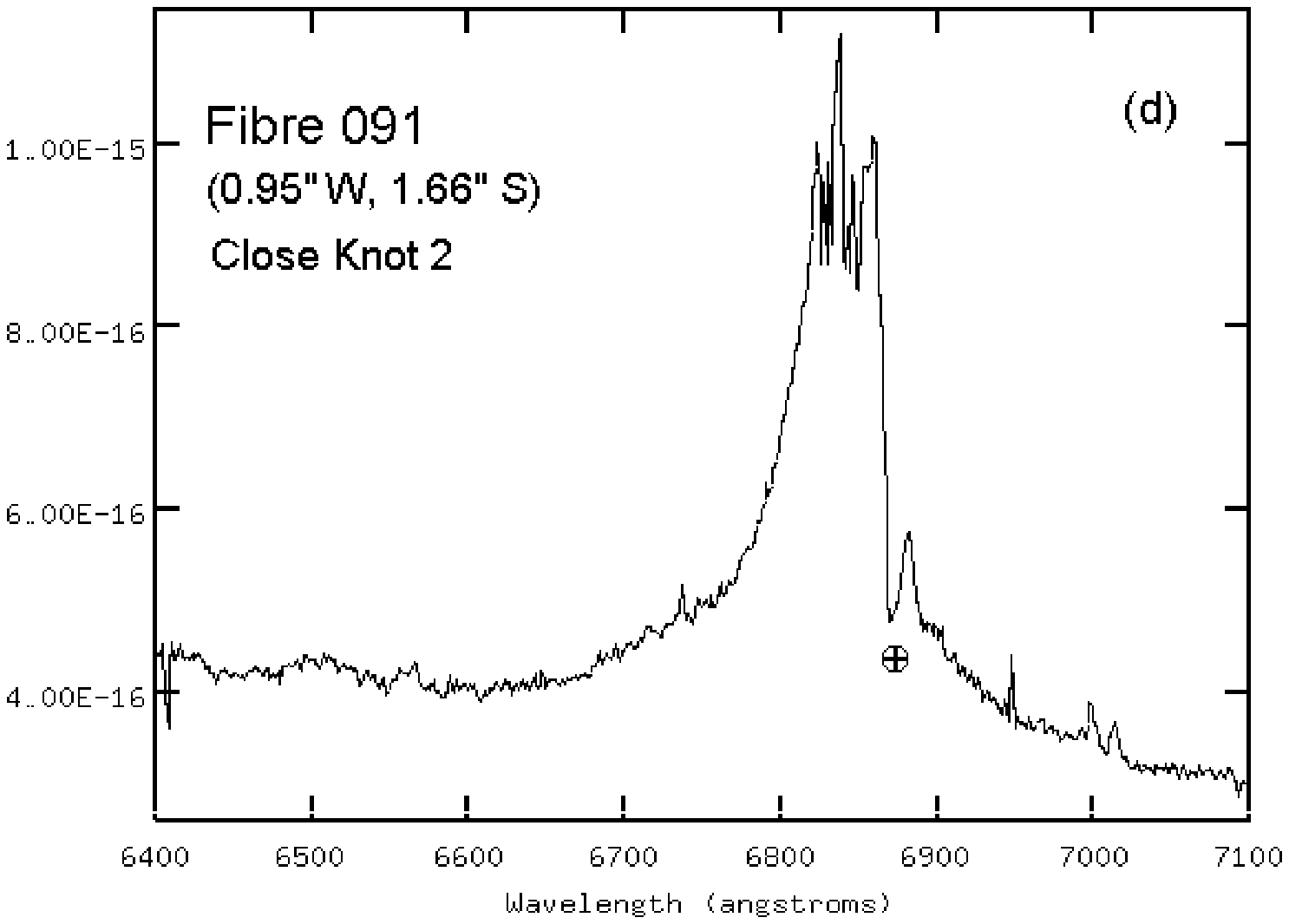} \cr
\end{tabular}
\vspace{6.0 cm}
\caption {
WHT + INTEGRAL 2D spectra (for individual fibres) in the main structures
of the nuclear and circumnuclear regions of Mrk 231
(showing several OF components in the shells).
The scales of flux are given in units of [erg $\times$ cm$^{-2}$ $\times$
s$^{-1}$ $\times$ \AA$^{-1}$].
}
\label{fig6}
\end{figure*}

\clearpage

\begin{figure*}
\vspace{12.0 cm}
\begin{tabular}{cc}
\includegraphics{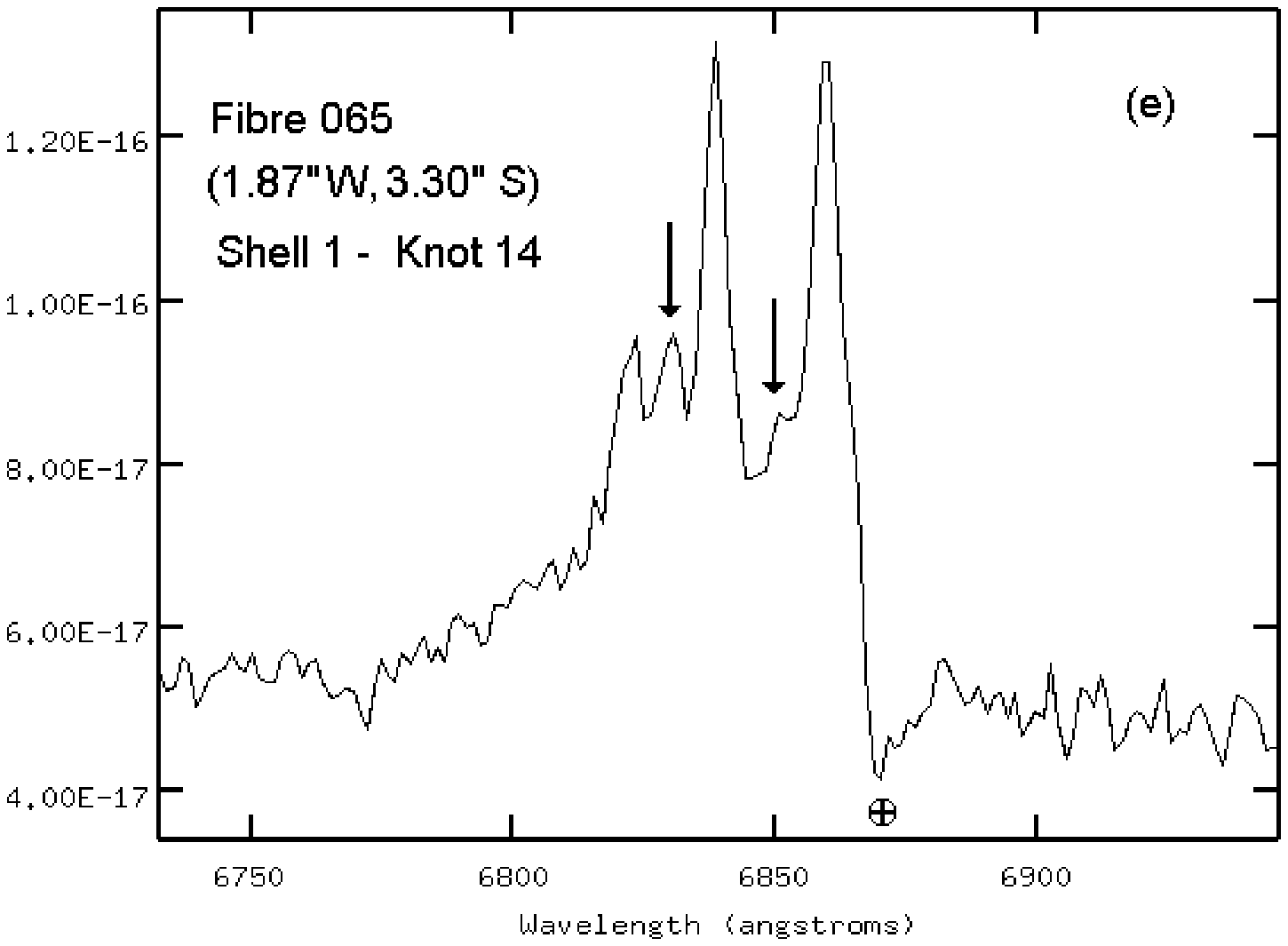}&
\includegraphics{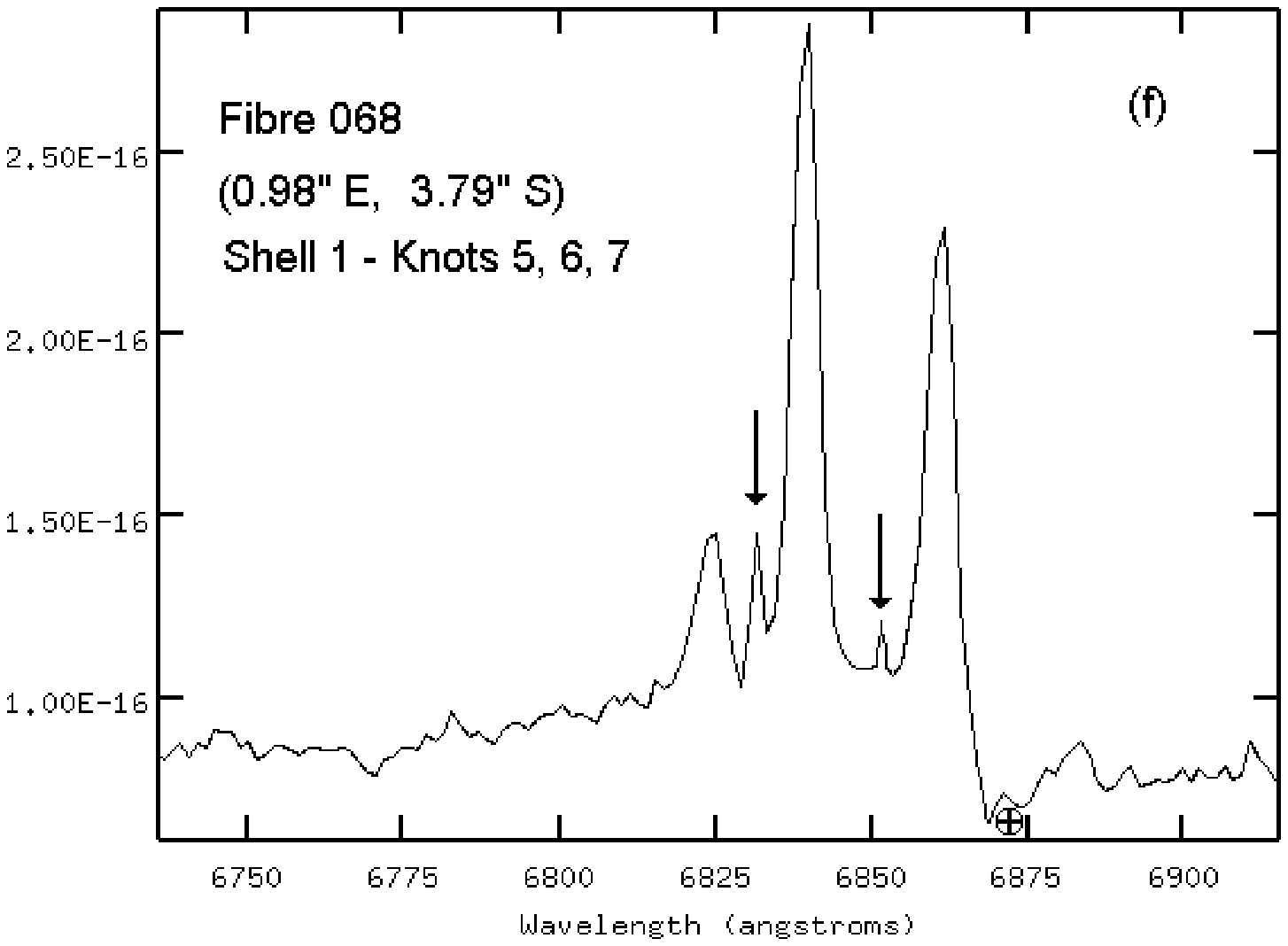} \cr
\includegraphics{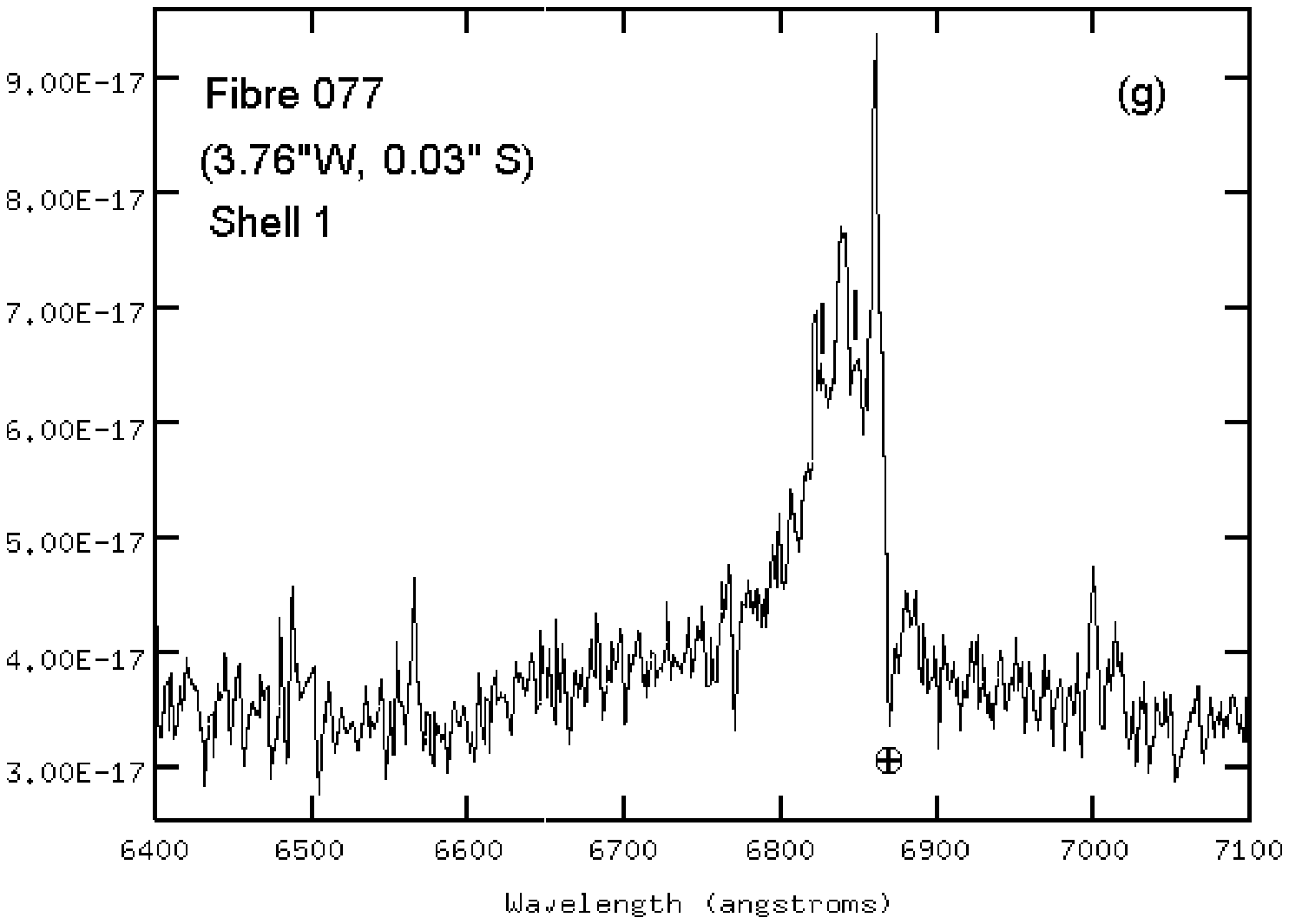}&
\includegraphics{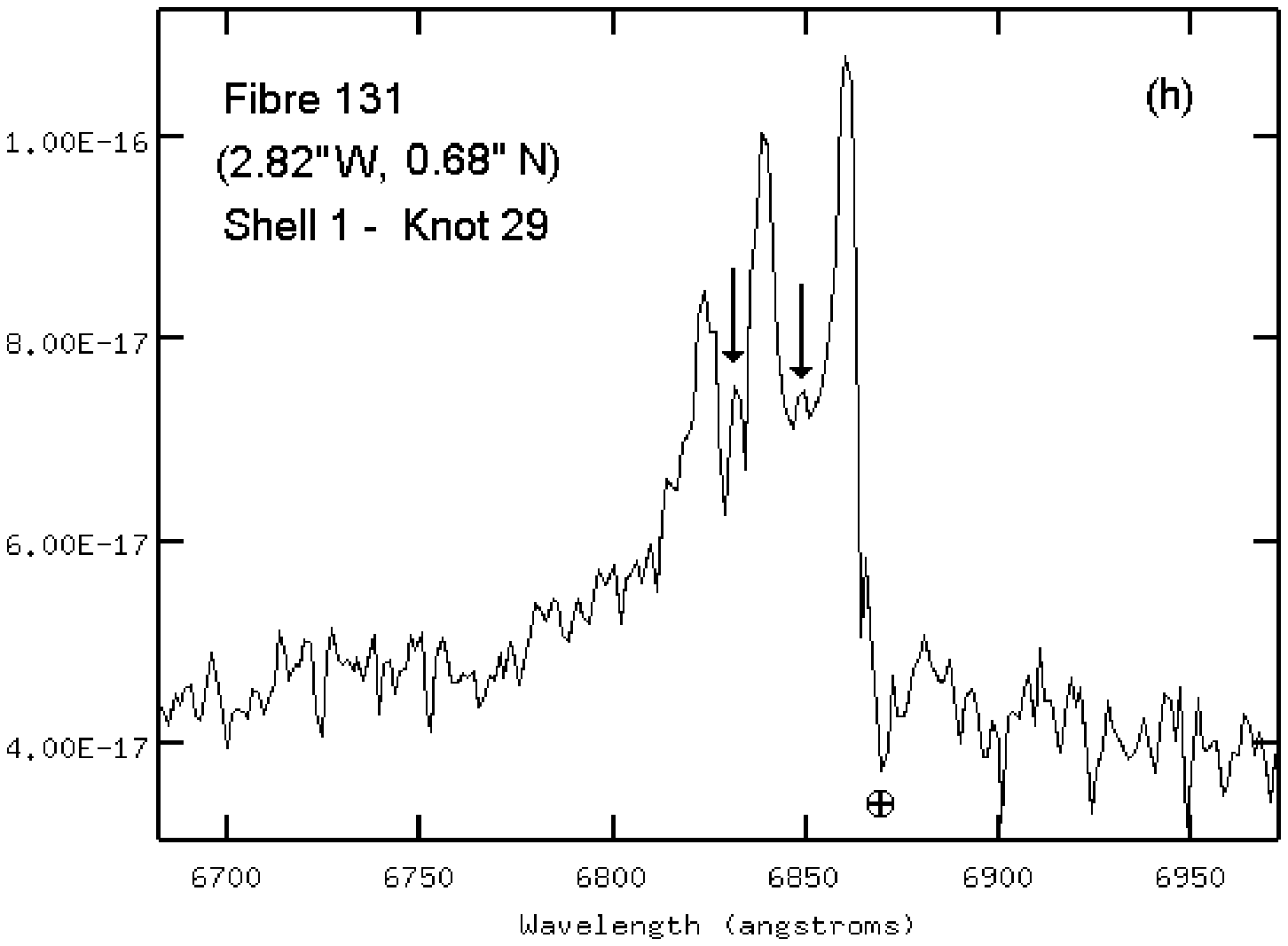} \cr
\end{tabular}
\vspace{6.0 cm}
\addtocounter{figure}{-1}
\caption {Continued
}
\label{fig12c}
\end{figure*}

\clearpage

\begin{figure*}
\vspace{12.0 cm}
\begin{tabular}{c}
\includegraphics{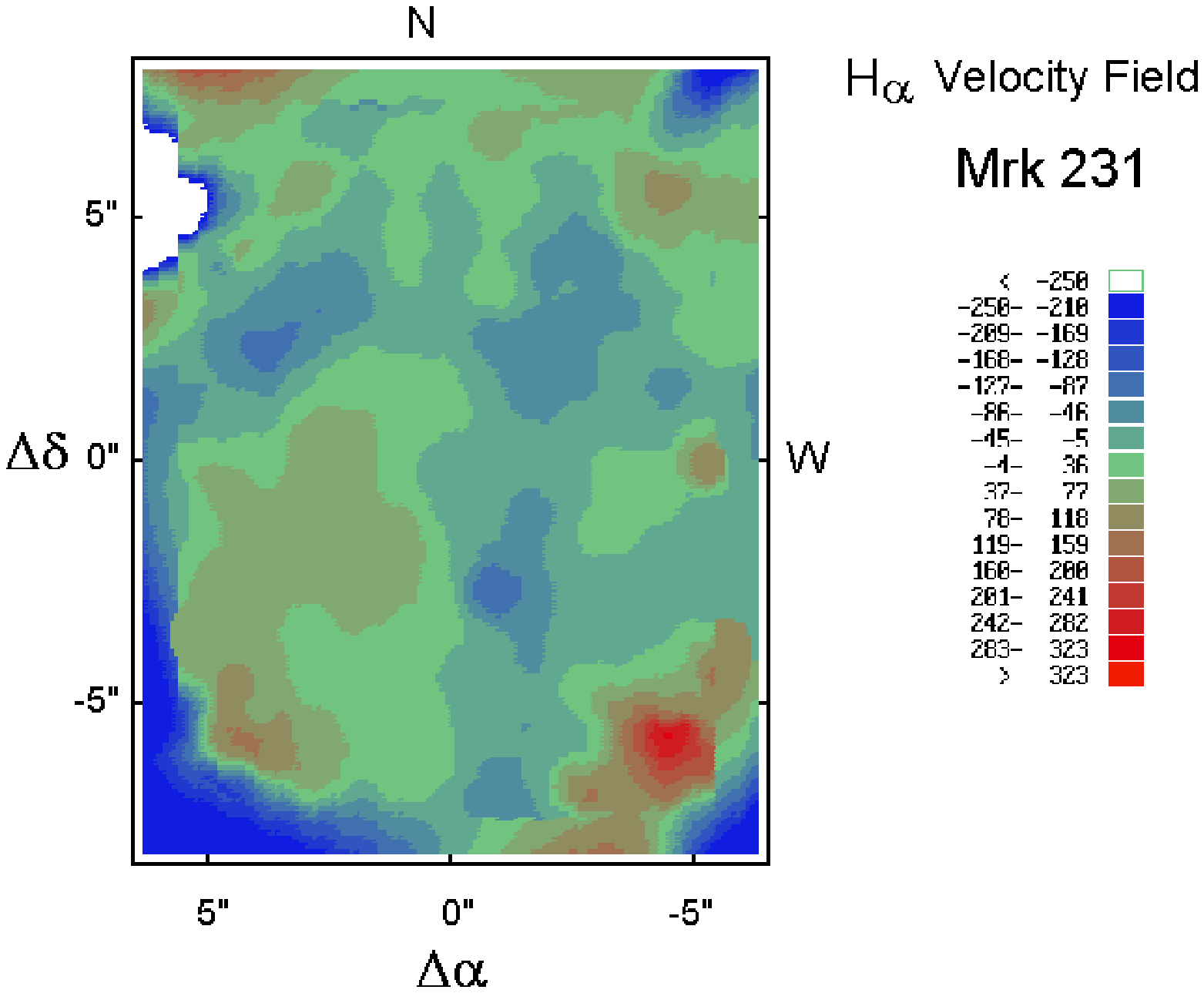}\cr
\end{tabular}
\vspace{6.0 cm}
\caption {WHT+INTEGRAL kinematics map of the ionized gas (H$\alpha$),
for the central region of Mrk 231.
}
\label{f72dkinha}
\end{figure*}

\clearpage
\begin{figure*}
\vspace{12.0 cm}
\begin{tabular}{c}
\includegraphics{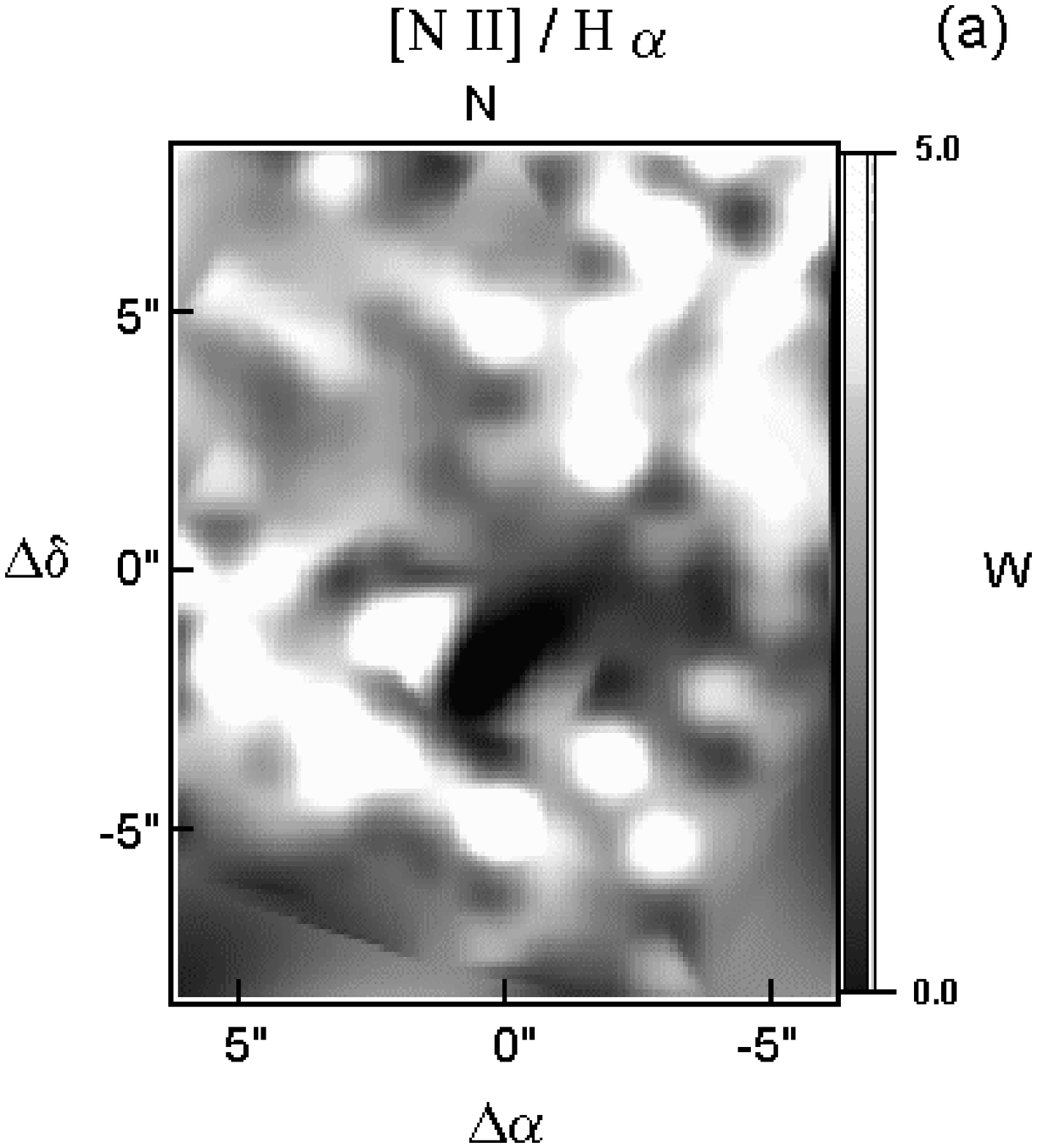} \cr
\includegraphics{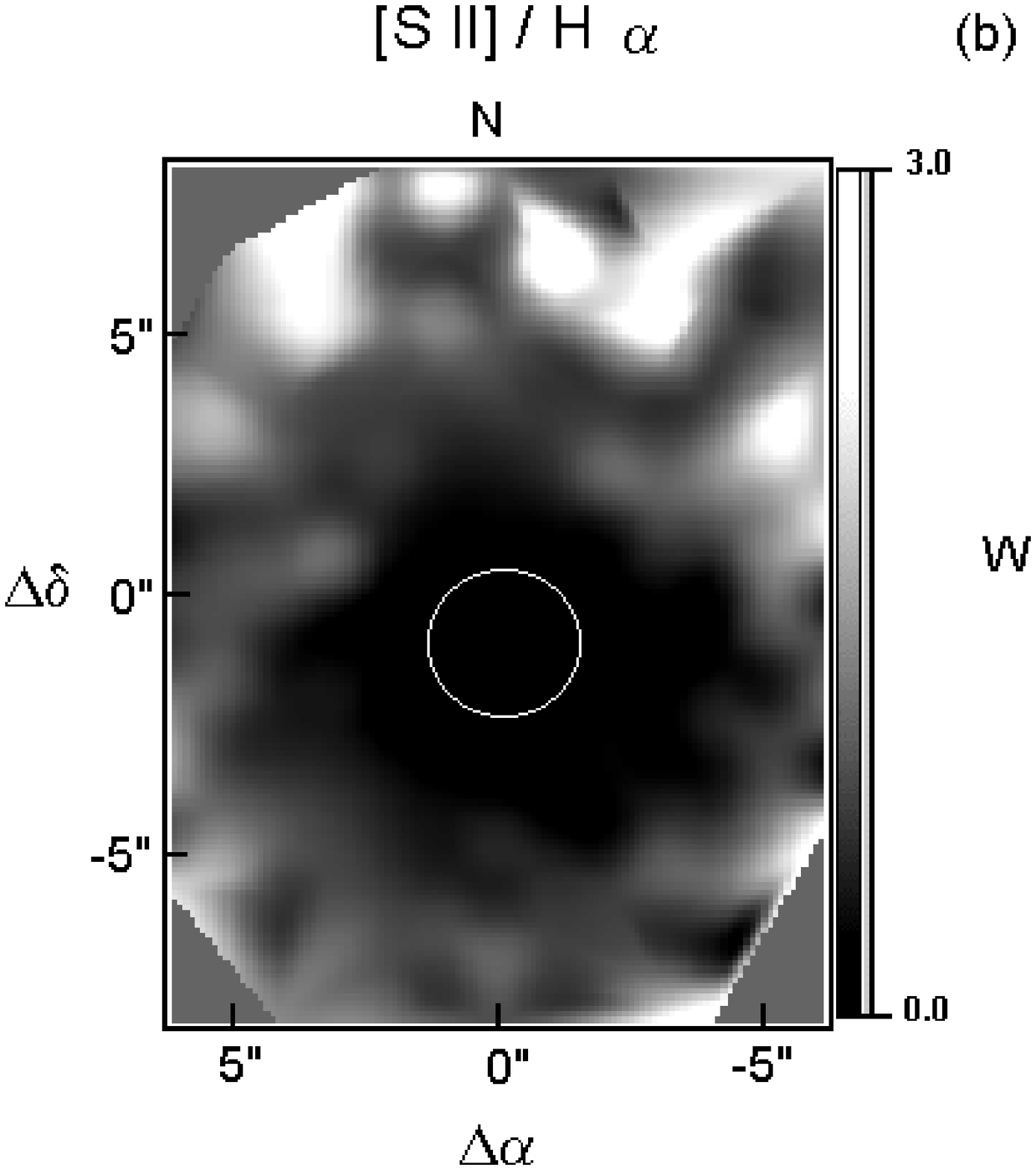} \cr
\end{tabular}
\vspace{8.0 cm}
\caption {
WHT+INTEGRAL maps of the emission line ratios for the central region of the
main body of Mrk 231. The circle in panel (b) shows the nuclear region,
of Mrk 231 (in the field of WHT+INTEGRAL), 
where the narrow component of emission lines [S {\sc ii}] are very weak.
 }
\label{f82delr1}
\end{figure*}

\clearpage
\begin{figure*}
\vspace{12.0 cm}
\begin{tabular}{c}
\includegraphics{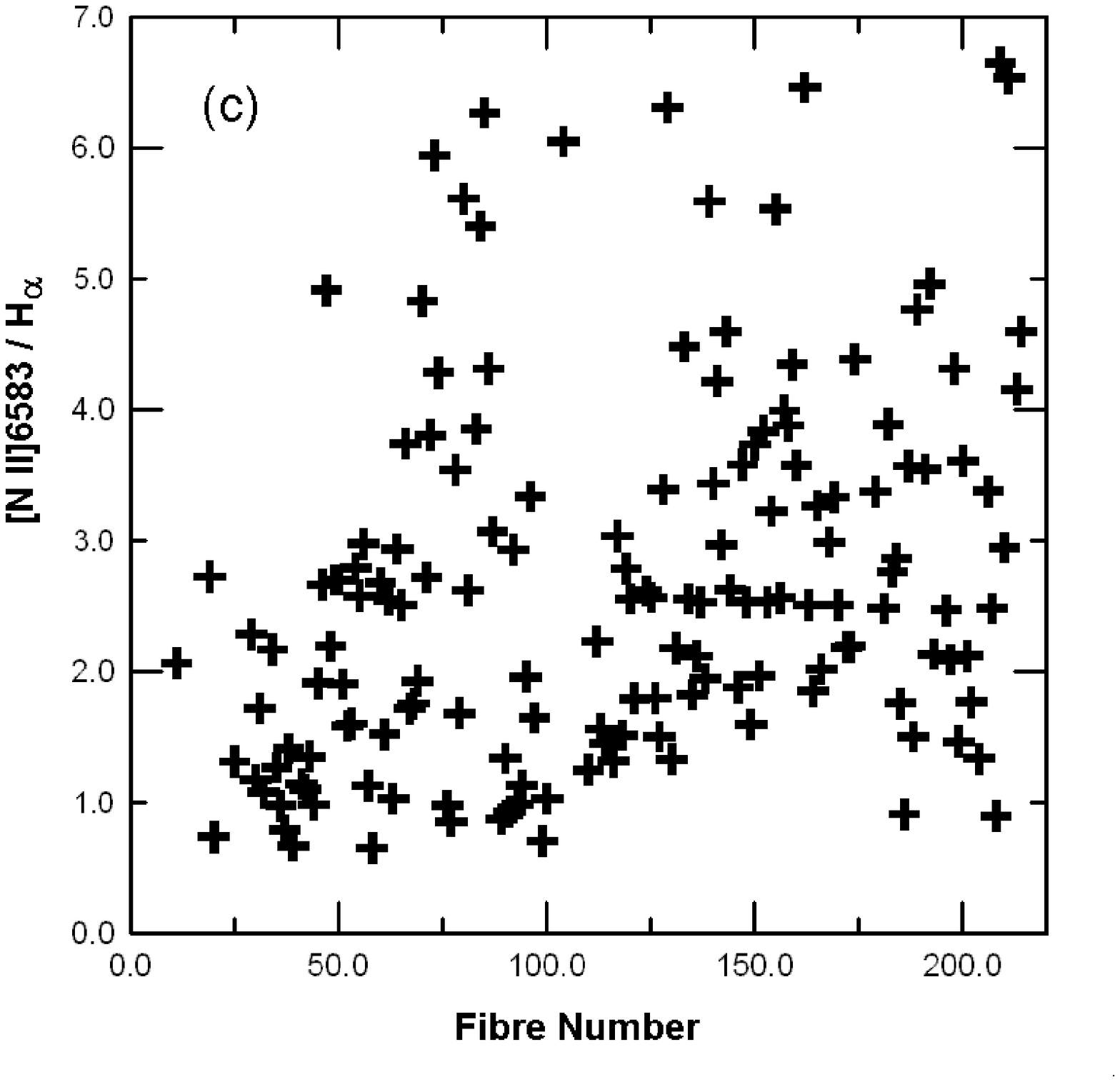} \cr
\includegraphics{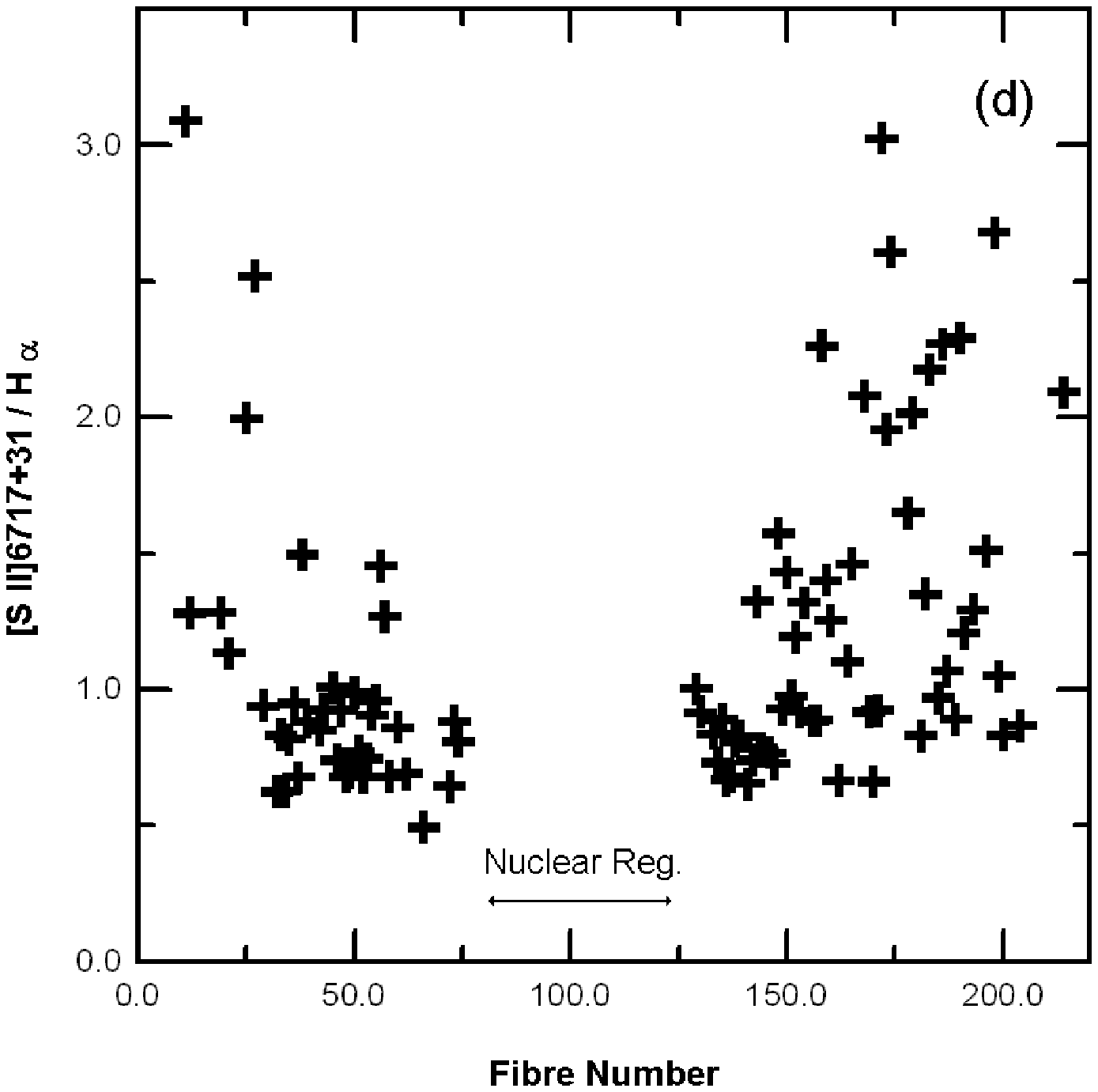} \cr
\end{tabular}
\vspace{8.0 cm}
\addtocounter{figure}{-1}
\caption {Continued
 }
\label{f82delr2}
\end{figure*}

\clearpage

\begin{figure*}
\vspace{12.0 cm}
\begin{tabular}{c}
\includegraphics{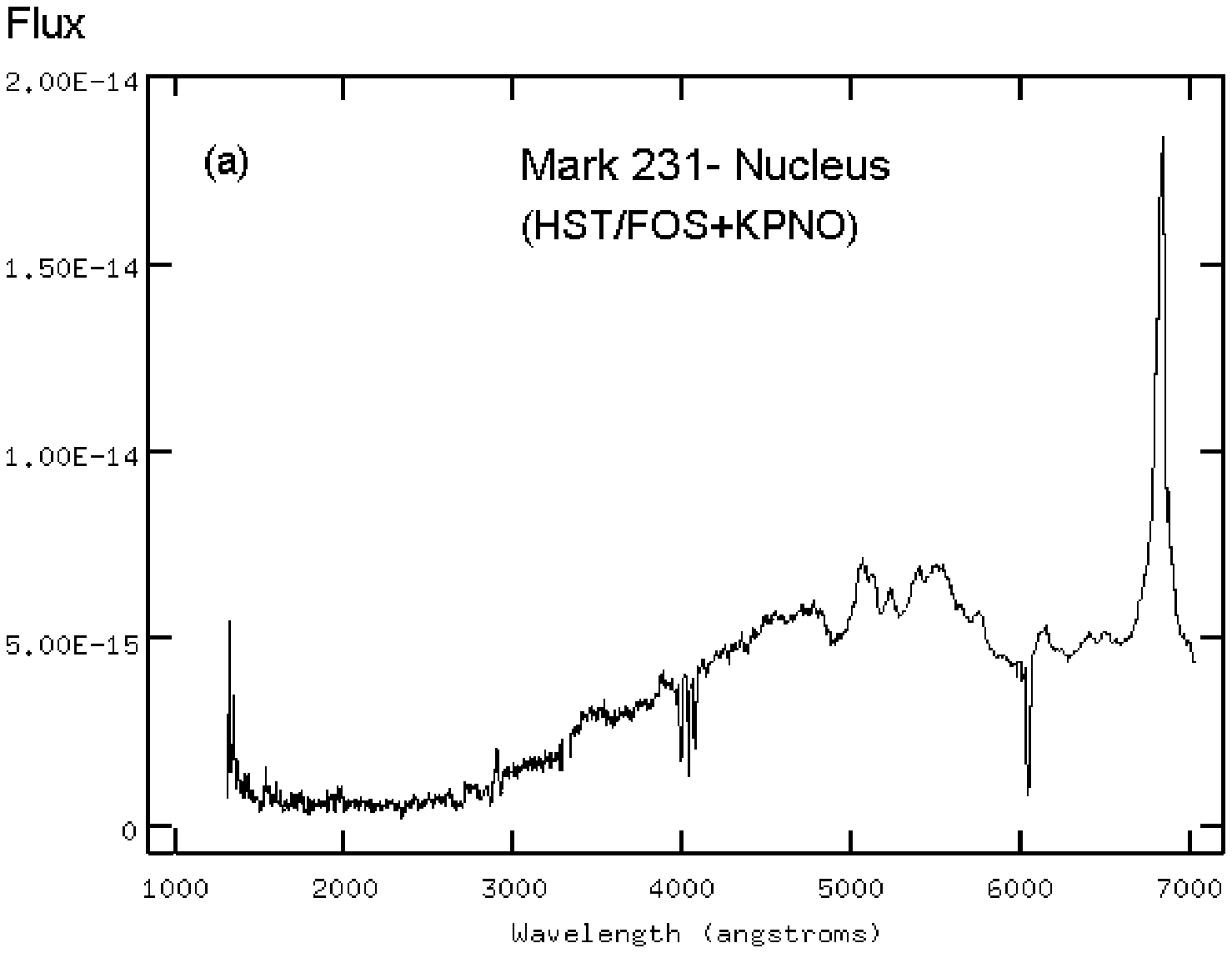} \cr
\includegraphics{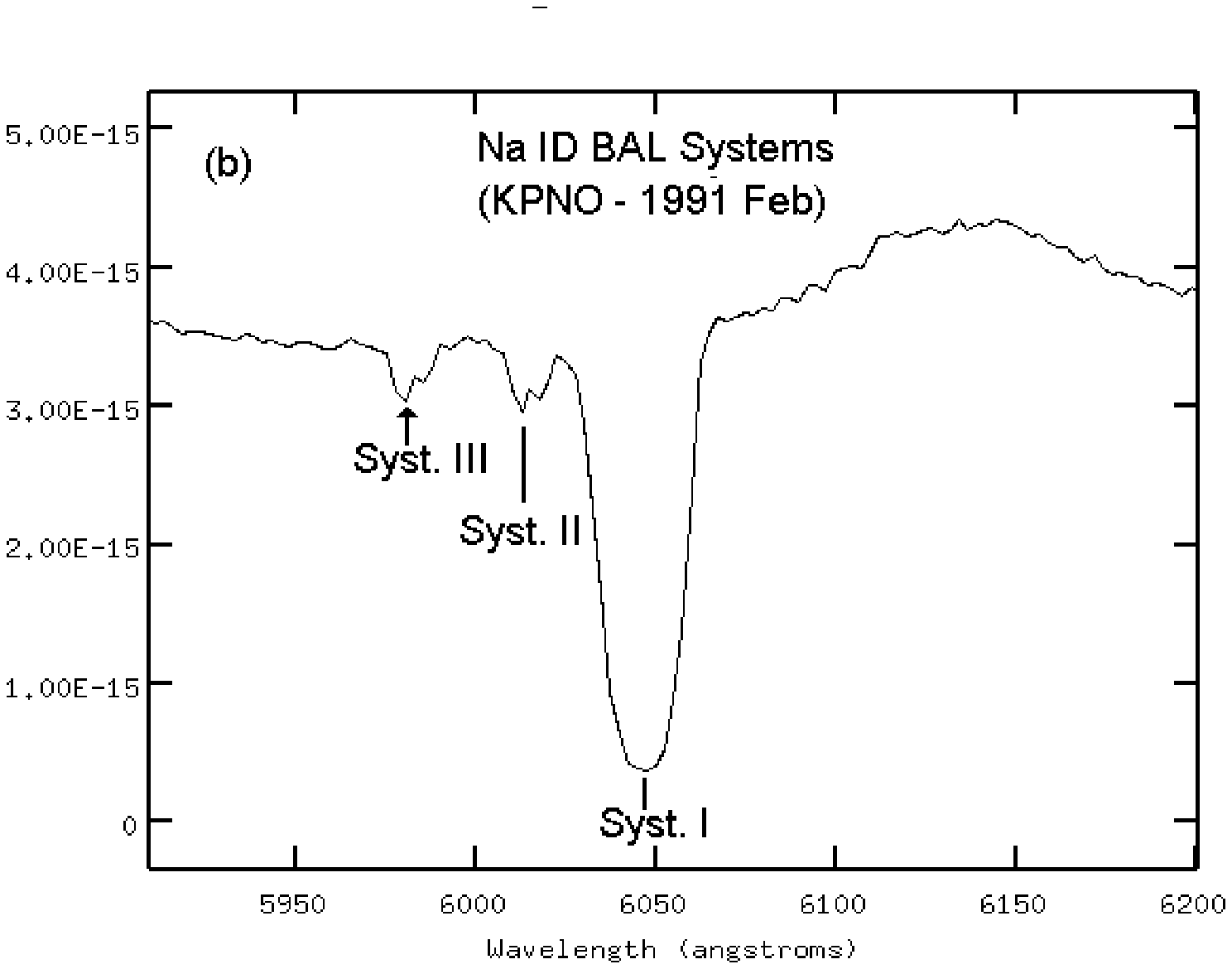} \cr
\includegraphics{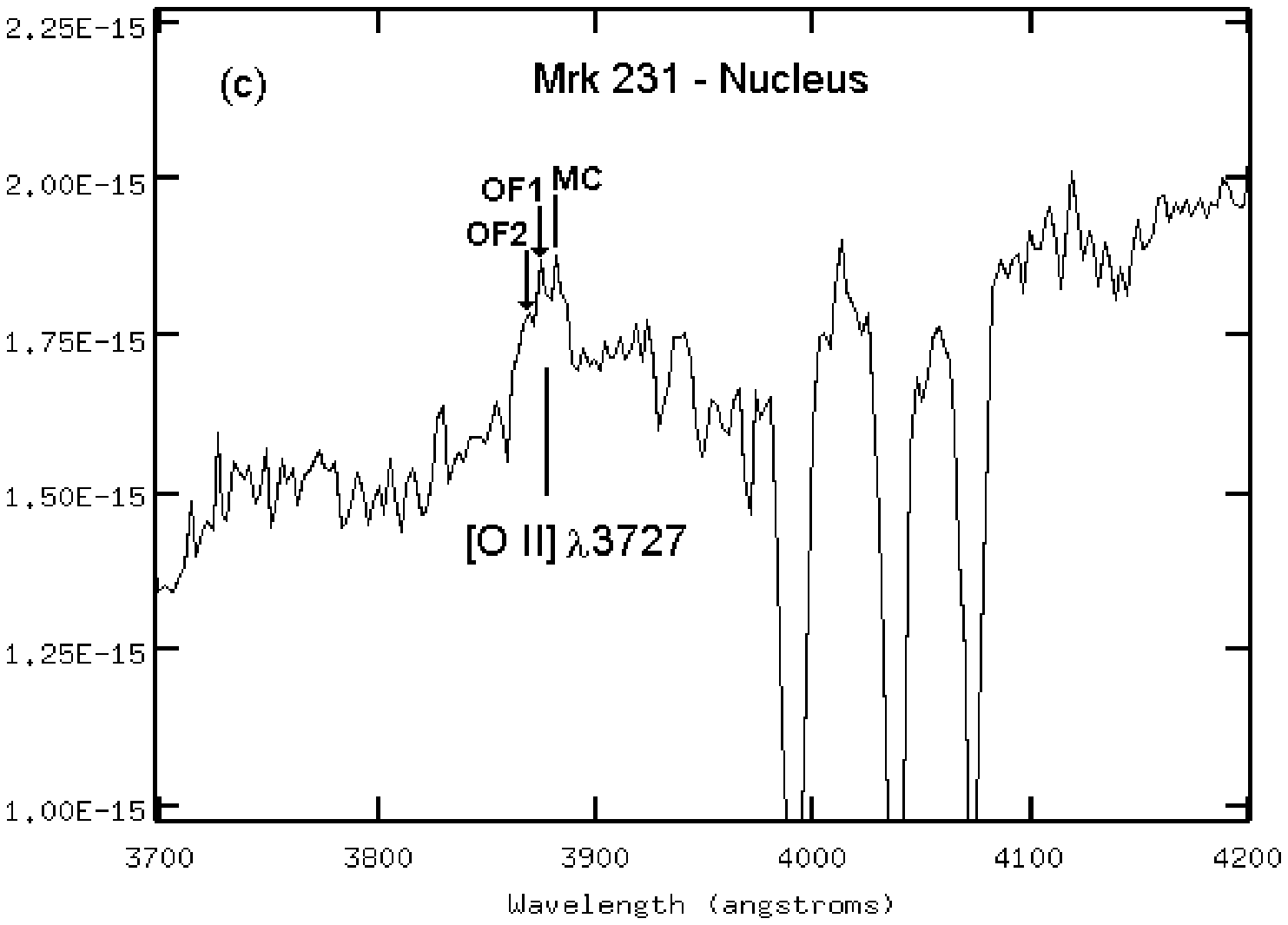} \cr
\end{tabular}
\vspace{8.0 cm}
\caption {KPNO 1D optical spectra of Mrk 231.
The scales of flux are given in units of [erg $\times$ cm$^{-2}$ $\times$
s$^{-1}$ $\times$ \AA$^{-1}$] and wavelength in [$\AA$].
 }
\label{f91doptico}
\end{figure*}

\clearpage

\clearpage
\begin{figure*}
\vspace{12.0 cm}
\begin{tabular}{c}
\includegraphics{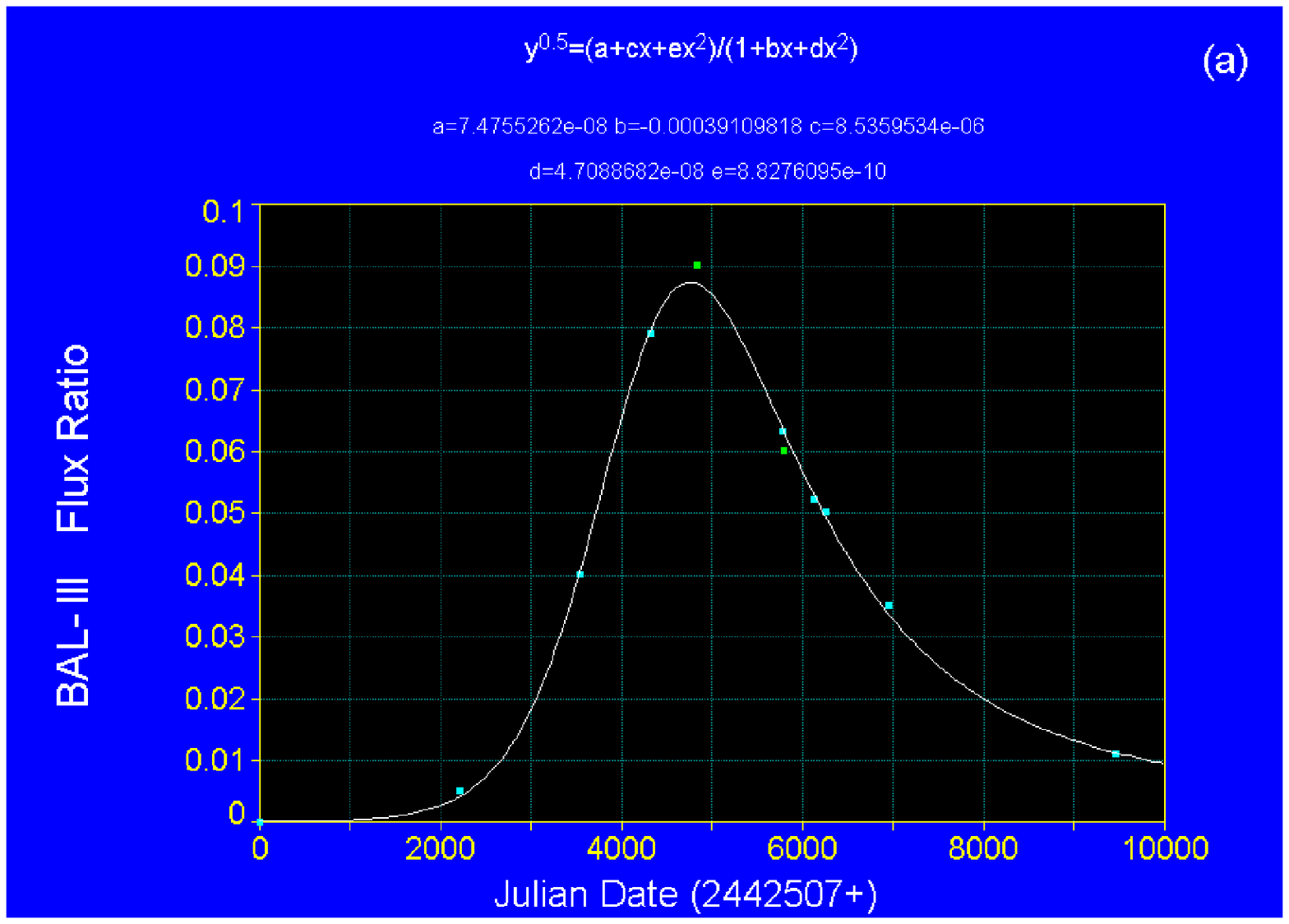} \cr
\includegraphics{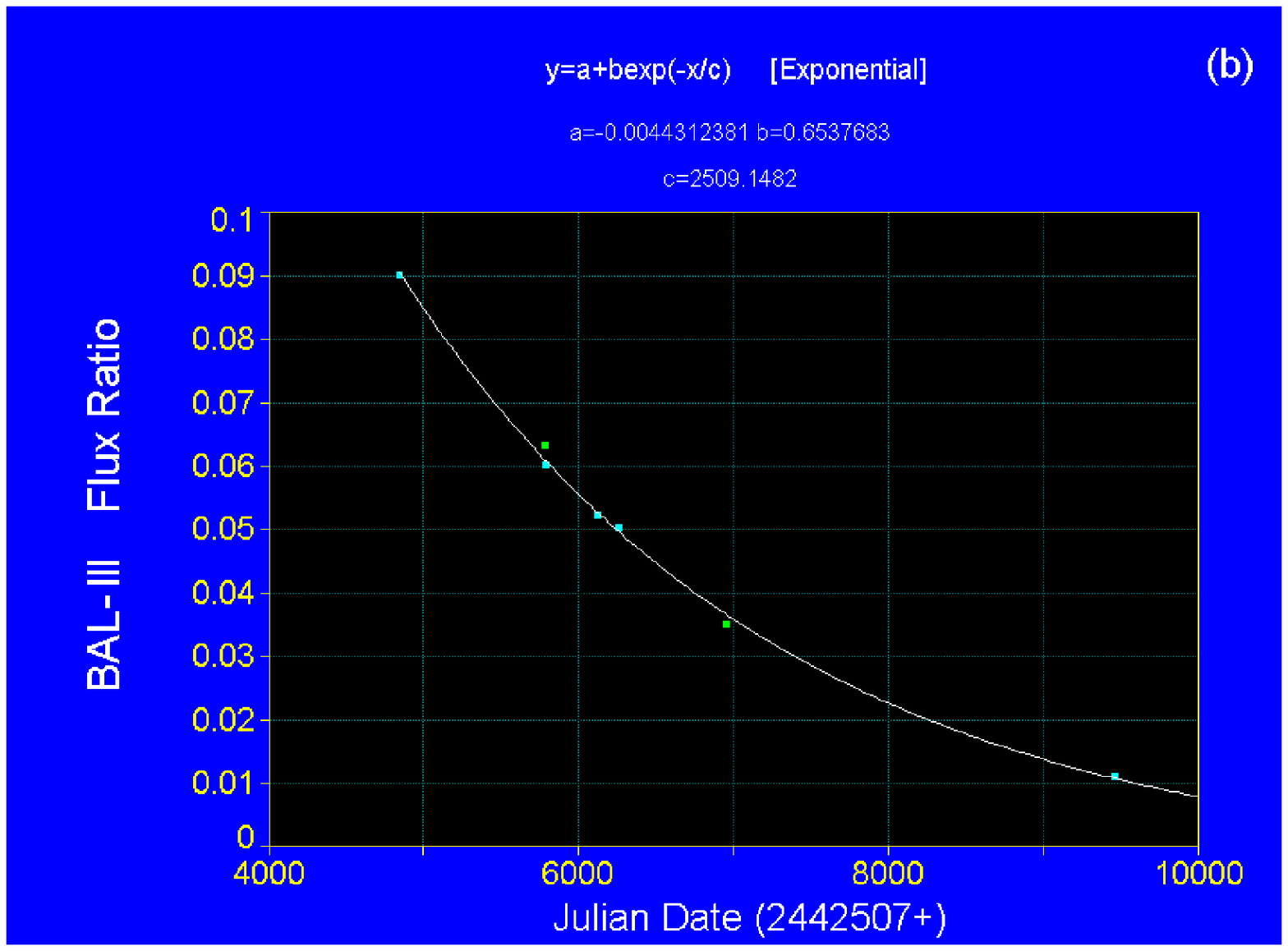} \cr
\end{tabular}
\vspace{5.0 cm}
\caption {Light curve variability of the Na ID BAL III system, between the
years 1975 and 2000 (a), and between 1988-2000 (b).
Showing also the best fit of the LC. For the fall of the LC the best fit is
an exponential function.
 }
\label{f10bal4var}
\end{figure*}

\clearpage

\begin{figure*}
\vspace{12.0 cm}
\begin{tabular}{c}
\includegraphics{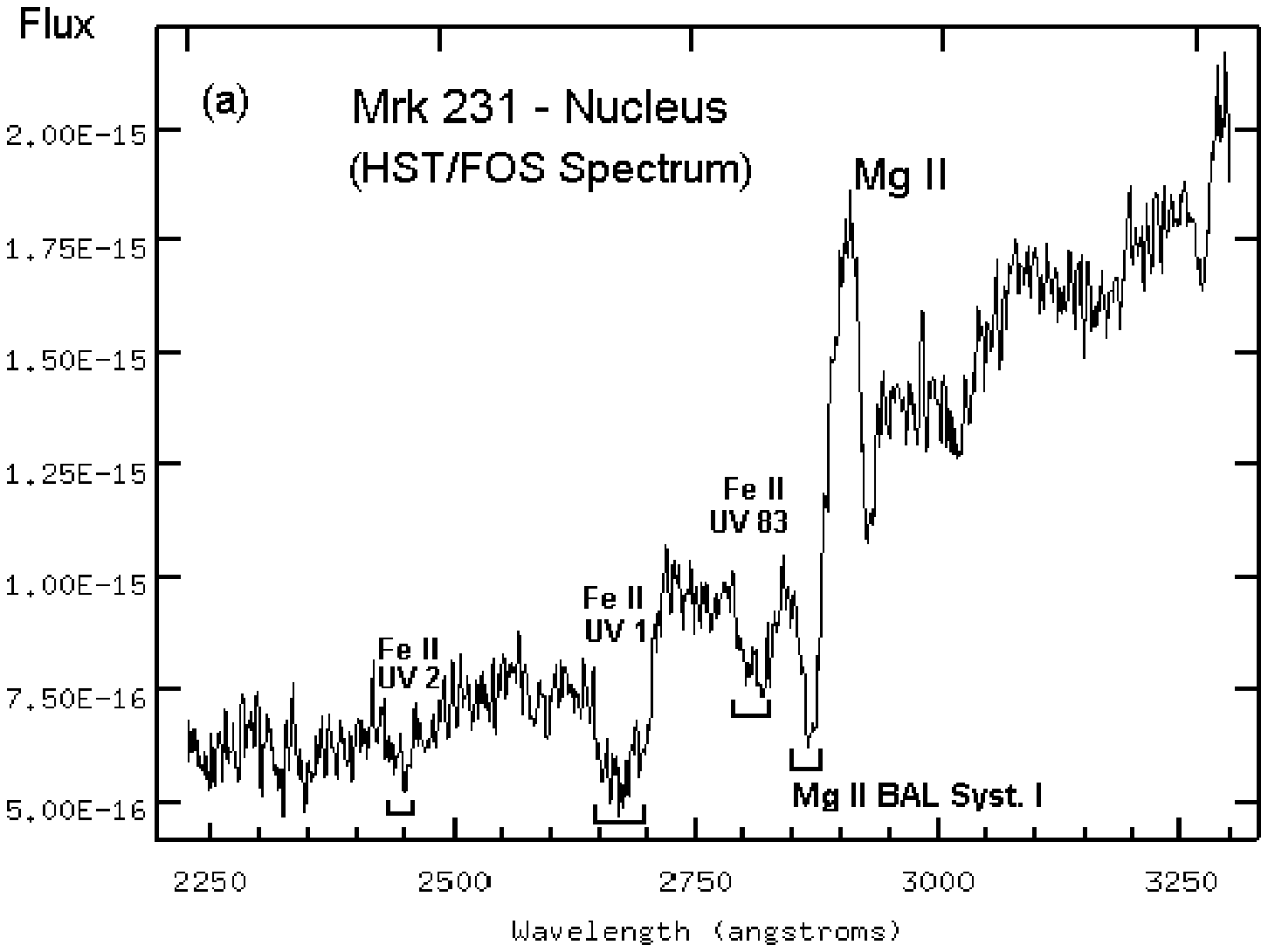} \cr
\includegraphics{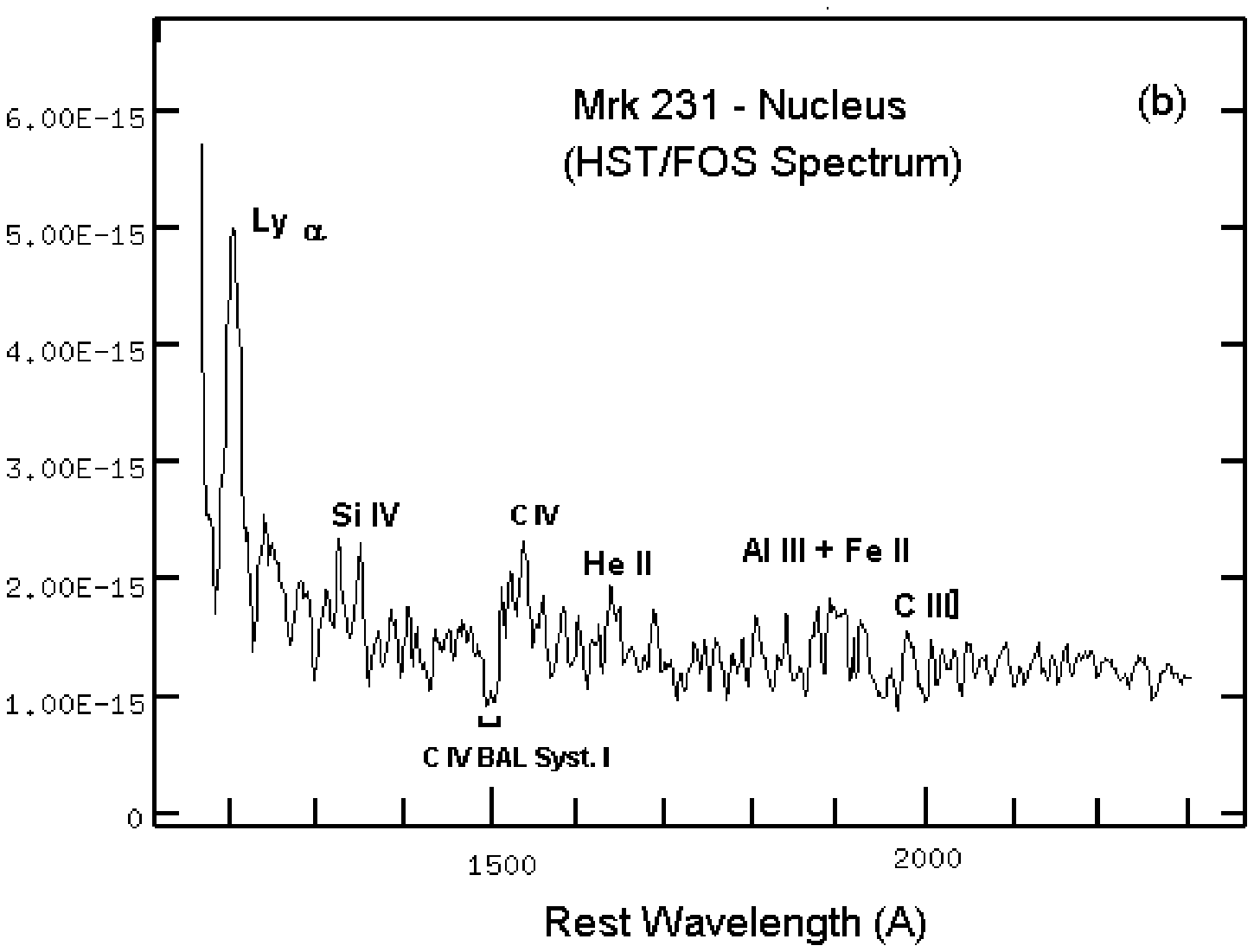} \cr
\end{tabular}
\vspace{6.0 cm}
\caption {UV 1D HST FOS spectra of Mrk 231.
 }
\label{f11uv1d}
\end{figure*}

\clearpage

\begin{figure*}
\vspace{12.0 cm}
\begin{tabular}{cc}
\includegraphics{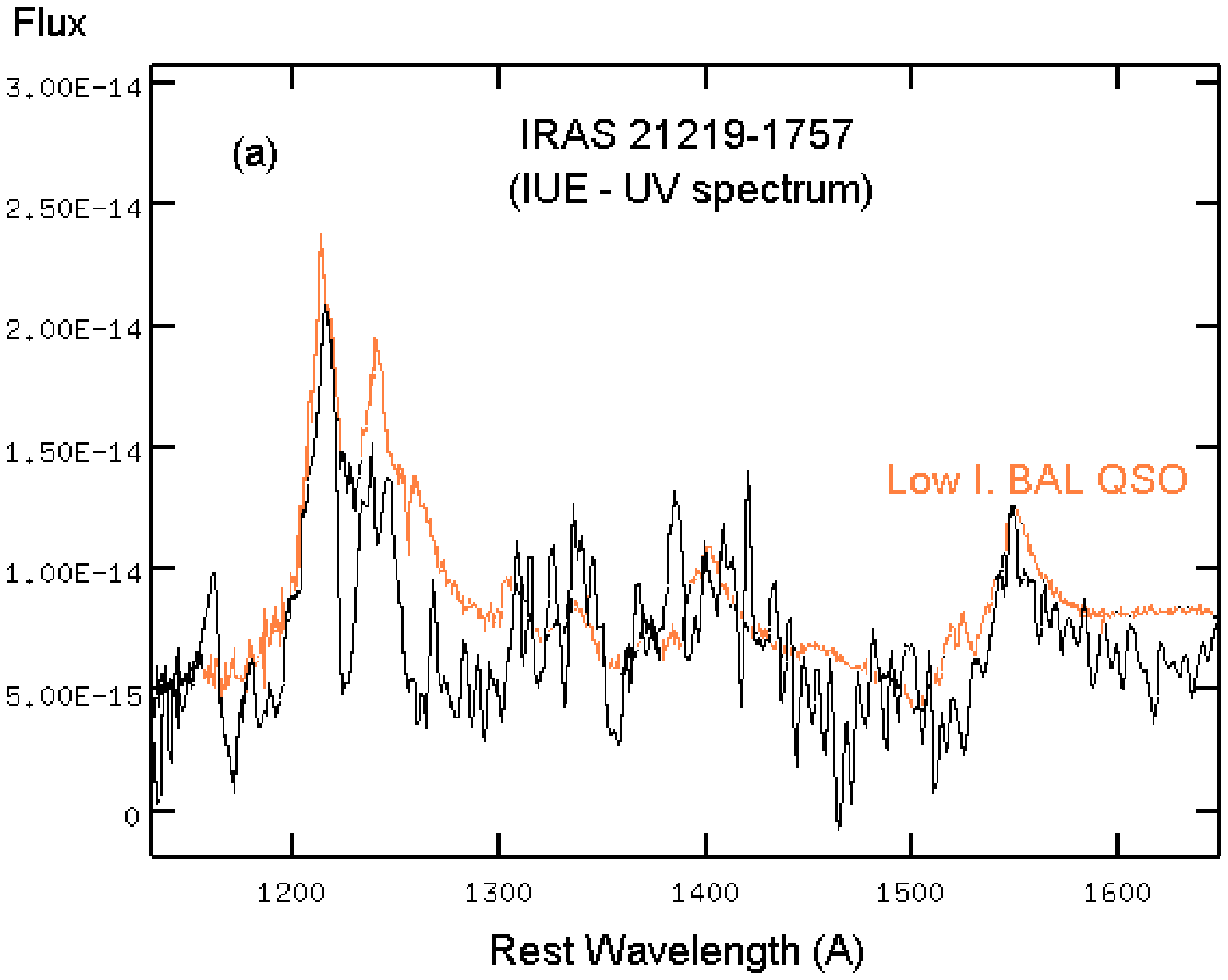}&
\includegraphics{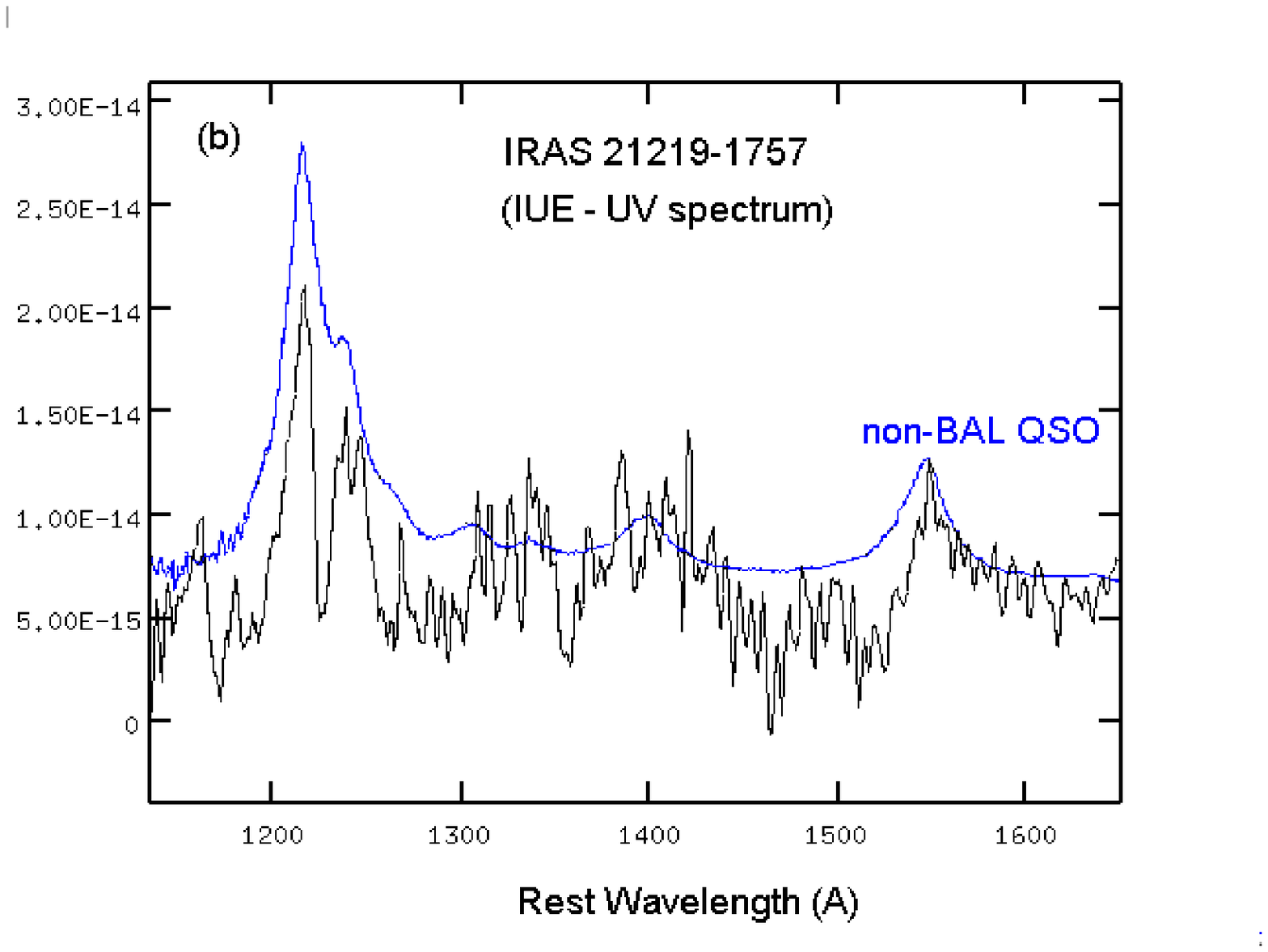} \cr
\includegraphics{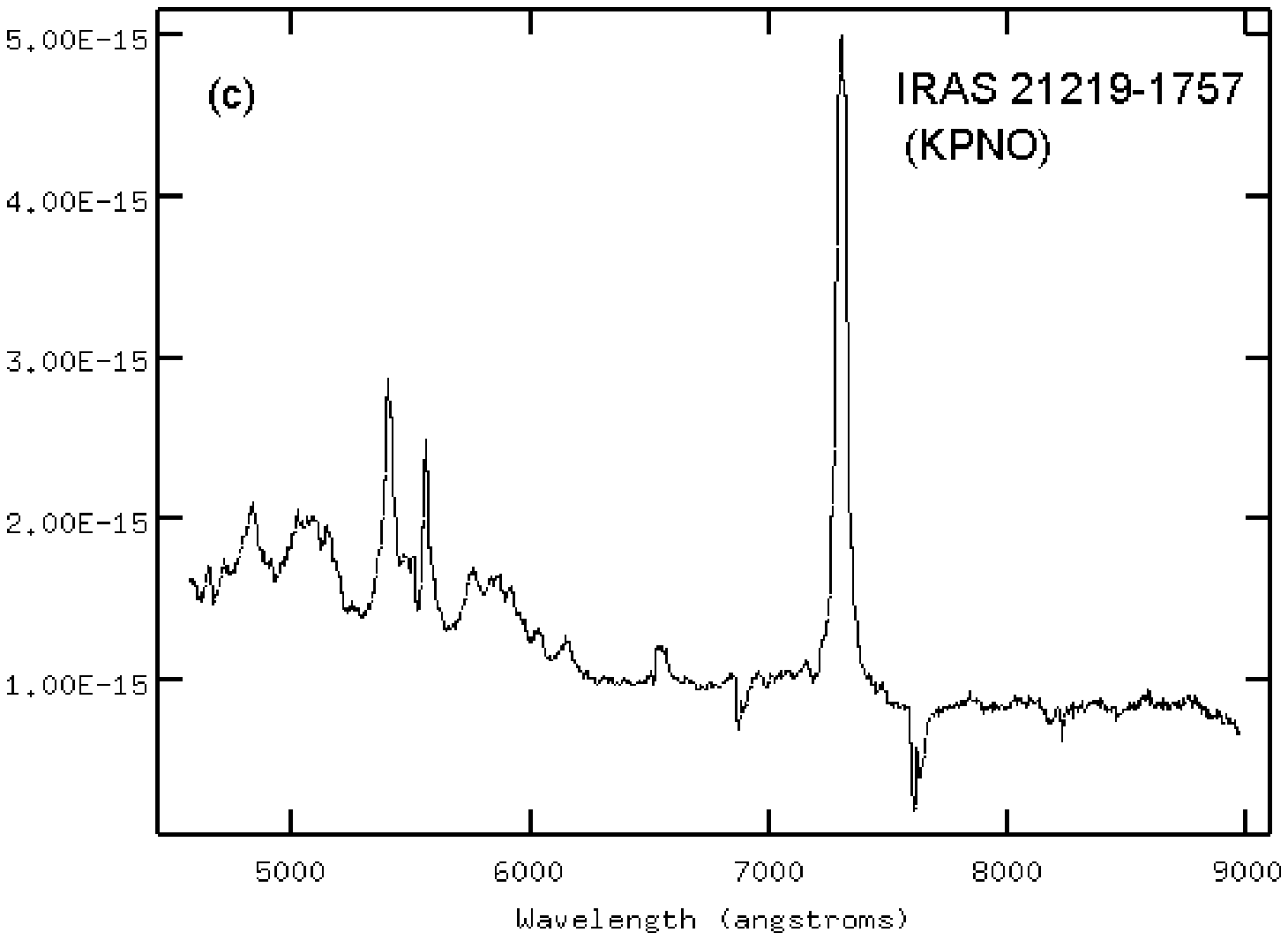}&
\includegraphics{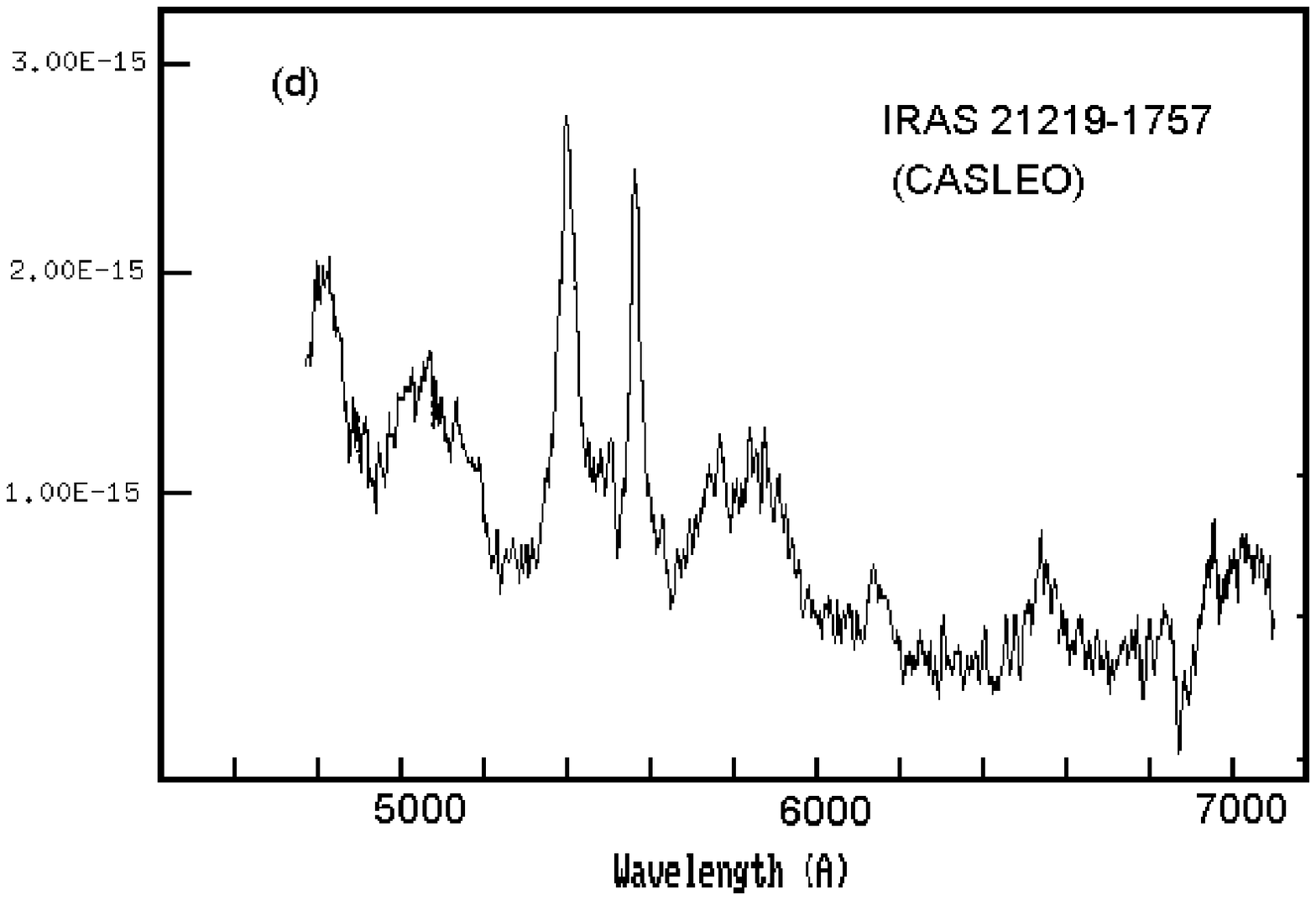} \cr
\end{tabular}
\vspace{6.0 cm}
\caption {
UV and optical spectra of the IR + GW/OF + Fe {\c ii}  QSO IRAS 21219-1757
showing UV BAL systems.
}
\label{fig12}
\end{figure*}

%\vspace{6.0 cm}
%\addtocounter{figure}{-1}
%\caption {Continued

\clearpage

\begin{figure*}
\vspace{12.0 cm}
\begin{tabular}{cc}
\includegraphics{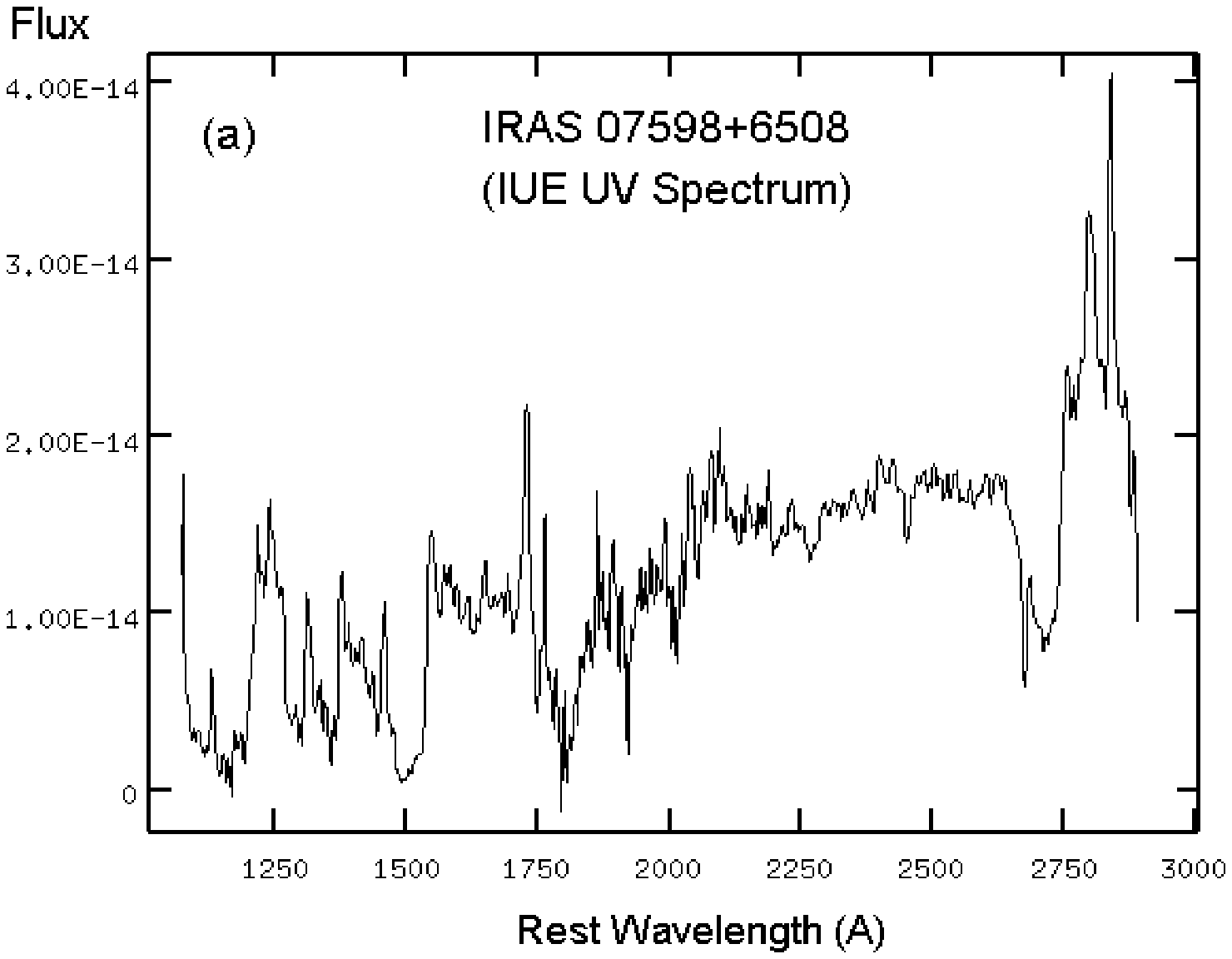}& 
\includegraphics{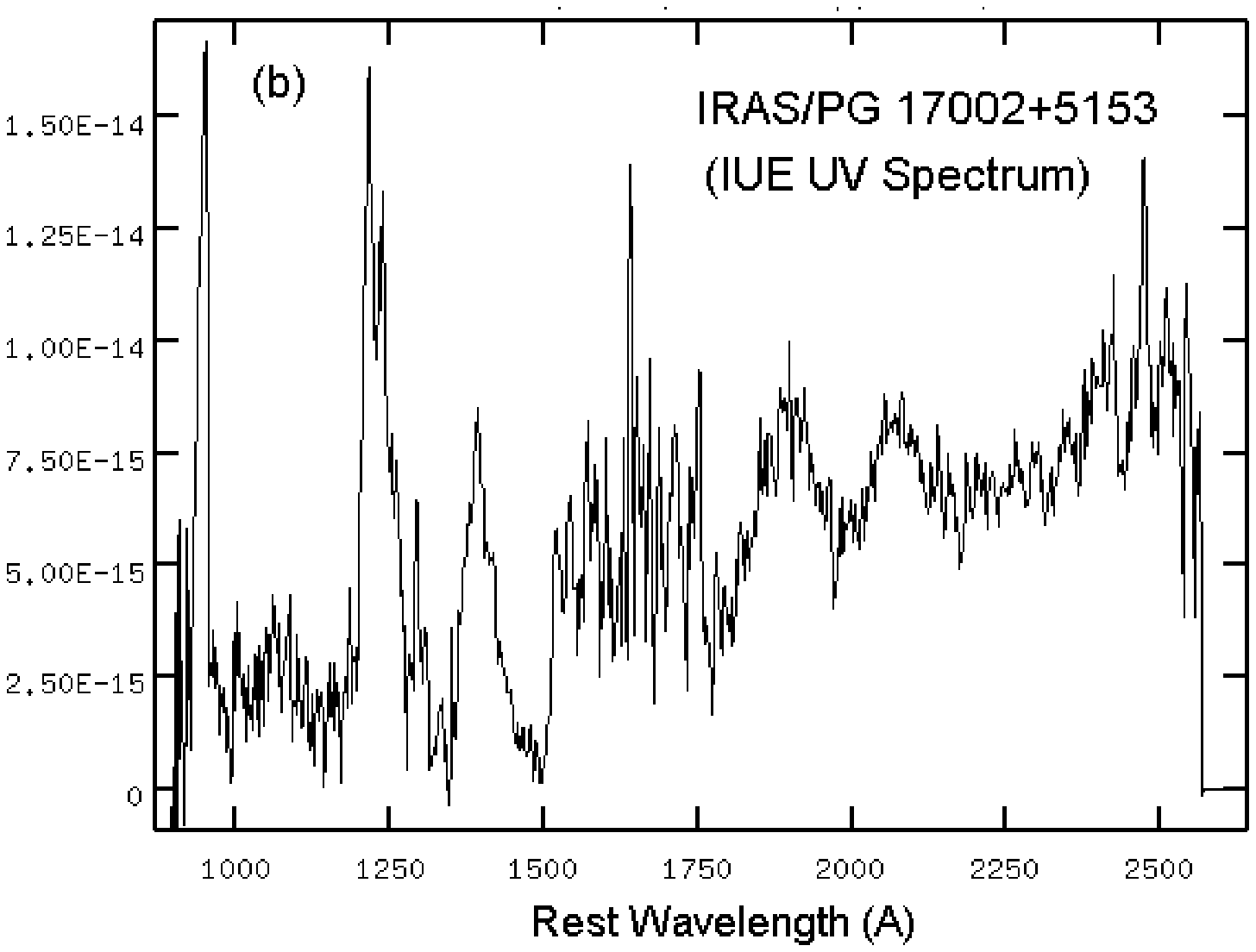} \cr
\includegraphics{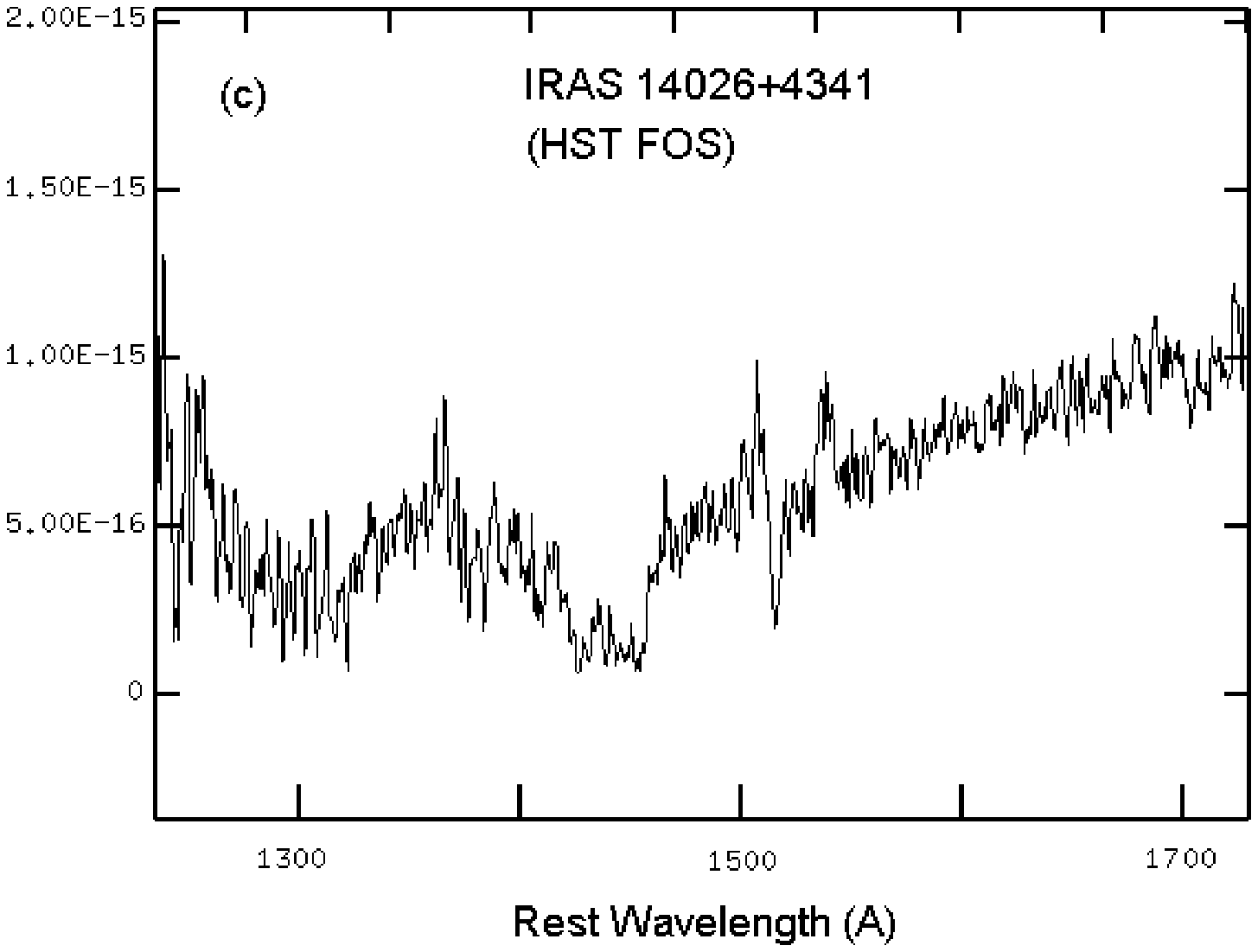}&
\includegraphics{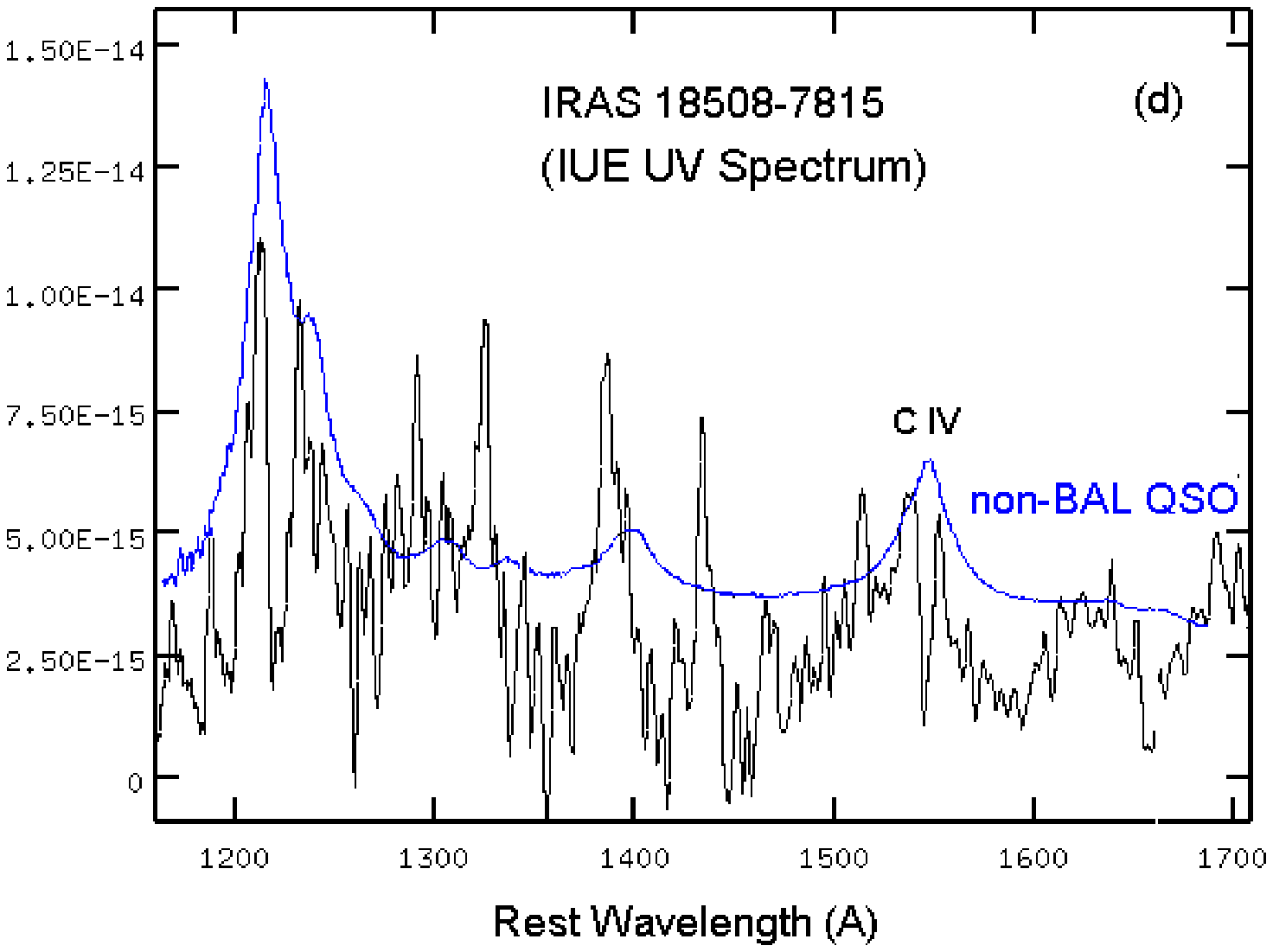} \cr
\includegraphics{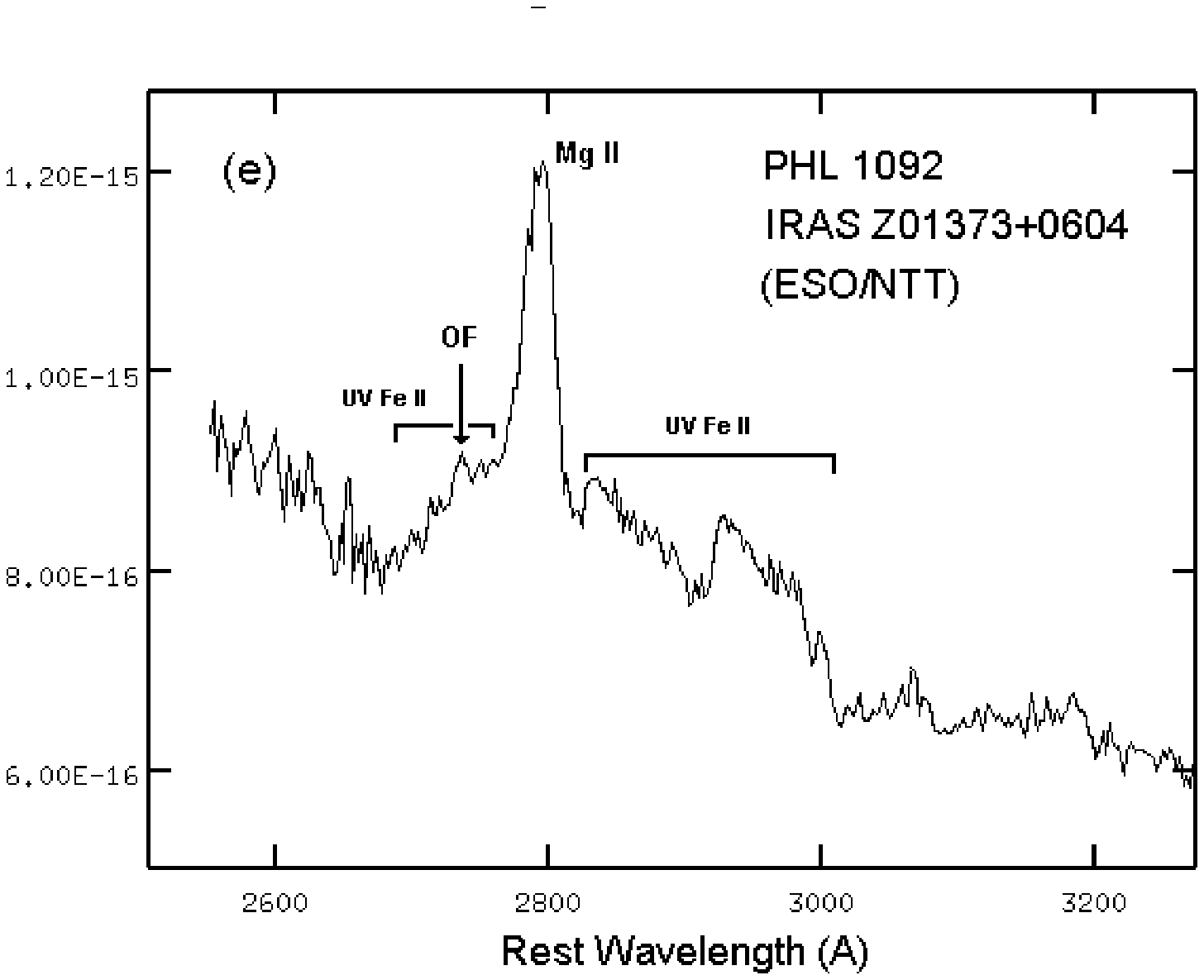}& 
\includegraphics{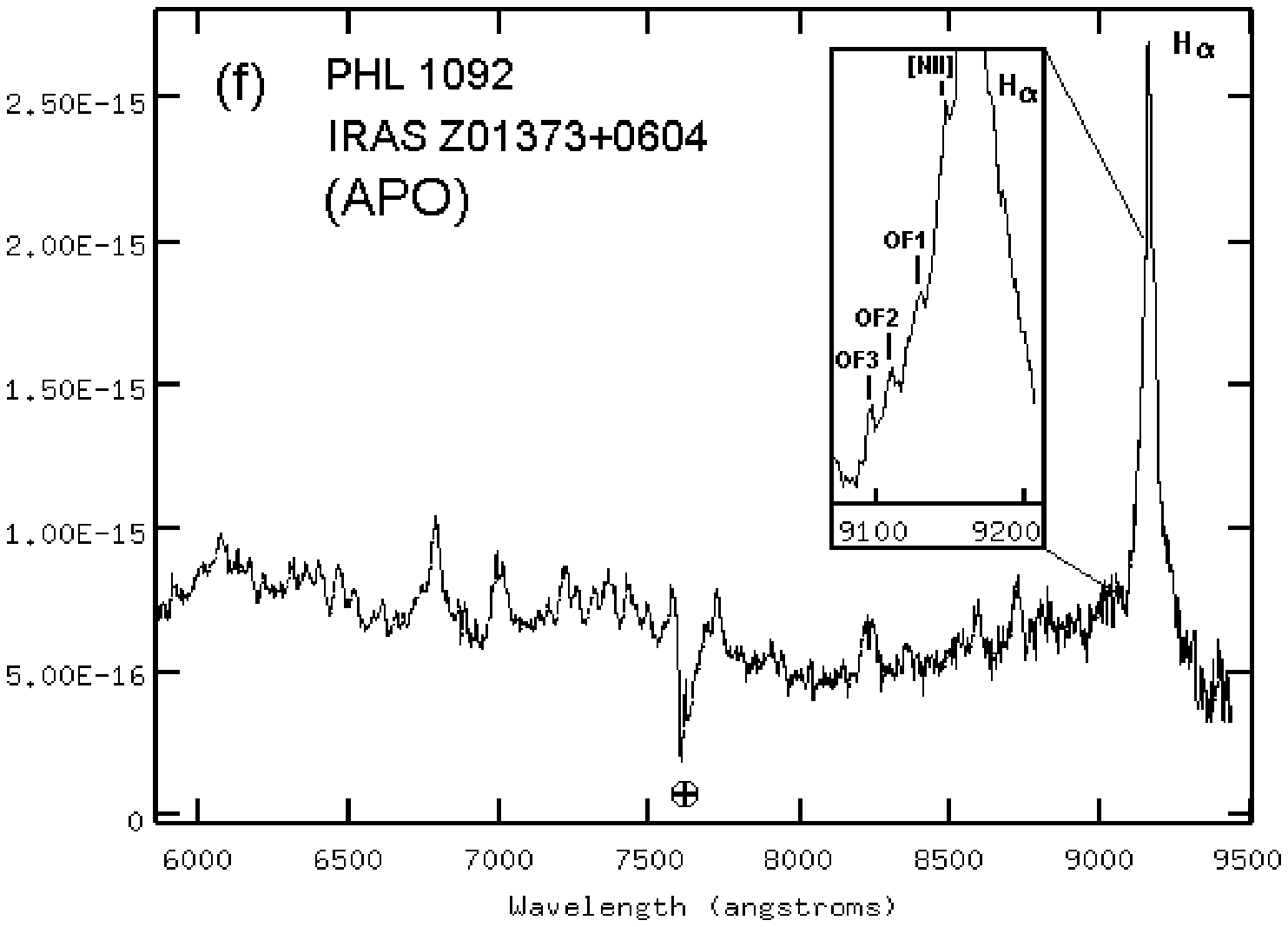} \cr
\end{tabular}
\vspace{8.0 cm}
\caption {UV and optical spectra of IR + GW/OF + Fe {\sc ii} QSOs (including
specially the BAL IR QSOs).
 }
\label{f13balirqso}
\end{figure*}

\clearpage

\begin{figure*}
\vspace{12.0 cm}
\begin{tabular}{c}
\includegraphics{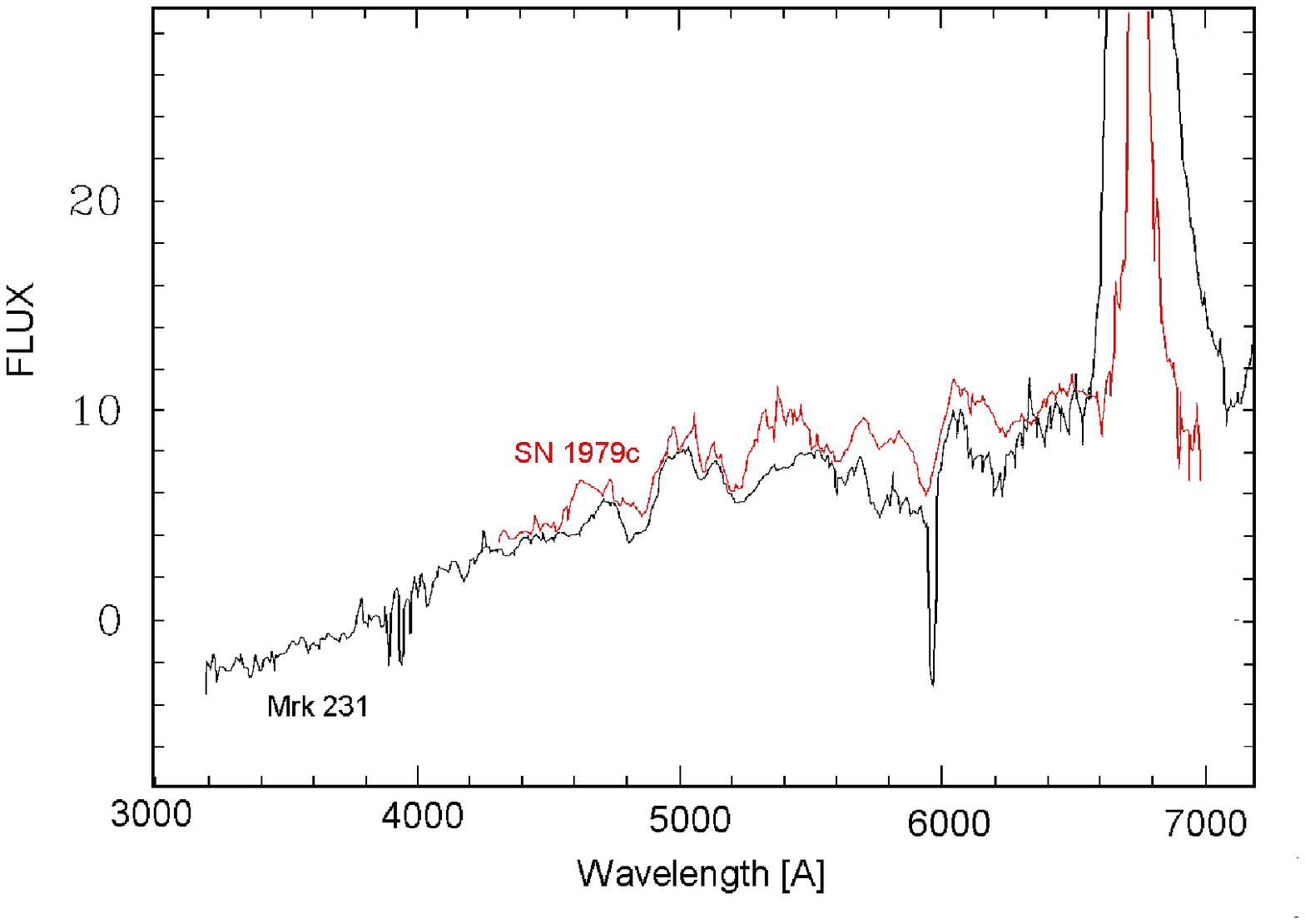}\cr
\end{tabular}
\vspace{6.0 cm}
\caption {Optical spectrum of Mrk 231 superposed to the spectrum of the
SN 1979c (Branch et al. 1981).
}
\label{f14sn}
\end{figure*}

\clearpage

\begin{figure*}
\vspace{12.0 cm}
\begin{tabular}{c}
\includegraphics{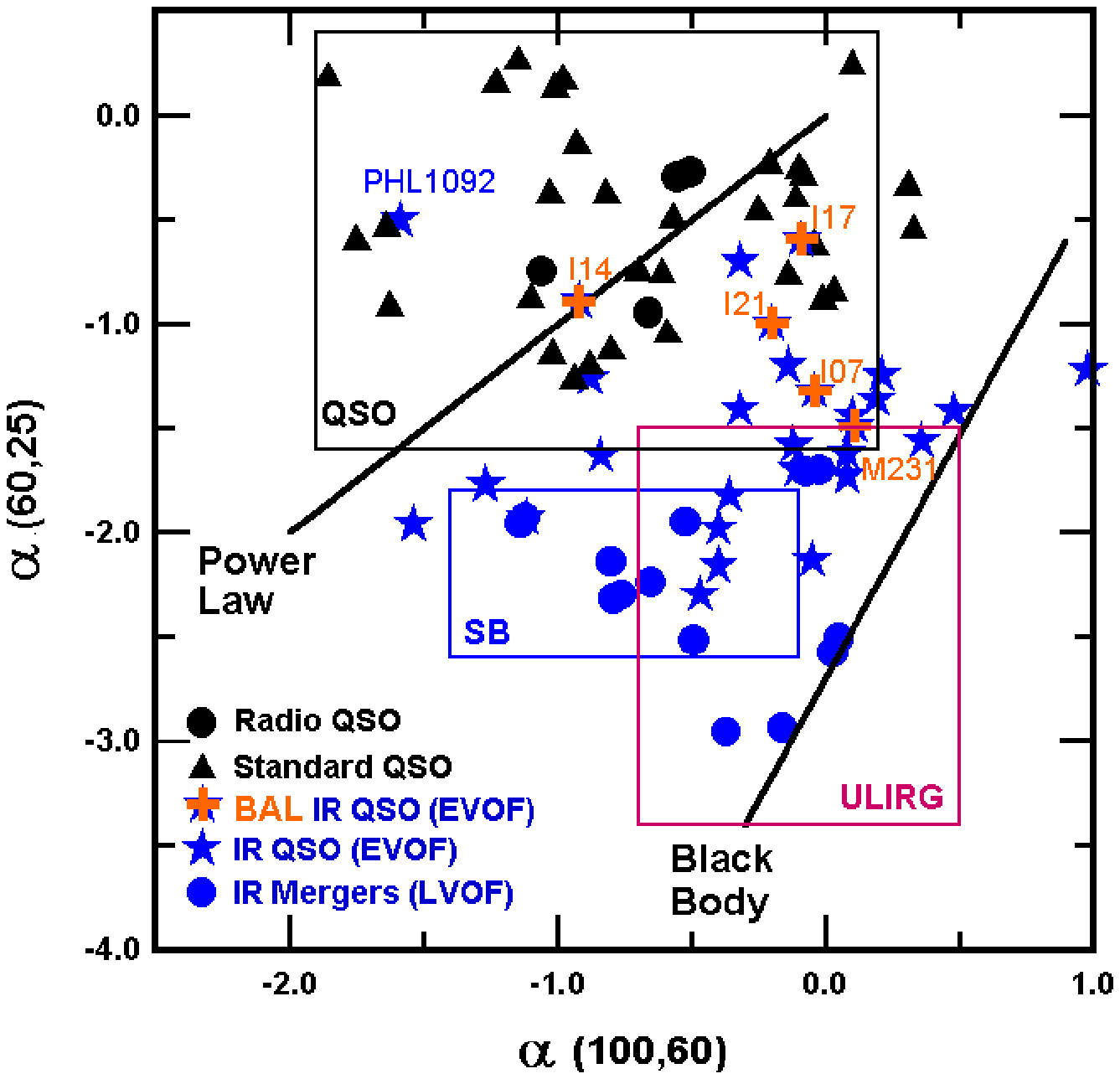}\cr
\end{tabular}
\vspace{8.0 cm}
\caption {IR colour--colour diagram for IR  mergers/QSOs with galactic
winds (from Table 4 of this paper) and for standard QSOs (from the PG sample;
Boroson \& Green 1992).
 }
\label{f15ircc}
\end{figure*}


\begin{thebibliography}{99}

\bibitem[1972a]{ada72}Adams T. F., 1972, ApJ, 176, L1

\bibitem[1972b]{ada72b}Adams T. F., Weedman D. W., 1972, ApJ, 172, L19

\bibitem[2002]{aji02}Ajiki M., et al., 2002, ApJ, 576, L25

\bibitem[1971]{ara71}Arakelian M. A., Debai E., Esipov V., Markarian B.,
1971, Astrophyzika, 7, 177

\bibitem[1988]{arm88}Armus L., Heckman T.M., Miley G., 1988, ApJ, 326, L45

\bibitem[1994]{arm94}Armus L., et al., 1994, AJ, 108, 76

\bibitem[1998]{arr98}Arribas S. et al. 1998, SPIE, 3355, 821

\bibitem[1993]{art93}Artymowicz P, Lin D., Wampler E., 1993, ApJ, 409, 592

%\bibitem[1989]{bar89}Barnes J., 1989, Nat, 338, 123

\bibitem[1992]{bar92}Barnes J., Hernquist L., 1992, ARA\&A, 30, 705

\bibitem[2003]{bar03}Barth A., Martini P., Nelson C., Ho L., 2003, ApJ, 594, L95

\bibitem[1986]{bei86}Beichman C. A., et al., 1986, ApJ, 308, L1

%\bibitem[1998]{bek98}Bekki K., Shioya Y., 1998, ApJ, 497, 108

%\bibitem[1985]{ber85}Bergvall N., Johansson L., 1985, A\&A, 149, 475

\bibitem[2003a]{ber03a}Bertoldi F. et al., 2003a, A\&A, 409, L47

\bibitem[2003b]{ber03b}Bertoldi F. et al., 2003b, A\&A, 406, L55

\bibitem[1991]{ber91}Berman B., Suchkov A., 1991, Ap\&SS, 184, 169

%\bibitem[1969]{bev69}Bevington P., 1969, Data Reduction and Error Analysis for
%the Physical Sciences (McGraw-Hill, New York)

\bibitem[1977]{bok77}Boksemberg A., Carswell R., Allen D., Fosbury R.,
Penston M., Sargent W., 1977, MNRAS, 178, 451

\bibitem[1992]{bor92}Boroson T., Green R., 1992, ApJS, 80, 109

\bibitem[1992]{borm92}Boroson T., Meyers K., 1992, ApJ, 397, 442

\bibitem[1991]{bor91}Boroson T., Meyers K., Morris S., Persson S., 1991,
ApJ, 370, L19

%\bibitem[1996]{boy96}Boyce P.J., et al., 1996, ApJ, 473, 760

\bibitem[1981]{bra81}Branch D., Falk S., McCall M., Rybski P., Uomoto A.,
1981, ApJ, 244, 780

\bibitem[1996]{bry96}Bryant P., Scoville N., 1996, ApJ, 457, 678

\bibitem[1999]{bro99}Bromm V., Coppi P., Larson R., 1999, ApJ, 527, L5

\bibitem[2003]{bro03}Bromm V., Loeb A., 2003, Nat preprint (astro--ph 0310622)

\bibitem[1997]{can97}Canalizo G., Stockton A., 1997, ApJ, 480, L5

\bibitem[1998]{can98}Canalizo G., Stockton A., Roth K., 1998, AJ, 115, 890

\bibitem[2001]{can01}Canalizo G.,  Stockton A., 2001, ApJ, 555, 719 

\bibitem[2002]{cap02}Cappeti A.,
2002, in Henney W. Saffe W., Raga A., Binette L. eds., Rev. Mex. Astron.
Astrofis. Serie Conf. Vol. 13, Emission Lines from Jets flows, UNAM Mexico,
p. 163

\bibitem[1999]{car99}Carilli C., Menten K., Yun M., 1999, ApJ, 521, L25

\bibitem[2000]{car00}Carilli C., et al., 2000, ApJ, 533, L13

\bibitem[2002]{car02}Carilli C., Blain A., 2002, ApJ, 569, 605

\bibitem[2004a]{car04a}Carilli C., et al., 2004a, AJ, 128, 997

\bibitem[2004b]{car04b}Carilli C., Bertoldi F., Walter F., Menten K., Beelen A.,
Cox P., Omont A., 2004b, preprint (astro-ph 0402573)

\bibitem[2002]{cec02}Cecil G., Ferruit P., Veilleux S., 
2002, in Henney W. Saffe W., Raga A., Binette L. eds., Rev. Mex. Astron.
Astrofis. Serie Conf. Vol. 13, Emission Lines from Jets flows, UNAM Mexico,
p. 170

\bibitem[1985]{che85}Chevalier R., Clegg A., 1985, Nat, 317, 44

\bibitem[1996]{cle96}Clements D., et al., 1996, MNRAS, 279, 459

\bibitem[2002]{cox01}Cox P., et al., 2002, A\&A, preprint
(astro--ph 0203355)

\bibitem[1991a]{col91a}Colina L., L\'{\i}pari S.L., Macchetto F., 1991a,
ApJ, 379, 113

%\bibitem[1991b]{col91b}Colina L., L\'{\i}pari S.L., Macchetto F., 1991b,
%ApJ, 382, L63

\bibitem[1999]{col99}Colina L., Arribas S., Borne K., 1999, ApJ, 527, L13

\bibitem[1999]{col99}Collin S., Zahn P., 1999, A\&A, 344, 433

\bibitem[2000]{col00}Collin S., Joly M., 2000, New Astron. Rev. 44, 531

\bibitem[1999]{col99}Collin S., Zahn J., 1999, A\&A, 344, 433

\bibitem[1991]{con91}Condon J., Huang Z., Yin Q., Thuan T., 1991, ApJ, 378, 65

%\bibitem[1991]{con91}Conti P. S., 1991, ApJ, 377, 115

%\bibitem[1994]{con94}Conti P. S., Vacca W. D., 1994, ApJ, 423, L97

\bibitem[1984]{cut84}Cutri R., Rieke G., Lebofsky M., 1984, ApJ, 287, 566

\bibitem[2002]{daw02}Dawson S., Spirand H., Stern D., Day A., van Breugel W.,
Vries W., Reuland M., 2002, ApJ, 570, 92

\bibitem[2001]{dek01}de Kool M., Arav N., Becker R., Gregg M., White R.,
Laurent-Muehleisen S., Price T., Korista K., 2001, ApJ, 548, 609

\bibitem[2001]{dek01}de Kool M., Becker R., Arav N., Gregg M., White R.,
 2002, ApJ, 570, 514

\bibitem[1985]{deg85}de Gripj M., Miley G., Lub J., Joung T., 1985, Nat, 314, 240

\bibitem[1987]{deg87}de Gripj M., Miley G., Lub J. , 1987, A\&AS, 70, 95

\bibitem[1986]{dek86}Dekel A., Silk J., 1986, ApJ, 303, 39

\bibitem[2002]{die02}Dietrich M., Appenzeller I., Vestergaard M., Wagner S.,
2002, ApJ, 564, 581

\bibitem[1994]{dop94}Dopita M., 1994, in Bicknell G., Dopita M., Quin P., eds.,
 The Physics of Active Galaxies, (ASP Conf. Series Vol. 54), p. 287

\bibitem[1995]{dop95}Dopita M., Sutherland R., 1995, ApJ, 455, 468

\bibitem[1998]{dow98}Downes D., Solomon P.M., 1998, ApJ, 507, 615

\bibitem[1992]{dys92}Dyson J., Perry J., Williams R., 1992, in Testing the AGN
Paradigm, eds. S. Holt, S. Neff, M. Urry (AIP, New York) 548

%\bibitem[1996]{ega96}Egami E., et al., 1996, AJ, 112, 73

%\bibitem[1994]{els94}Elston R., Thompson K., Hill J., 1994, Nat, 367, 250

\bibitem[2001]{far01}Farrah D. et al., 2001, MNRAS, 326, 1333 

\bibitem[2000]{fer00}Ferrarese L., Merrit D., 2000, ApJ, 539, L9 

\bibitem[1995]{for95}Forster K., Rich M., McKarthy J., 1995, ApJ, 450, 74

\bibitem[2003]{fre03}Freudling W., Corbin M., Korista K., 2003, ApJ, 587, L67

\bibitem[2002]{fry02}Frye B., Broadhurst T., Benitez N., 2002, ApJ, 568, 558

%\bibitem[1998]{fri98}Friaca A., Terlevich R., 1998, MNRAS, 298, 399

\bibitem[2000]{gab00}Gebhardt K. et al., 2000, ApJ, 539, L13

\bibitem[2002]{gal02}Gallagher E., Brandt W., Chartas G., Garmire G.,
Sambruna R., 2002, ApJ, 569, 655

\bibitem[1998]{gen98}Genzel R., et al., 1998, ApJ, 498, 579

\bibitem[2001]{gen01}Genzel R., Tacconi L., Rigopuulou D., Lutz D., Tecza M.,
2001, ApJ, 563, 527

\bibitem[1987]{ham87}Hamilton D., Keel W., 1987, ApJ, 321, 211

\bibitem[1984]{haz84}Hazard C., Morton D., Terlevich R., McMahon R., 1984,
ApJ, 282, 33

\bibitem[1979]{hei79}Heiles C., 1979, ApJ, 229, 533

%\bibitem[1999]{hin99}Hines D., et al., 1999, ApJ, 512, 140

%\bibitem[1995]{hin95}Hines D., Wills B., 1995, ApJ, 448, L69

\bibitem[1980]{hec80}Heckman T.M., 1980, A\&A, 87, 152

\bibitem[1987]{hec87}Heckman T.M., 1987, in E. Ye. Khachikian, K. Fricke, J.
Melnick, eds., Observational Evidence of Activity in Galaxies, (Dordrecht:
Reidel), p. 421

\bibitem[1996]{hec96}Heckman T.M., 1996, in M. Eracleous, A. Koratkar, C.
Leitherer, L. Ho, eds., The Physics of LINERs in View of Recent Observations,
(ASP Conf. Series Vol. 103), p. 241

\bibitem[1987]{hec87}Heckman T.M., Armus L.,  Miley G., 1987, AJ, 93, 276

\bibitem[1990]{hec90}Heckman T.M., Armus L.,  Miley G., 1990, ApJS, 74, 833

\bibitem[1996]{hec96}Heckman T.M., Dahalem M., Eales S., Fabbiano G., Weaver
K., 1996, ApJ, 457, 616

\bibitem[2000]{hec00}Heckman T.M., Lehnert M., Strickland D., Armus L., 2000,
ApJS, 129, 493

%\bibitem[1987]{hei87}Heiles C., 1987, ApJ, 315, 555

%\bibitem[1992]{hei92}Heiles C., 1992, in Proc.Evolution of ISM and
%Dynam. of Galax., eds. J.Palous, W. Burton, and P. Lindblad
%(Cambridge: Camb. Univ. Press), 12

\bibitem[1987]{hut87}Hutching J., Neff S., 1987, AJ, 93, 14

\bibitem[1981]{ike81}Ikeuchi S., 1981, PASJ 33, 211

\bibitem[1986]{ike86}Ikeuchi S., Ostriker J., 1986, ApJ, 301, 522

\bibitem[2002]{iwa02}Iwamuro F., Mtohara K., Maihara T., Kimura M., Yoshi Y.,
Doi M., 2002, ApJ, 565, 63

\bibitem[1985]{jos85}Joseph R.D., Wright G., 1985, MNRAS, 214, 87

\bibitem[1998]{kau98}Kauffmann G., Charlot S., 1998, MNRAS, 294, 705

\bibitem[2003]{kaw03}Kawakatu M., Umemura M., Mori M., 2003, ApJ, 583, 85

%\bibitem[1998]{kims98}Kim D., Sanders D., 1998, ApJS, 119, 41

\bibitem[2002]{kim02}Kim D., Sanders D., 2002, ApJS, 119, 41

%\bibitem[1995]{kimv95}Kim D., Sanders D., Veilleux S., Mazzarella J.,
%Soifer B., 1995, ApJS, 98, 129

\bibitem[1997]{kra97}Krabbe A., Colina L., Thatte N., Kroker H., 1997, ApJ,
476, 98

\bibitem[1992]{kol92}Kollatschny W., Dietrich M., Hagen H., 1992, A\&A, 264,
L5

\bibitem[2000]{kor00}Kormendy J., 2000, Sci., 289, 1484

\bibitem[1992]{kor95}Kormendy J., Richstone D., 1995, ARA\&A, 33, 581 

\bibitem[1992]{kor92}Kormendy J., Sanders D., 1992, ApJ, 390, L53

\bibitem[1974]{lar74}Larson R., 1974, MNRAS, 166, 585

\bibitem[1998]{lar98}Larson R., 1998, MNRAS, 301, 569

\bibitem[2003]{lar03}Larson R., 2003, preprint (astro-ph 0306595) 

%\bibitem[1988]{law88}Lawrence A., et al., 1988, MNRAS, 235, 261

\bibitem[1997]{law97}Lawrence A., et al., 1997, MNRAS, 285, 879

\bibitem[1994]{lip94}L\'{\i}pari S.L., 1994, ApJ, 436, 102

\bibitem[1994]{lcm94}L\'{\i}pari S.L., Colina L.,  Macchetto F., 1994, ApJ, 427, 174

\bibitem[2000a]{lip00a}L\'{\i}pari S.L., Diaz R., Taniguchi Y., Terlevich R.,
Dottori H., Carranza G., 2000a, AJ, 120, 645

\bibitem[2000b]{lip00b}L\'{\i}pari S.L., et al., 2000b, preprint
(astro-ph/0007316)

\bibitem[2003]{lip03}L\'{\i}pari S.L., Terlevich R., Diaz R., Taniguchi Y.,
Zheng W., Tsvetanov Z., Carranza G., Dottori H., 2003, MNRAS, 340, 289

\bibitem[2004]{lip04a}L\'{\i}pari S.L., Mediavilla E, Diaz R., Garcia-Lorenzo B,
Acosta-Pulido J., Ag\"uero M.,Terlevich R., 2004a, MNRAS, 348, 369

\bibitem[2004]{lip04b}L\'{\i}pari S.L., et al., 2004b, in T. Storchi Bergmann,
L. Ho, H. Schmitt, eds., The Interplay among Black Hole Stars and IGM in
Galactic Nuclei, IAU Symp. No. 222, (ASP Conf. Series), p. 529

\bibitem[2004]{lip04c}L\'{\i}pari S.L. et al., 2004c, MNRAS, 354, L1

\bibitem[2004]{lip04d}L\'{\i}pari S.L. et al., 2004d, MNRAS, 355, 641

\bibitem[1991]{lip91a}L\'{\i}pari S.L., Macchetto F.,  Golombeck D., 1991a,
ApJ, 366, L65

\bibitem[1991]{lip91b}L\'{\i}pari S.L., Bonatto Ch.,  Pastoriza M., 1991b,
MNRAS, 253, 19

\bibitem[1992]{lip92a}L\'{\i}pari S.L.,  Macchetto F., 1992a, ApJ, 387, 522

\bibitem[1992]{lip92b}L\'{\i}pari S.L.,  Macchetto F., 1992b, ApJ, 397, 126

\bibitem[1993]{lip93}L\'{\i}pari S.L., Terlevich R.,  Macchetto F., 1993, ApJ, 406, 451

%\bibitem[1997]{lip97}L\'{\i}pari S.L., Tsvetanov Z.,  Macchetto F., 1997, ApJS, 111, 369

\bibitem[2003]{lon03}Lonsdale C.J., Lonsdale C.J., Smith H., Diamond P.,
2003, ApJ, 592, 804

\bibitem[1988]{low88}Low F., Cutri R., Huchra J., Kleinmann S., 1988, ApJ, 327, L41

\bibitem[1989]{low89}Low F., Cutri R., Kleinmann S., Huchra J., 1989, ApJ, 340, L1

\bibitem[1999]{lut99}Lutz D., Veilleux S., Genzel R., 1999, ApJ, 517, L13

\bibitem[1998]{mag98}Magorrian J. et al., 1998, AJ, 115, 2285

\bibitem[2003]{mai03}Maiolino R., Juarez Y., Mujica R., Nagar N., Oliva E.,
2003, ApJ, 596, L155

\bibitem[2004a]{mai04a}Maiolino R., Oliva E., Ghinassi F., PedaniM., Mannucci F.,
Mujica R., Juarez Y., 2004a, A\&A, preprint (astro-ph 0312402)

\bibitem[2004b]{mai04b}Maiolino R., Schneider R., Oliva E., Bianchi S. Ferrara A.
Mannucci F., Pedani M., Roca Sogorb M., 2004b, Nat, preprint (astro-ph 0409577)

\bibitem[1969]{mar69}Markarian B. E., 1969, Astrophyzika, 5, 286

\bibitem[1999]{mar99}Martin C., 1999, ApJ, 513, 156

\bibitem[1987]{mcc87}McCarthy P., Heckman T., van Breugel W., 1987, AJ, 93, 264

\bibitem[1989]{mac89}Mac Low M., McCray R., Norman M., 1989, ApJ, 337, 141

%\bibitem[1989]{mat89}Mathews W., 1989, AJ, 97, 42

\bibitem[1971]{mat71}Mathews W., Baker J., 1971, ApJ, 170, 241

%\bibitem[1992]{mat92}Mathews W., Dones J., 1992, Lick Obs. Bull., preprint

%\bibitem[1987]{mat87}Matteuchi F., Tornambe A., 1987, A\&A, 185, 51

\bibitem[1980]{mea80}Meaburn J., 1980, MNRAS, 192, 365

\bibitem[1990]{mel90}Melnick J.,  Mirabel I.F., 1990, A\&A, 231, L19

\bibitem[2001]{mur01}Murphy T., Soifer B., Matthews K., Armus L., 2001, ApJ, 559, 201

\bibitem[1995]{mur95}Murray N.,  Chiang J., Grossman S., Voit G., 1996, ApJ, 464, 641

\bibitem[1977]{mck77}McKee C., Ostriker J., 1977, ApJ, 218, 148

\bibitem[1990]{mel90}Melnick J., Mirabel I.F., 1990, A\&A, 231, L19

\bibitem[1998]{men98}Mendes de Olivera C., et al., 1998, ApJ, 507, 691

\bibitem[2001]{mer01}Merrit D., Ferrarese L., 2001, MNRAS, 320, L30

\bibitem[1994a]{mih94a}Mihos C., Hernquist L., 1994a, ApJ, 425, L13

\bibitem[1994b]{mih94b}Mihos C., Hernquist L., 1994b, ApJ, 431, L9

\bibitem[1996]{mih96}Mihos C.,  Hernquist L., 1996, ApJ, 464, 641

\bibitem[1992]{mih92}Mihos C.,  Richstone D., Bothun G., 1992, ApJ, 400, 153

\bibitem[2003]{mor03}Morganti R., Oosterloo T., Emonts B., van der Hulst J.,
Tadhunter C., 2003, ApJ, 593, L69

\bibitem[1992]{mos92}Moshier M. et al, 1992, Explanatory Supplement to the
Faint IRAS Source Survey, Version 2 (Pasadena, Jet Propulsion Laboratory)

\bibitem[1995]{mur95}Murray N., et al., 1995, ApJ, 451, 498

\bibitem[1988]{nef88}Neff S., Ulvestad J., 1988, AJ, 96, 841

\bibitem[1989]{nor89}Norman C., Ikeuchi S., 1989, ApJ, 395, 372

\bibitem[1988]{nor88}Norman C., Scoville N., 1988, ApJ, 332, 124

\bibitem[2001]{omo01}Omont A., et al., 2001, A\&A, preprint
(astro--ph 0107005)

\bibitem[1981]{ost81}Ostriker J., Cowie L., 1981, ApJ, 243, L127

\bibitem[1992]{per92}Perry J., 1992, in Relationships Between AGN and Starburst
Galaxies, ed. A. Filippenko (ASP Conf.S.31, San Francisco) 169

\bibitem[1992]{perd92}Perry J., Dyson R., 1992,
in Testing the AGN Paradigm, eds. S. Holt, S. Neff, M. Urry
(AIP, New York) 553

\bibitem[1983]{pre83}Preuss E., Fosbury R., 1983, MNRAS, 204, 783

\bibitem[20001]{pet01}Pettini M., Shapley A., Steidel C., Cuby J., Dickinson
M., Moorwood A., Adelberg K., Giavalisco M., 2001, ApJ, 554, 981

\bibitem[2003]{rei03}Reichard T., et al., 2003, AJ, 125, 1711

%\bibitem[1977]{ree77}Rees M.,, 1977, ARA\&A, 42, 471

\bibitem[1977]{reeo77}Rees M., Ostriker J., 1977, MNRAS, 179, 541

\bibitem[1972]{rie72}Rieke G., Low F. J., 1972, ApJ, 176, L95

\bibitem[1975]{rie75}Rieke G., Low F. J., 1975, ApJ, 200, L67

\bibitem[1985]{rie85}Rieke G., et al., 1985, ApJ, 290, 116

\bibitem[1999]{rig99}Rigopuulou D., Spoon H., Genzel R., Lutz D.,
Moorwood A., Tran Q., 1999, AJ, 118, 2625
                                     
\bibitem[1989]{row89}Rowan-Robinson M., Crawford J., 1989, MNRAS, 238, 523

\bibitem[2002]{rup02}Rupke D., Veilleux S., Sanders D., 2002, ApJ, 570, 588

\bibitem[1996]{san96}Sanders D.B.,  Mirabel F., 1996, ARA\&A, 34, 749
     
\bibitem[1987]{san87}Sanders D.B., Scoville N., Soifer B., Young J.,
Danielson G., 1987, ApJ, 312, L5

\bibitem[1991]{san91}Sanders D.B., Scoville N., Soifer B.T., 1991, ApJ, 370, 158

\bibitem[1988a]{san88a}Sanders D.B., Soifer B.T., Elias J.H., Madore B.F.,
Matthews K., Neugebauer G., Scoville N.Z., 1988a, ApJ, 325, 74

\bibitem[1988b]{san88b}Sanders D.B., Soifer B.T., Elias J.H., Neugebauer G.,
  Matthews K., 1988b, ApJ, 328, L35

%\bibitem[1999]{sch99}Schaerer D., Contini T.   Pindao M., 1999, A\&AS 136, 35

%\bibitem[1998]{sch98}Schinnerer A., Eckart A., Tacconi L., 1998, ApJ, 500, 147

\bibitem[1985]{sch85}Schmidt G., Miller J., 1985, ApJ, 290, 517

\bibitem[1992]{sco92}Scoville N.Z., 1992, in Relationships Between AGN
and Starburst Galaxies, ed. A. Filippenko (ASP Conf.S.31, San
Francisco) 159

\bibitem[1996]{sco96}Scoville N.Z., Norman C., 1996, ApJ, 451, 510

\bibitem[1991]{sco91}Scoville N.Z., Sargent A., Sanders D.,
Soifer B., 1991, ApJ, 366, L5

\bibitem[1991]{scos91}Scoville N.Z., Soifer B.T., 1991, in Massive
Stars in Starbursts, eds. C. Leitherer, N. Walborn, T.M. Heckman,
C. Norman (Cambridge Univ. Press) 233

\bibitem[1987]{sek87}Sekiguchi K., 1987, ApJ, 316, 645

\bibitem[1977]{sil77}Silk J., 1977, ApJ, 211, 638

\bibitem[1998]{shi98}Shier L., Fischer J., 1998, ApJ, 497, 163

\bibitem[1980]{schw80}Schweizer F., 1980, ApJ, 237, 303

\bibitem[1982]{schw82}Schweizer F., 1982, ApJ, 252, 455

\bibitem[1996]{schw96}Schweizer F., et al., 1996, AJ, 111, 109

\bibitem[1995]{smi95}Smith P. S., Schmidt G., Allen R., Angel J., 1995, ApJ, 444, 146

\bibitem[1998]{smi98}Smith H., Lonsdale C., Lonsdale C., Diamond P., 1998, ApJ, 493, L17

\bibitem[2003]{sol03}Solomon P., Vaden Bout P., Guelin M., 2003, Nat, 426, 636

\bibitem[1992]{spr92}Sprayberry D., Foltz C., 1992, ApJ, 390, 39

\bibitem[1998]{sto98}Stockton A., Canalizo G., Close L., 1998, ApJ, 500, L121

\bibitem[2000]{str00}Strickland, D, Stevens, I., 2000, MNRAS, 314, 511

\bibitem[2000]{sul00}Sulentic J., Marziani P., Dultzin--Hacyan D., 2000, ARA\&A,
38, 521

\bibitem[1994]{suc94}Suchkov A., Balsara D., Heckman T., Leitherer C., 1994, ApJ, 430, 511

\bibitem[1998a]{sur98a}Surace J., 1998, PhD Thesis, Univ. of Hawaii

\bibitem[1998]{sur98}Surace J., et al., 1998, ApJ, 492, 116

%\bibitem[1994]{tan94}Taniguchi Y. et al., 1994, AJ, 107, 1668

%\bibitem[1999a]{tan99a}Taniguchi Y. et al., 1999a, ApJ, 514, 660

%\bibitem[1999b]{tan99b}Taniguchi Y., Ikeuchi S., Shioya K., 1999b, ApJ, 514, L9

\bibitem[2000]{tan00}Taniguchi Y., Shioya K., 2000, ApJ, 532, L12

\bibitem[1999]{tay99}Taylor G., Silver C., Ulvestad J., Carrilli C., 1999,
ApJ, 519, 185

\bibitem[1988]{ten88}Tenorio-Tagle G., Bodenheimer P., 1988, ARA\&A, 26, 145

\bibitem[1990]{ten90}Tenorio-Tagle G., Rozyczka M., Bodenheimer P., 1990,
A\&A, 237, 207

\bibitem[1992]{ter92}Terlevich R., et al., 1992, MNRAS, 255, 713

\bibitem[2002]{ter02}Terlevich R., L\'{\i}pari S., Sodre L., 2002, in preparation

\bibitem[1993]{ter93}Terlevich R. et al., 1993, in First Light in the Universe:
Star or QSOs, eds. B. Rocca-Volmerange, M. Dennefeld, B. Guiderdoni, 
Tran Thanh Van (Editions Frontieres), p 261

\bibitem[1977]{too77}Toomre A., 1977, in The Evolution of Galaxies and
Stellar Population, eds.B. Tinsley   R. Larson (Yale University Observatory, New Haven) 401

\bibitem[1972]{too72}Toomre A.   Toomre J., 1972, ApJ, 179, 623

\bibitem[1988]{tom88}Tomisaka K., Ikeuchi S., 1988, ApJ, 330, 695

\bibitem[1999]{tur99}Turner T., 1999, ApJ, 511, 142

\bibitem[1997]{tur97}Turnsheck D., Monier E., SirolaC., Espey B., 1997, ApJ,
476, 40

\bibitem[1972]{ulr72}Ulrich M., 1972, ApJ, 178, 113

\bibitem[1999a]{ulv99a}Ulvestad J., Wrobel J., Carilli C., 1999a, ApJ, 516, 127

\bibitem[1999b]{ulv99b}Ulvestad J., Wrobel J., Roy A., Wilson A., Falcke H.,
Krichbaun T., 1999b, ApJ, 517, L81

\bibitem[1986]{vad87}Vader J.P., 1986, ApJ, 305, 669

\bibitem[1987]{vad87}Vader J.P., 1987, ApJ, 317, 128

%\bibitem[1994]{vei94}Veilleux S., Bland-Hawthorn J., Tully R., Filippenko A.,
%Sargent W., 1994, ApJ, 433, 48

%\bibitem[1995]{vei95}Veilleux S., Kim D., Sanders D., Mazzarella J.,
%Sifer B., 1995, ApJS, 98, 171

\bibitem[1997]{vei97}Veilleux S., Bland--Hawthorn J., 1997, ApJ, 479, L105

\bibitem[1999]{vei99}Veilleux S., Kim D., Sanders D., 1999, ApJ, 522, 113

\bibitem[2002a]{vei02a}Veilleux S., Cecil G., Bland--Hawthorn J., Shopell P.,
2002a, in Henney W., Steffe W., Raga A., Binette L., eds., Rev. Mex. Astron.
Astrofis. Serie Conf. Vol.. 13, Emission Lines from Jets FGlows. UNAM, Mexico,
222

\bibitem[2002b]{vei02b}Veilleux S., Kim D., Sanders D., 2002b, APJS, 143, 315

%\bibitem[1999]{vie99}Viegas S., Contini M.   Contini T., 1999, A\&A, 347, 112

\bibitem[1993]{voi93}Voit G., Weymann R.,  Korista K., 1993, ApJ, 413, 95

\bibitem[1997]{wan97}Wang J., Heckman T., Weaver K., Armus L., 1997, ApJ, 474, 659

\bibitem[1973]{wee73}Weedman D., 1973, ApJ, 183, 29

%\bibitem[1983]{wee83}Weedman D., 1983, ApJ, 266, 479

%\bibitem[1991]{wey91}Weymann R., et al., 1991, ApJ, 373, 23

%\bibitem[1999]{whi99}Whitmore B., et al., 1999, AJ, 118, 1551

%\bibitem[1992]{wil92}Wills B., et al., 1992, ApJ, 400, 96

\bibitem[1990]{wri90}Wright G. et al., 1990, Nat, 344, 417

\bibitem[2002]{zhe02}Zheng X., Xia X., Mao S., Deng Z., 2002, AJ, 124, 18


\end{thebibliography}
\end{document}